\documentclass[english,fleqn]{article}
\usepackage[T1]{fontenc}
\usepackage[latin9]{inputenc}
\usepackage{array}
\usepackage{rotating}
\usepackage{subfigure}
\usepackage{float}
\usepackage{color}
\usepackage{graphicx}
\usepackage{amssymb}
\usepackage{esint}

\makeatletter

\providecommand{\tabularnewline}{\\}

\newcommand{\lyxaddress}[1]{
\par {\raggedright #1
\vspace{1.4em}
\noindent\par}
}

\makeatother

\usepackage{babel}

\begin{document}

\title{The Carter Constant for Inclined Orbits About a Massive Kerr Black
Hole: I. circular orbits}

\author{P. G. Komorowski{*}, S. R. Valluri{*}\#, M. Houde{*}}

\maketitle

\lyxaddress{{*}Department of Physics and Astronomy, \#Department of Applied Mathematics,
University of Western Ontario, London, Ontario}

\begin{abstract}
In an extreme binary black hole system, an orbit will increase its
angle of inclination (${\iota}$) as it evolves in Kerr spacetime.
We focus our attention on the behaviour of the Carter constant (${Q}$)
for near-polar orbits; and develop an analysis that is independent
of and complements radiation reaction models. For a Schwarzschild
black hole, the polar orbits represent the abutment between the prograde
and retrograde orbits at which ${Q}$ is at its maximum value for
given values of latus rectum (${\tilde{l}}$) and eccentricity (${e}$).
The introduction of spin (${\tilde{S}}={\left|\mathbf{J}\right|}/{M}^{2}$)
to the massive black hole causes this boundary, or abutment, to be
moved towards greater orbital inclination; thus it no longer cleanly
separates prograde and retrograde orbits. 

To characterise the abutment of a Kerr black hole (KBH), we first
investigated the last stable orbit (LSO) of a test-particle about
a KBH, and then extended this work to general orbits. To develop a
better understanding of the evolution of ${Q}$ we developed analytical
formulae for ${Q}$ in terms of ${\tilde{l}}$, ${e}$, and ${\tilde{S}}$
to describe elliptical orbits at the abutment, polar orbits, and last
stable orbits (LSO). By knowing the analytical form of $\partial{Q}/\partial{\tilde{l}}$
at the abutment, we were able to test a 2PN flux equation for Q. We
also used these formulae to numerically calculate the $\partial{\iota}/\partial{\tilde{l}}$
of hypothetical circular orbits that evolve along the abutment. From
these values we have determined that $\partial{\iota}/\partial{\tilde{l}}=-\left(122.7{\tilde{S}}-36{\tilde{S}}^{3}\right){\tilde{l}}^{-11/2}-\left(63/2\,{\tilde{S}}+35/4\,{\tilde{S}}^{3}\right){\tilde{l}}^{-9/2}-15/2\,{\tilde{S}}{\tilde{l}}^{-7/2}-9/2\,{\tilde{S}}{\tilde{l}}^{-5/2}$.
By taking the limit of this equation for ${\tilde{l}}\rightarrow\infty$,
and comparing it with the published result for the weak-field radiation-reaction,
we found the upper limit on $\left|\partial{\iota}/\partial{\tilde{l}}\right|$
for the full range of ${\tilde{l}}$ up to the LSO. Although we know
the value of $\partial{Q}/\partial{\tilde{l}}$ at the abutment, we
find that the second and higher derivatives of ${Q}$ with respect
to ${\tilde{l}}$ exert an influence on $\partial{\iota}/\partial{\tilde{l}}$.
Thus the abutment becomes an important analytical and numerical laboratory
for studying the evolution of ${Q}$ and ${\iota}$ in Kerr spacetime
and for testing current and future radiation back-reaction models
for near-polar retrograde orbits.
\end{abstract}
\setlength{\mathindent}{2.5cm}

\section{Introduction}

In his landmark work of 1968, Brandon Carter derived a new constant
of motion that pertained to orbital motion in the gravitational field
of a Kerr black hole (KBH) \cite{1968PhRv..174.1559C}. In due course,
this constant became known as the Carter constant, which joins the
set of important constants of motion: orbital angular momentum (${L}_{z}$,
${z}$-axis projection), orbital energy (${E}$), and finally the
Carter constant (${Q}$). These constants of motion can be developed
rigorously from the Hamilton-Jacobi equation \cite{1968PhRv..174.1559C,1973blho.conf...57C,H.:2005zl}. 

An extreme mass ratio binary black hole system is composed of a secondary
object (which may be a compact object (CO) of several solar masses)
in orbit around a primary object (which is a massive black hole (MBH)
of several million solar masses). Extreme mass ratio inspirals (EMRIs)
are expected to emit gravitational wave radiation (GW) of sufficiently
high energy and in the appropriate frequency band for detection by
the Laser Interferometer Space Antenna (LISA) to be feasible \cite{PhysRevD.61.084004,2006PhRvD..73f4037G,2006PhRvD..73h3006M,2007PThPh.117.1041G}.
The emission of GW causes the constants of motion to evolve, which
in turn affects the GW power spectrum. Therefore some useful methods
have been developed to describe this evolution. For example, the quadrupole
formalism \cite{PhysRev.131.435,PhysRev.136.B1224,Thorne:1987fk,Barack:2004uq}
and the Teukolsky equation \cite{1973ApJ...181..513P,1973ApJ...185..635T,1973ApJ...185..649P}
have yielded important results. The analytical description of the
evolution of ${Q}$ has been more difficult to achieve than it has
for the other two constants of motion \cite{PhysRevD.55.3444}; although
the use of the Teukolsky equation has shown great promise \cite{1995PhLA..202..347O,PhysRevD.55.3444,PhysRevD.61.084004,2006PhRvD..73f4037G,2007PThPh.117.1041G}
in this endeavour. 

As the CO inspirals, the gravitational radiation reaction causes the
value of ${Q}$ to change \cite{1995PhLA..202..347O,Ryan:1996ly,1996PhRvD..53.4319K,PhysRevD.55.3444,PhysRevD.61.084004,2001PhRvD..64f4004H,2002PhRvD..66d4002G,2002PhRvD..66f4005G,2006PhRvD..73f4037G,2007PhRvD..76d4007B,2007PThPh.117.1041G}.
Therefore a non-equatorial orbit lists as its angle of inclination,
${\iota}$, increases with respect to time; a near-polar prograde
orbit becomes polar, and ultimately retrograde \cite{Ryan:1995ix,Ryan:1996ly}.
Such listing behaviour of an inclined orbit has been studied and confirmed
using the most current Teukolsky-based fluxes \cite{2007PhRvD..76d4007B}.
It is our goal to develop an analytical and numerical methodology
for testing and improving radiation reaction models for predicting
orbit listing and inspirals for near-polar orbits.

Our interest lies in studying KBH systems; yet, the Schwarzschild
black hole (SBH) is an important datum. An infinitesimal amount of
spin angular momentum (${\delta}{\tilde{S}}\ll1$) may be imparted
to an MBH such that, for practical purposes, it can be regarded as
an SBH (by virtue of its minuscule effect on the surrounding spacetime);
and yet, the spherical symmetry of the system has been broken and
a ${z}$-axis defined. Then an SBH can be considered to have a prograde
or retrograde inclined orbit; and the set of polar orbits define the
abutment, at which ${Q}$ will be at its maximum value (${Q}$ is
non-negative for any bound orbit), for given values of ${\tilde{l}}$,
${e}$, and ${\tilde{S}}$ (\textit{ceteris paribus}).

In section \ref{sec:The-Motion-of-a-Test-Particle} the motion of
a test-particle in an inclined orbit is analysed from first principles
\cite{1973blho.conf..215B,1983bhwd.book.....S,1983mtbh.book.....C,Schutz:2005fk}
to yield the effective radial potential and an analytical expression
of ${\tilde{L}}_{z}$ for a last stable orbit (LSO). In section \ref{sec:Analysis-of-the-Trajectory-Equations}
we continue our analysis to find the roots of the effective radial
and polar-angle potentials and use them to derive analytical expressions
for ${\tilde{E}}$ and ${X}^{2}$ (where ${X}={\tilde{L}}_{z}-{\tilde{S}}{\tilde{E}}$).
The concept of the abutment is then refined. In section \ref{sec:The-characteristics-of-the-Carter-Constant-Equations},
we derive a set of critical formulae that express ${Q}$ at the LSO,
along the abutment, and for the set of polar orbits. The interrelationships
between these formulae are examined and a map of admissible values
of ${Q}$, with respect to ${\tilde{l}}$ and ${e}$ is drawn. In
section \ref{sec:The-Analysis-of-the-Carter-Constant}, this map is
used to better understand the path in the ${Q}$-${\tilde{l}}$ plane
that is followed by an evolving circular orbit. We demonstrate the
importance of the first and second derivatives of ${Q}$ (on the abutment)
with respect to ${\tilde{l}}$ for understanding the rate of change
of ${\iota}$ as the orbit lists.

We shall use geometrical units by \textcolor{black}{setting the speed
of light and gravitational constant to unity (i.e. ${c}=1$ and ${G}=1$);
therefore, mass-energy is in units of time (seconds).} In addition,
many of the parameters we use will be normalised with respect to the
mass of the black hole (${M}$) or with respect to the test-particle
mass (${m}$). In appendix A, the symbols used in this paper are tabulated.
By emphasising normalised variables, the analytical equations and
numerical formalism are much better handled.

\section{\label{sec:The-Motion-of-a-Test-Particle}The Motion of a Test-Particle
in an Inclined Orbit}

A sound mathematical analysis can be made on the assumption that the
secondary body is of infinitesimal mass (i.e. a test-particle). In
such a case, the background metric of the MBH dominates. In the case
of EMRIs, the small ratio of the CO mass to the MBH mass (${\eta}\lessapprox10^{-5}$)
warrants our use of idealised test-particle calculations \cite{PhysRevD.61.084004,2000PhRvD..62l4022O,2006PhRvD..73f4037G,2007PThPh.117.1041G}.

\subsection{\label{sub:Basic-orbital-equations}Basic orbital equations}

We begin by considering a test-particle in orbit about a KBH of arbitrary
spin, ${\tilde{S}}$, for which the four-momentum can be given the
general definition \cite{Komorowski:2008wc},\begin{eqnarray}
P_{\gamma} & = & \biggl[-{m}{\tilde{E}},\: m\frac{{\tilde{\Sigma}}}{{\tilde{\Delta}}}\left({\it \frac{dR}{d\tilde{\tau}}}\right),\:{m}{M}{\tilde{L}}_{\theta},\:{m}{M}{\tilde{L}}_{z}\biggr],\label{eq:FourMomentumNormalised}\end{eqnarray}
where\begin{eqnarray}
{\tilde{\Delta}} & = & \left({R}^{2}-2\, R+{\tilde{S}}^{2}\right)\end{eqnarray}
and

\begin{eqnarray}
{\tilde{\Sigma}} & = & {\left(R^{2}+\tilde{S}^{2}{\cos^{2}\left(\theta\right)}\right)}.\label{eq:Sigma_R_Theta}\end{eqnarray}
Unlike the analysis in Komorowski et al. \cite{Komorowski:2008wc},
we shall use normalised variables at the outset and offer a more thorough
treatment. Because we are now considering inclined elliptical orbits,
one cannot simplify the four-momentum by setting ${\tilde{L}}_{\theta}=0$.
But by knowing the Carter constant in terms of normalised variables
(i.e. obtained by dividing through by ${m}{M}$), \begin{eqnarray}
{Q} & = & {\frac{\cos^{2}\left(\theta\right){\tilde{L}_{{z}}}^{2}}{\sin^{2}\left(\theta\right)}}+{\tilde{L}_{{\theta}}}^{2}+\cos^{2}\left(\theta\right){\tilde{S}}^{2}\left(1-{\tilde{E}}^{2}\right),\end{eqnarray}
one can obtain the component of orbital angular momentum, ${\mathbf{L}}$,
projected upon the equatorial plane of the KBH,

\begin{eqnarray}
{\tilde{L}}_{\theta} & = & \sqrt{{Q}-{\frac{\cos^{2}\left(\theta\right){\tilde{L}_{{z}}}^{2}}{\sin^{2}\left(\theta\right)}}-\cos^{2}\left(\theta\right){\tilde{S}}^{2}\left(1-{\tilde{E}}^{2}\right)},\label{eq:Theta}\end{eqnarray}
and substitute it into the expression for the four-momentum:\setlength{\mathindent}{0cm}

\noindent \begin{eqnarray}
P_{\gamma} & =\label{eq:FourMomentum_XQ}\\
 & \biggl[-{m}{\tilde{E}},m\frac{{\tilde{\Sigma}}}{{\tilde{\Delta}}}\left({\it \frac{dR}{d\tilde{\tau}}}\right),{m}{M}{\sqrt{{Q}-{\frac{\cos^{2}\left(\theta\right){\tilde{L}_{{z}}}^{2}}{\sin^{2}\left(\theta\right)}}-\cos^{2}\left(\theta\right){\tilde{S}}^{2}\left(1-{\tilde{E}}^{2}\right)}},{m}{M}{\tilde{L}}_{z}\biggr].\nonumber \end{eqnarray}
\setlength{\mathindent}{2.5cm}The invariant quantity,

\begin{eqnarray}
\vec{P}\cdot\vec{P} & = & \left.P_{\gamma}P_{\delta}g^{\delta\gamma}\right|_{Kerr}=-{m}^{2},\label{eq:PdotP_orbital_element}\end{eqnarray}
is calculated tensorially using the inverse Kerr metric (see Appendix
B) and used to develop the radial orbital equation for a test-particle:\begin{eqnarray}
{\tilde{\Sigma}}^{2}{\left(\frac{dR}{d\tilde{\tau}}\right)}^{2} & = & -\left({1-{\tilde{E}}^{2}}\right){R}^{4}+2\,{R}^{3}\nonumber \\
 & - & \left({\tilde{L}_{z}}^{2}+{\tilde{S}}^{2}{\left(1-{\tilde{E}}^{2}\ \right)}+Q\right){R}^{2}\nonumber \\
 & + & 2\,\left(\left({\tilde{L}_{z}-{\tilde{S}}{\tilde{E}}}\right)^{2}+Q\right)R-{Q{\tilde{S}}^{2}}.\label{eq:Orbital_Equation_Q}\end{eqnarray}
By setting ${Q}=0$, equation (\ref{eq:Orbital_Equation_Q}) reduces
to the equation for an equatorial orbit (see \cite{2002PhRvD..66d4002G}) 

At the radial turning points, $d{R}/d{\tilde{\tau}}=0$. Equation
(\ref{eq:Orbital_Equation_Q}) becomes:\begin{eqnarray}
0 & = & {R}^{4}-2\,{\frac{{R}^{3}}{1-{\tilde{E}}^{2}}}\nonumber \\
 & + & {\frac{\left({\tilde{L}_{z}}^{2}+{\tilde{S}}^{2}{\left(1-{\tilde{E}}^{2}\ \right)}+Q\right){R}^{2}}{1-{\tilde{E}}^{2}}}\nonumber \\
 & - & 2\,{\frac{\left(\left({\tilde{L}_{z}-{\tilde{S}}{\tilde{E}}}\right)^{2}+Q\right)R}{1-{\tilde{E}}^{2}}}+{\frac{Q{\tilde{S}}^{2}}{1-{\tilde{E}}^{2}}}.\label{eq:Orbital_Equation_Q_Turning_Point}\end{eqnarray}

In the limit ${\tilde{S}}\rightarrow0$ (set ${\tilde{S}}=0$ while
retaining a non-zero value for ${Q}$) equation (\ref{eq:Orbital_Equation_Q_Turning_Point})
becomes\begin{eqnarray}
0 & = & {R}^{4}-2\,{\frac{{R}^{3}}{1-{\tilde{E}}^{2}}}+{\frac{\left({\tilde{L}_{z}}^{2}+Q\right){R}^{2}}{1-{\tilde{E}}^{2}}}-2\,{\frac{\left({\tilde{L}_{z}}^{2}+Q\right)R}{1-{\tilde{E}}^{2}}}.\label{eq:Orbital_Equation_Q_S_is_zero}\end{eqnarray}
Thus, the square of the total orbital angular momentum, ${{\tilde{L}}^{2}}={\tilde{L}_{z}}^{2}+Q$,
confirms that, for the specific case of an SBH, ${Q}$ represents
the square of the component of angular momentum projected on $x-y$
plane of the coordinate system (see Appendix B in Schmidt \cite{2002CQGra..19.2743S}
and Appendix C in this paper for a more detailed treatment).

Some important research \cite{2002PhRvD..66f4005G,2008PhRvD..77l4050S}
has been performed by working with the orbital inclination angle,
${\iota}$, instead of ${Q}$; but in our study, the value of ${Q}$
will be taken as a system parameter. If ${Q}$ is chosen to be zero,
then the orbital plane coincides with the equatorial plane of the
KBH (i.e. ${\iota}=0$ and ${\theta}\equiv\pi/2$) and for a test-particle
in a polar orbit (i.e. ${\iota}=\frac{\pi}{2}$ and $0\leq\theta\leq\pi$)
${\tilde{L}}_{z}$ must vanish. The choice of working directly with
the Carter constant, ${Q}$, as a system parameter is consistent with
the approach taken by Carter \cite{1968PhRv..174.1559C} and more
recently emphasised by others \cite{2006PhRvD..73b4027D,2006PhRvD..73f4037G,2007PhRvD..76d4007B}.

\subsection{Effective Radial Potential}

To proceed, we use a version of equation (\ref{eq:Orbital_Equation_Q}),
which is quadratic in, ${\tilde{E}}$, \begin{eqnarray}
-R\left({R}^{3}+R{\tilde{S}}^{2}+2\,{\tilde{S}}^{2}\right){\tilde{E}}^{2}+4\, R\tilde{L}_{{z}}{\tilde{S}}{\tilde{E}}\nonumber \\
+R\left(R-2\right)\left(Q+{\tilde{L}_{{z}}}^{2}\right)+{R}^{2}\left({R}^{2}-2\, R+{\tilde{S}}^{2}\right)+Q{\tilde{S}}^{2} & = & 0.\label{eq:QuadraticInE}\end{eqnarray}
The roots of this equation can be used to determine the effective
potential of the test-particle (${\tilde{V}}_{\pm}$):\begin{eqnarray}
{\tilde{V}}_{\pm} & = & \left(2\, R{\tilde{L}}_{{z}}\tilde{S}\pm\sqrt{{R}{Z}{\tilde{\Delta}}}\right)\nonumber \\
 & \times & \left(R\left({R}^{3}+{\tilde{S}}^{2}R+2\,{\tilde{S}}^{2}\right)\right)^{-1}\label{eq:EffectivePotential}\end{eqnarray}
where\begin{eqnarray*}
{Z} & = & {R}^{5}+\left(Q+{\tilde{S}}^{2}+{\tilde{L}_{{z}}^{2}}\right){R}^{3}+2\,{R}^{2}{\tilde{S}}^{2}+{\tilde{S}}^{2}RQ+2\, Q{\tilde{S}}^{2}.\end{eqnarray*}
When the last stable orbit (LSO) is reached, ${\tilde{E}}$ corresponds
to a local maximum of ${\tilde{V}}_{+}$ closest to the event horizon.
Therefore one calculates the derivative of ${\tilde{V}}_{+}$ with
respect to ${R}$ and equates it to zero, i.e. 

\begin{eqnarray}
\frac{d{\tilde{V}}_{+}}{d{R}} & =- & \Biggl(2R\tilde{S}\tilde{L}_{{z}}\left(3\,{R}^{2}+{\tilde{S}}^{2}\right)\sqrt{{R}{Z}{\tilde{\Delta}}}+{R}^{3}{Z}_{1}{\tilde{L_{z}}^{2}}\nonumber \\
\nonumber \\ &  & +\left({R}^{3}+{\tilde{S}}^{2}R+2\,{\tilde{S}}^{2}\right){Z}_{2}\Biggr)\nonumber \\
 & \times & \Biggl({R}\sqrt{{R}{Z}{\tilde{\Delta}}}{\left({R}^{3}+{\tilde{S}}^{2}R+2\,{\tilde{S}}^{2}\right)^{2}}\Biggr)^{-1}\nonumber \\
 & = & 0\label{eq:diffVbyR}\end{eqnarray}
where\begin{eqnarray*}
{Z}_{1} & ={R}^{5}-3\,{R}^{4}+{\tilde{S}}^{2}{R}^{3} & -3\,{R}^{2}{\tilde{S}}^{2}+6\,{\tilde{S}}^{2}R-2\,{\tilde{S}}^{4},\end{eqnarray*}
and\begin{eqnarray*}
{Z}_{2} & = & -{R}^{6}+{R}^{5}Q-\left(2\,{\tilde{S}}^{2}+3\, Q\right){R}^{4}+\left(2\, Q{\tilde{S}}^{2}+4\,{\tilde{S}}^{2}\right){R}^{3}\\
 & - & \left({\tilde{S}}^{4}+2\, Q{\tilde{S}}^{2}\right){R}^{2}+{\tilde{S}}^{4}QR+{\tilde{S}}^{4}Q.\end{eqnarray*}

\subsection{Orbital Angular Momentum at the Last Stable Orbit}

We can now develop an equation for the ${\tilde{L}}_{z}$ of a test-particle
in an inclined orbit about a KBH (in a manner similar to that described
in \cite{Komorowski:2008wc}) and extend the concept to general orbits.
It should be noted, the value of ${\tilde{L}}_{z}$ considered here
is not valid for general orbits, but pertains to the LSO. After eliminating
the square root in equation (\ref{eq:diffVbyR}) to yield a new equation
that is quadratic in, ${{\tilde{L}}_{z}}^{2}$, the solution is found
to be:

\begin{eqnarray}
{{\tilde{L}}_{z}}^{2}=\Bigl\{-{R}^{8}+\left(3+Q\right){R}^{7}+\left(-2\,{\tilde{S}}^{2}-6\, Q\right){R}^{6}\nonumber \\
+\left(\left(-6+2\, Q\right){\tilde{S}}^{2}+9\, Q\right){R}^{5}+\left(-{\tilde{S}}^{4}+\left(-10\, Q+12\right){\tilde{S}}^{2}\right){R}^{4}\nonumber \\
+\left(\left(-5+Q\right){\tilde{S}}^{4}+6\, Q{\tilde{S}}^{2}\right){R}^{3}-6\,{\tilde{S}}^{4}{R}^{2}Q\nonumber \\
+5\,{\tilde{S}}^{4}{R}Q-2\,{\tilde{S}}^{6}Q\nonumber \\
\pm2{\tilde{S}}{\left(3\,{R}^{2}+{\tilde{S}}^{2}\right)}{\tilde{\Delta}}\sqrt{\left({R}^{5}-{R}^{4}Q+3\,{R}^{3}Q+{Q}^{2}{\tilde{S}}^{2}\right)}\Bigr\}\nonumber \\
\biggl({R}^{4}\left({R}^{3}+9\, R-4\,{\tilde{S}}^{2}-6\,{R}^{2}\right)\biggr)^{-1},\label{eq:LzSquared}\end{eqnarray}
which provides a relationship between ${\tilde{L}}_{z}$ and ${R}$
(which here, represents the pericentre radius) of the test-particle
LSO. This result is independent of whether one begins the calculation
with ${\tilde{V}}_{+}$ or ${\tilde{V}}_{-}$. 

\begin{figure}[H]
\includegraphics[scale=0.5]{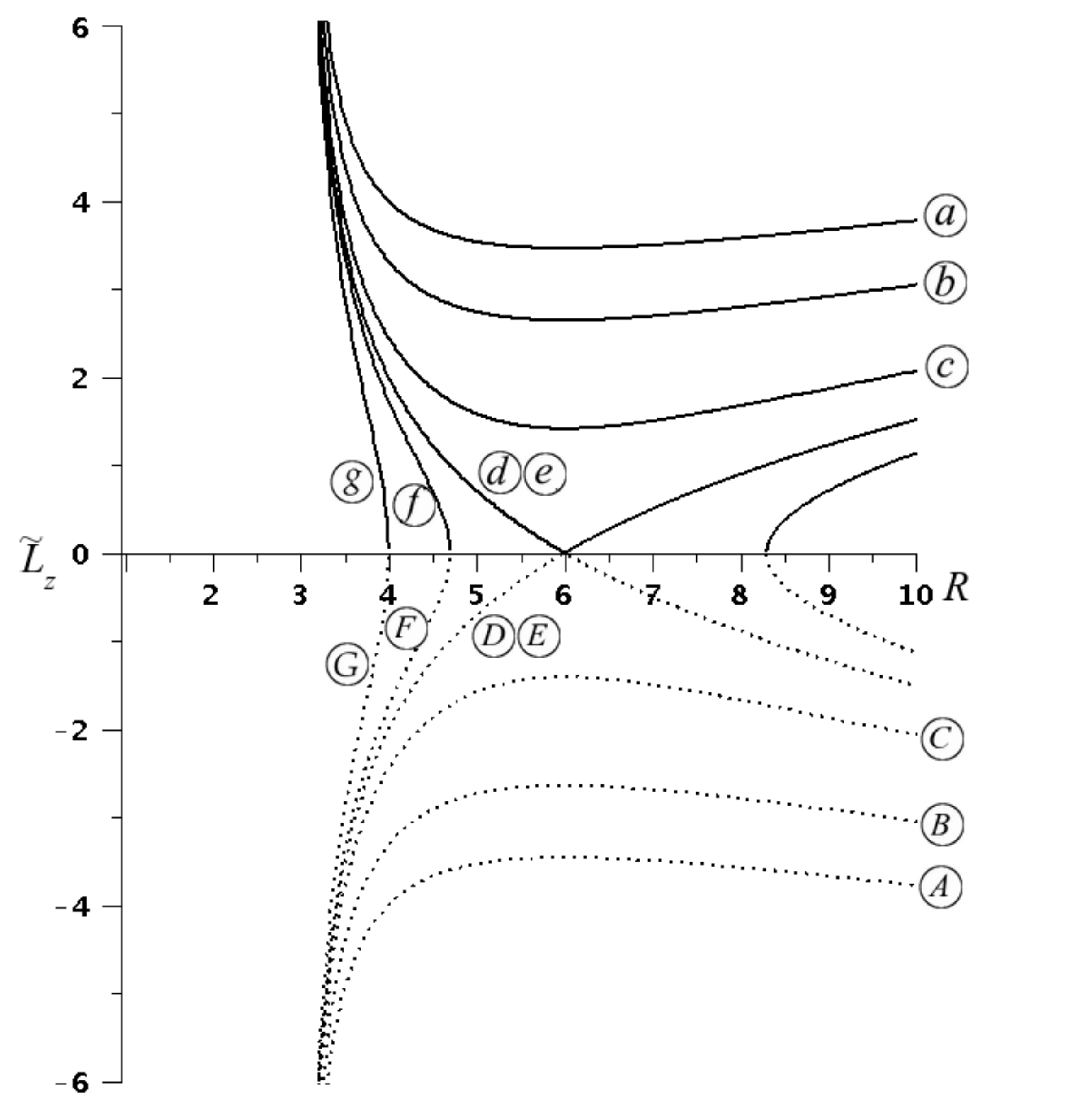}

\caption{\label{fig:SBH}The relationship between ${\tilde{L}}_{z}$ and pericentre,
${R}$, for an SBH. Various values of ${Q}$ are depicted. The polar
LSO and abutment are superimposed (${d},{e}$). In table \ref{tab:SBH}
some values of ${R}_{LSO}$ for this SBH system are listed.}

\end{figure}

\begin{figure}[H]
\includegraphics[scale=0.5]{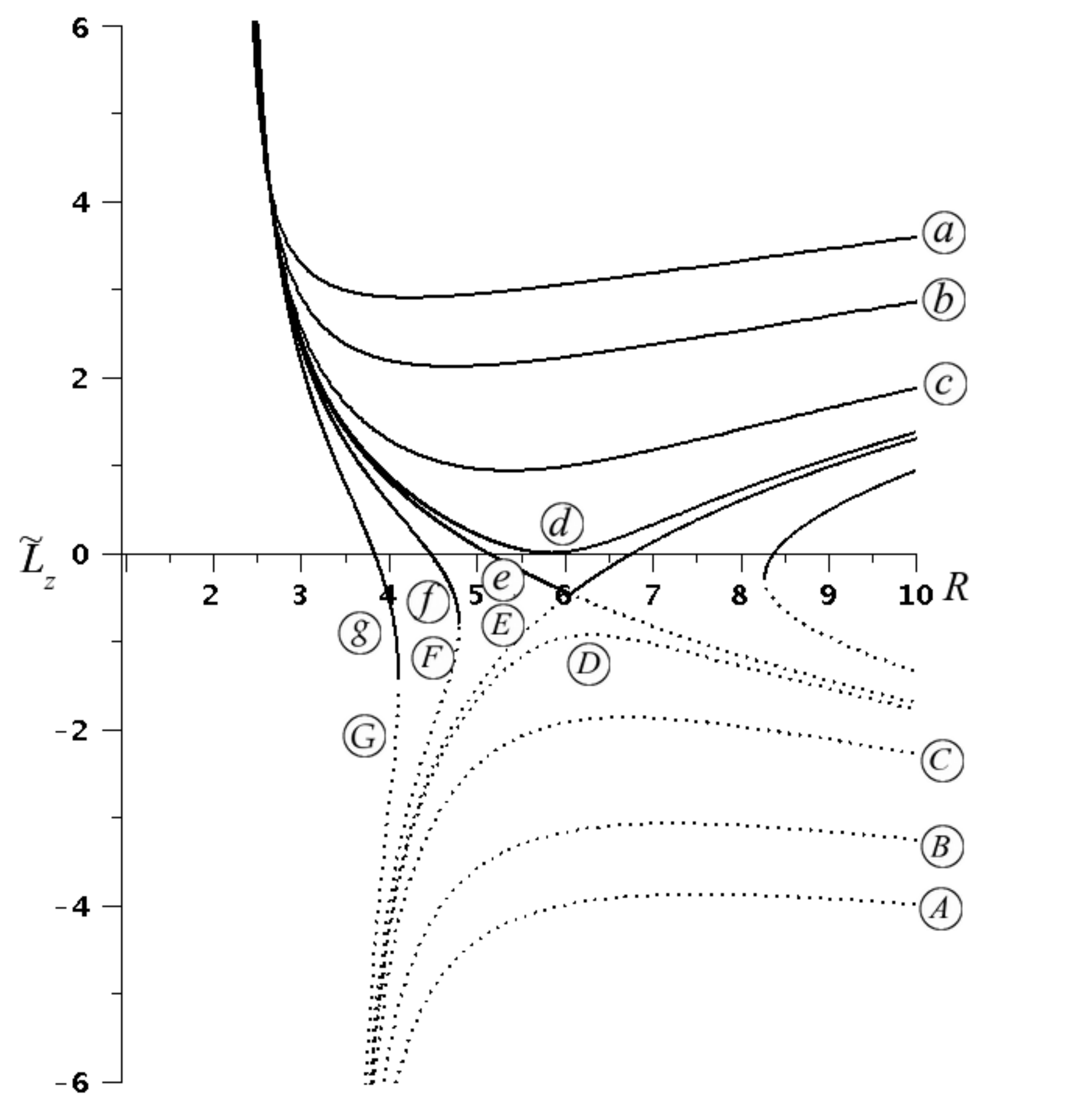}

\caption{\label{fig:KBH_S_0.5}The relationship between ${\tilde{L}}_{z}$
and pericentre, ${R}$, for a KBH with ${\tilde{S}}=0.50$. Various
values of ${Q}$ are depicted. The polar LSO (${d}$) and the LSO
at the abutment (${e}$) are distinct. In table \ref{tab:KBH50} some
values of ${R}_{LSO}$ for this KBH system are listed.}

\end{figure}

\begin{figure}[H]
\includegraphics[scale=0.5]{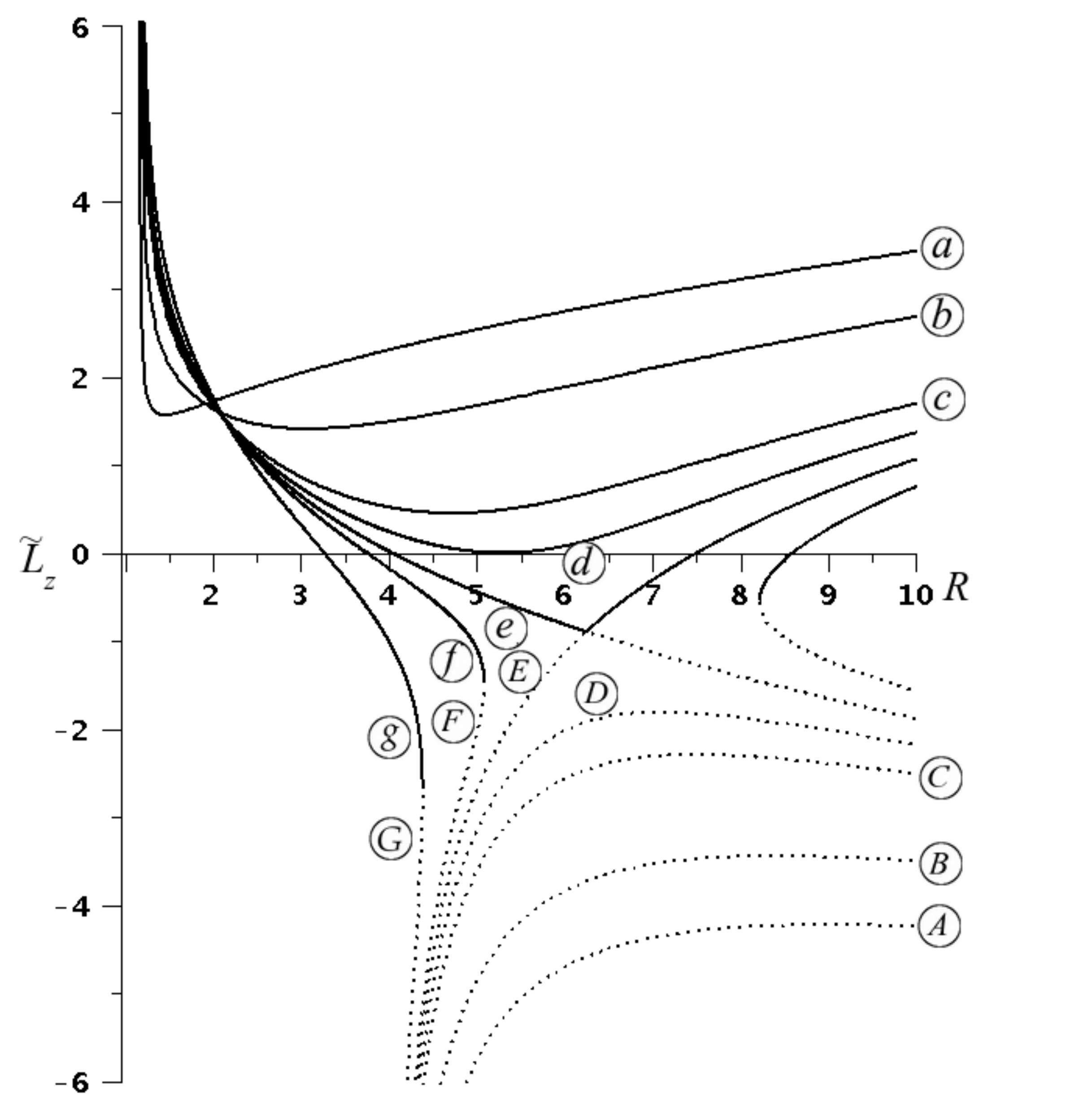}

\caption{\label{fig:KBH_S_0.99}The relationship between ${\tilde{L}}_{z}$
and pericentre, ${R}$, for a KBH with ${\tilde{S}}=0.99$. Various
values of ${Q}$ are depicted. The separation of the polar LSO (${d}$)
and the LSO at the abutment (${e}$) is increased with the higher
value of ${\tilde{S}}$. In table \ref{tab:KBH99} some values of
${R}_{LSO}$ for this KBH system are listed.}

\end{figure}

One can now plot ${\tilde{L}}_{z}$ with respect to the value of the
pericentre (the point of closest approach, ${R}_{p}$) for an LSO
for the cases where ${\tilde{S}}=0.0$ (Figure \ref{fig:SBH}), ${\tilde{S}}=0.5$
(Figure \ref{fig:KBH_S_0.5}) and ${\tilde{S}}=0.99$ (Figure \ref{fig:KBH_S_0.99}).
The values of ${\tilde{L}}_{z}$ calculated for an SBH are plotted
in Figure \ref{fig:SBH} for the range of ${Q}$ values $0.0$ to
${16.0}$.

For an SBH, the prograde (plus) and retrograde (minus) formulae for
${\tilde{L}}_{z}^{2}$ (equation (\ref{eq:LzSquared})) are reflections
of one another about the ${R}$ axis (Figure \ref{fig:SBH}). But
when ${\tilde{S}}>0$, the plus equations are pulled below the ${R}$
axis and this symmetry is lost (Figures \ref{fig:KBH_S_0.5} and \ref{fig:KBH_S_0.99}).
There now exists a set of retrograde LSOs which are governed by the
plus form of equation (\ref{eq:LzSquared}). The importance of this
fact is revealed when we set the quantity beneath the square root
in equation (\ref{eq:LzSquared}) to zero; i.e.\begin{eqnarray}
{R}^{5}-{Q}{R}^{4}+3{Q}{R}^{3}+{Q}^{2}{S}^{2} & = & 0.\label{eq:R_LSO_Abutment}\end{eqnarray}
The polynomial describes the boundary at which the plus and minus
equations for ${\tilde{L}}_{z}^{2}$ are equal and it offers an insight
into the behaviour of ${Q}$ for LSOs that are nearly polar (${\iota}\simeq\pi/2$).
If ${\tilde{S}}=0$, then for an elliptical LSO \cite{PhysRevD.50.3816,Barack:2004uq,Komorowski:2008wc},
\begin{equation}
{R}_{p}=2\left(3+{e}\right)/\left(1+{e}\right).\label{eq:R_pericentre}\end{equation}
By substituting equation (\ref{eq:R_pericentre}) into equation (\ref{eq:R_LSO_Abutment})
and solving for ${Q}$, one obtains:\begin{eqnarray}
{Q} & = & 4\left(3+{e}\right)^{2}\left[\left(1+{e}\right)\left(3-{e}\right)\right]^{-1}.\label{eq:Q_Lz_Schwarzschild}\end{eqnarray}
This result applies to LSOs at the boundary and specifies an upper
limit on ${Q}$ for orbits around an SBH. Now we must develop these
ideas for general orbits about a KBH that have not yet reached their
LSO.

\section{\label{sec:Analysis-of-the-Trajectory-Equations}Analysis of the
Trajectory Equations}

\subsection{Introduction}

There exist four trajectory equations \cite{1968PhRv..174.1559C,Bardeen:1972fi,2002CQGra..19.2743S}
in two categories:

\subsubsection*{category (a)}

(those that are periodic in radius, ${R}$, or polar angle, ${\theta}$)

\begin{equation}
{\tilde{\Sigma}}\frac{{dR}}{{d\tilde{\tau}}}=\pm\sqrt{{\tilde{V}}_{R}\left(R\right)}\label{eq:rho-drbydt}\end{equation}
\begin{equation}
{\tilde{\Sigma}}\frac{{d\theta}}{{d\tilde{\tau}}}=\pm\sqrt{{\tilde{V}}_{\theta}\left(\theta\right)}\label{eq:rho-dthetabydt}\end{equation}

\subsubsection*{category (b)}

(those that are monotonically increasing in azimuthal angle, ${\varphi}$,
or coordinate time, ${t}$)

\begin{equation}
{\tilde{\Sigma}}\frac{{d\phi}}{{d\tilde{\tau}}}={\tilde{V}}_{\phi}\label{eq:dPhidTao}\end{equation}
\begin{equation}
{\tilde{\Sigma}}\frac{{d\tilde{t}}}{{d\tilde{\tau}}}={\tilde{V}}_{t}.\label{eq:dtdtao}\end{equation}
See Appendix B to see equations for functions ${\tilde{V}}_{R}\left({R}\right)$,
${\tilde{V}}_{\theta}$, ${\tilde{V}}_{\phi}$, and ${\tilde{V}}_{t}$.
We have already developed equation (\ref{eq:rho-drbydt}) in section
\ref{sub:Basic-orbital-equations} (equation (\ref{eq:Orbital_Equation_Q})).
And equation (\ref{eq:rho-dthetabydt}) can also be developed by a
similar method (see Appendix C). 

One obtains, \textit{viz.} equations (\ref{eq:rho-drbydt}) and (\ref{eq:rho-dthetabydt}),
the following condition\begin{equation}
\frac{{dR}}{{\sqrt{\tilde{V}_{R}\left(R\right)}}}=\frac{{d\theta}}{{\sqrt{\tilde{V}_{\theta}\left(\theta\right)}}},\label{eq:equality-on-geodesic}\end{equation}
on the geodesic of the test-particle, which is a general form of the
equation specified by Schmidt (equation (16) in \cite{2002CQGra..19.2743S}).
Given equation (\ref{eq:rho-dthetabydt}) \cite{Bardeen:1972fi},
one can find the proper time of the orbit:

\begin{eqnarray}
\tilde{\tau} & = & {\int_{{\theta}_{1}}^{{\theta}_{2}}}{\frac{{\tilde{\Sigma}d\theta}}{{\sqrt{{\tilde{V}}_{\theta}}}}}\nonumber \\
 & = & {\int_{{r}_{1}}^{{r}_{2}}}{\frac{{R^{2}dR}}{{\sqrt{{\tilde{V}}_{R}}}}}+{\tilde{S}^{2}}{\int_{{\theta}_{1}}^{{\theta}_{2}}}\frac{{\cos^{2}\left(\theta\right)}}{{\sqrt{{\tilde{V}}_{\theta}}}}d\theta,\label{eq:Tao-Integral}\end{eqnarray}
where the integral has been split into its separate ${R}$ and $\theta$
integral terms \textit{viz.} equation (\ref{eq:equality-on-geodesic}).
The same result is found when starting with equation (\ref{eq:rho-drbydt})
instead. Two other important integrals that can be calculated for
coordinate time and azimuthal angle are given in Schmidt (equations
(14) and (15) \cite{2002CQGra..19.2743S}) in which a detailed analysis
is made on the basis of the Hamiltonian. Equation (\ref{eq:Tao-Integral})
can be solved to yield elliptic functions \cite{2002PhRvD..66d4002G};
therefore, the roots of ${\tilde{V}}_{R}$ and ${V}_{\theta}$ contain
information necessary for deriving analytical formulae for ${Q}$
in terms of ${\tilde{l}}$ and ${e}$ (for given ${\tilde{S}}$) and
${\iota}$ as a function of ${Q}$. This will be shown in sections
\ref{sub:Roots-of-the-Polar-Angle-Equation} and \ref{sec:The-characteristics-of-the-Carter-Constant-Equations}.

\subsection{Roots of the Radial Equation}

We introduce ${X}={\tilde{L}}_{z}-{\tilde{S}}{\tilde{E}}$ and convert
equation (\ref{eq:Orbital_Equation_Q_Turning_Point}) to the form:\begin{eqnarray}
0 & = & {R}^{4}-2\,\frac{{R}^{3}}{\left({1-{\tilde{E}}^{2}}\right)}\nonumber \\
 & + & \frac{\left({X}^{2}+{\tilde{S}}^{2}+2{\tilde{S}}{\tilde{E}}{X}+Q\right){R}^{2}}{\left({1-{\tilde{E}}^{2}}\right)}\nonumber \\
 & - & \frac{2\,\left({X}^{2}+Q\right){R}}{\left({1-{\tilde{E}}^{2}}\right)}+\frac{{Q{\tilde{S}}^{2}}}{\left({1-{\tilde{E}}^{2}}\right)}.\label{eq:Orbital_Equation_Q_X}\end{eqnarray}
This substitution is consistent with the approach in \cite{Komorowski:2008wc}
and that undertaken by Glampedakis and Kennefick \cite{2002PhRvD..66d4002G};
and it will help us to derive the latus rectum of the LSO. Analytically,
the use of ${X}^{2}$ in this case offers an advantage over the use
of ${L}_{z}^{2}$.

\subsubsection{Elliptical Orbits.}

By finding the four roots of equation (\ref{eq:Orbital_Equation_Q_X})
one can derive analytical formulae for ${X}$ and ${\tilde{E}}$,
in terms of ${e}$, ${\tilde{l}}$, ${Q}$, and ${\tilde{S}}$, which
apply to general orbits (and are not limited to the LSO). Although
the roots are easily obtained in terms of the constants of motion:
${\tilde{L}}_{z}$, ${\tilde{E}}$, and ${Q}$; they are complicated
and as such not helpful. To simplify the analysis, we assume \textit{a
priori} that an inclined orbit can be characterised by an eccentricity,
${e}$ \cite{1987GReGr..19.1235S}. Therefore the radius of the orbit
at its pericentre is described by

\begin{eqnarray}
{r}_{p} & = & {\frac{{\tilde{l}}}{1+{e}}},\label{eq:Root_Pericentre}\end{eqnarray}
and correspondingly, at its apocentre\begin{eqnarray}
{r}_{a} & = & {\frac{{\tilde{l}}}{1-{e}}}.\label{eq:Root_Apocentre}\end{eqnarray}

To proceed, consider an expansion of the four possible roots $\left\{ {r}_{4}<{r}_{3}\leq{r}_{p}\leq{r}_{a}\right\} $:

\begin{eqnarray}
{R}^{4}\nonumber \\
-\left(r_{{3}}+r_{{4}}+r_{{a}}+r_{{p}}\right){R}^{3}\nonumber \\
+\left(r_{{4}}r_{{3}}+r_{{3}}r_{{a}}+r_{{3}}r_{{p}}+r_{{4}}r_{{a}}+r_{{4}}r_{{p}}+r_{{p}}r_{{a}}\right){R}^{2}\nonumber \\
-\left(r_{{4}}r_{{3}}r_{{a}}+r_{{4}}r_{{3}}r_{{p}}+r_{{3}}r_{{p}}r_{{a}}+r_{{4}}r_{{p}}r_{{a}}\right)R\nonumber \\
+r_{{4}}r_{{3}}r_{{p}}r_{{a}} & = & 0\label{eq:Polynomial_Expansion}\end{eqnarray}
By equating the coefficients of the two polynomials in equations (\ref{eq:Orbital_Equation_Q_X})
and (\ref{eq:Polynomial_Expansion}) one obtains the two independent
equations:

\begin{eqnarray}
{\frac{{\tilde{l}}\left(2\, r_{{4}}r_{{3}}+r_{{3}}{\tilde{l}}+r_{{4}}{\tilde{l}}\right)}{\left(1-{e}^{2}\right)}} & = & 2\,{\frac{{X}^{2}+Q}{1-{\tilde{E}}^{2}}},\label{eq:q1}\end{eqnarray}
and\begin{eqnarray}
{\frac{r_{{4}}r_{{3}}{\tilde{l}}^{2}}{1-{e}^{2}}} & = & {\frac{Q{\tilde{S}}^{2}}{1-{\tilde{E}}^{2}}},\label{eq:q2}\end{eqnarray}
which have been simplified \textit{viz.} equations (\ref{eq:Root_Pericentre})
and (\ref{eq:Root_Apocentre}). Let us solve equations (\ref{eq:q1})
and (\ref{eq:q2}) to obtain:

\begin{eqnarray}
{r}_{3} & = & {\frac{\left(1-{e}^{2}\right)}{{\tilde{l}}^{3}\left(1-{\tilde{E}}^{2}\right)}}\left[\left(Q\left(\tilde{l}-{\tilde{S}}^{2}\right)+{X}^{2}\tilde{l}\right)\pm\sqrt{{Z}_{3}}\right]\label{eq:r3}\end{eqnarray}
and \begin{eqnarray}
{r}_{4} & = & {\frac{Q{\tilde{S}}^{2}\left(1-{e}^{2}\right)}{{r}_{3}{\tilde{l}}^{2}\left(1-{\tilde{E}}^{2}\right)}}\label{eq:r4}\end{eqnarray}
where\begin{eqnarray*}
{Z}_{3} & = & \left({\tilde{l}}-{\tilde{S}}^{2}\right)^{2}{Q}^{2}\\
 & - & \tilde{l}\left({\frac{\left(1-{\tilde{E}}^{2}\right){\tilde{S}}^{2}{\tilde{l}}^{3}}{1-{e}^{2}}}-2\,{X}^{2}\tilde{l}+2\,{X}^{2}{\tilde{S}}^{2}\right)Q+{X}^{4}{\tilde{l}}^{2}.\end{eqnarray*}
If ${Q}=0$ then selecting the minus sign in equation (\ref{eq:r3})
yields ${r}_{3}=0$; therefore, the plus sign is the one taken as
physically meaningful. For ${Q}=0$, equation (\ref{eq:r3}) reduces
to \begin{equation}
{r}_{3}=2\,{\frac{{X}^{2}\left(1-{e}^{2}\right)}{{\tilde{l}}^{2}\left(1-{\tilde{E}}^{2}\right)}},\end{equation}
which applies to an equatorial orbit. The value of ${r}_{4}$ equals
zero when ${Q}=0$ as can be seen in equation (\ref{eq:r4}).

\subsubsection{Parabolic Orbits.}

Parabolic orbits have importance to the empirical study of the interaction
of stars with massive black holes (MBHs) \cite{2001ApJ...560L.143A,2004ApJ...610..707B}.
For parabolic orbits both ${e}=1$ and ${\tilde{E}}=1$. We refer
back to equation (\ref{eq:Orbital_Equation_Q_X}); and set ${\tilde{E}}=1$:\begin{eqnarray}
0 & = & {R}^{3}\nonumber \\
 & - & \frac{1}{2}\left({X}^{2}+{\tilde{S}}^{2}+2\,\tilde{S}X+Q\right){R}^{2}\nonumber \\
 & + & \left({X}^{2}+Q\right)R-\frac{1}{2}{Q{\tilde{S}}^{2}}.\label{eq:Parabolic_Orbital_Equation_Q}\end{eqnarray}
There are now three possible finite roots $\left\{ {r}_{4}<{r}_{3}\leq{r}_{p}\right\} $,
which can be used in a new general equation (${r}_{a}$ is infinite
in the case of a parabolic orbit): \begin{eqnarray}
{R}^{3}-\left(r_{{3}}+r_{{4}}+r_{{p}}\right){R}^{2}\nonumber \\
+\left(r_{{4}}r_{{3}}+r_{{3}}r_{{p}}+r_{{4}}r_{{p}}\right)R-r_{{4}}r_{{3}}r_{{p}} & = & 0.\label{eq:Parabolic_Polynomial_Expansion}\end{eqnarray}
The pericentre simplifies (\textit{viz.} ${{e}=1}$) to become,

\begin{eqnarray}
{r}_{p} & = & {\frac{{\tilde{l}}}{2}};\label{eq:Parabolic_Root_Pericentre}\end{eqnarray}
and correspondingly, at its apocentre,\begin{eqnarray}
{r}_{a} & \rightarrow & \infty.\label{eq:Parabolic_Root_Apocentre}\end{eqnarray}
We obtain the solutions for the additional roots:\begin{eqnarray}
{r}_{3} & = & {\frac{1}{{\tilde{l}}^{2}}}\left[\left(Q\left(\tilde{l}-{\tilde{S}}^{2}\right)+{X}^{2}\tilde{l}\right)\pm\sqrt{{Z}_{4}}\right]\label{eq:Parabolic_r3}\end{eqnarray}
and \begin{eqnarray}
{r}_{4} & = & {\frac{Q{\tilde{S}}^{2}}{{r}_{3}{\tilde{l}}}}\label{eq:Parabolic_r4}\end{eqnarray}
where\begin{eqnarray*}
{Z}_{4} & = & \left({\tilde{l}}-{\tilde{S}}^{2}\right)^{2}{Q}^{2}-\tilde{l}\left({{\tilde{S}}^{2}{\tilde{l}}^{2}}-2\,{X}^{2}\tilde{l}+2\,{X}^{2}{\tilde{S}}^{2}\right)Q+{X}^{4}{\tilde{l}}^{2}.\end{eqnarray*}

\subsection{\label{sub:Roots-of-the-Polar-Angle-Equation}Roots of the Polar-Angle
Equation}

Let us focus on the denominator of the second term in equation (\ref{eq:Tao-Integral}),
i.e. ${\sqrt{{\tilde{V}}_{\theta}}}$, to derive an analytical relationship
for the limits of integration, ${\theta}_{1}$ and ${\theta}_{2}$,
from which one may determine ${\iota}$. We shall work with ${\tilde{V}}_{\theta}$
in terms of ${\tilde{L}}_{z}$, i.e.

\begin{eqnarray}
\mathcal{I}=\nonumber \\
{\tilde{S}^{2}}{\int_{{\theta}_{1}}^{{\theta}_{2}}}\frac{{\sin\left(\theta\right)}{\cos^{2}\left(\theta\right)}d\theta}{{\sqrt{{Q}\sin^{2}\left(\theta\right)-\sin^{2}\left(\theta\right)\cos^{2}\left(\theta\right){\tilde{S}}^{2}\left(1-{\tilde{E}}^{2}\right)-\cos^{2}\left(\theta\right){\tilde{L}_{z}}^{2}}}}.\label{eq:Elliptic_Integral}\end{eqnarray}

Equation (\ref{eq:Tao-Integral}) is an elliptic integral; thus the
limits of integration correspond to the zeros of the denominator.
By making the substitution \begin{equation}
{u}=\cos\left({\theta}\right),\end{equation}
the integral, ${\mathcal{I}}$, becomes

\begin{eqnarray}
\mathcal{I} & = & {\tilde{S}^{2}}{\int_{{u}_{1}}^{{u}_{2}}}\frac{-{u^{2}}du}{{\sqrt{{\tilde{S}}^{2}\left(1-{\tilde{E}}^{2}\right){u^{4}}-\left({Q}+{\tilde{L_{z}}}^{2}+{\tilde{S}}^{2}\left(1-{\tilde{E}}^{2}\right)\right){u}^{2}+{Q}}}}\nonumber \\
 & = & {\tilde{S}^{2}}{\int_{{u}_{1}}^{{u}_{2}}}\frac{-{u^{2}}du}{{\tilde{S}}\sqrt{1-{\tilde{E}}^{2}}{\sqrt{\left({u^{2}-\beta_{+}}\right)\left({u^{2}-\beta_{-}}\right)}}},\end{eqnarray}
for which the roots, ${\beta}_{\pm}=\cos^{2}\left(\theta_{\pm}\right)$,
can be calculated and used to determine (\textit{viz.} $\cos\left(\theta\right)=\cos\left(\frac{\pi}{2}-\iota\right)=\sin\left(\iota\right)$)
the exact angle of inclination of an orbit for which the values of
${\tilde{S}}$, ${Q}$, and ${\tilde{L}_{z}}$ are known, when working
in Boyer-Lindquist coordinates. We will calculate ${\iota}$ using\setlength{\mathindent}{0cm}
\begin{equation}
\sin^{2}\left({\iota}\right)=\frac{1}{2}\,\frac{{Q+\tilde{L}_{z}^{2}+\tilde{S}^{2}\left({1-\tilde{E}^{2}}\right)-\sqrt{\left({Q+\tilde{L}_{z}^{2}+\tilde{S}^{2}\left({1-\tilde{E}^{2}}\right)}\right)^{2}-4\, Q\tilde{S}^{2}\left({1-\tilde{E}^{2}}\right)}}}{{\tilde{S}^{2}\left({1-\tilde{E}^{2}}\right)}},\label{eq:Theta-Roots-Equatorial}\end{equation}
\setlength{\mathindent}{2.5cm}which differs from the approximation
in \cite{2002PhRvD..66f4005G,2006PhRvD..73f4037G}. N.B.: in dealing
with results that are first-order and third-order in $\tilde{S}$,
one may consider the approximation to $\iota$ to be reasonably close
to equation (\ref{eq:Theta-Roots-Equatorial}) \cite{2011arXiv1101.0996K}.

\subsection{Orbital Energy and an Analytical Expression for ${X}^{2}$}

As outlined in \cite{2002PhRvD..66d4002G}, the next step will be
to develop a formula for orbital energy, ${\tilde{E}}$, in terms
of ${e}$, ${\tilde{l}}$, and ${X}^{2}$. By referring to equations
(\ref{eq:Orbital_Equation_Q_X}) and (\ref{eq:Polynomial_Expansion}),
this derivation proceeds by solving \begin{eqnarray}
r_{{4}}+r_{{3}}+r_{{p}}+r_{{a}} & = & 2\,\left(1-{\tilde{E}}^{2}\right)^{-1}\label{eq:4rs}\end{eqnarray}
to yield

\begin{eqnarray}
\tilde{E} & = & \pm\frac{1-{e}^{2}}{{\tilde{l}}^{2}}\sqrt{Q\left({\tilde{l}}-{\tilde{S}}^{2}\right)+{X}^{2}{\tilde{l}}+{\tilde{l}}^{3}\left({e}^{2}+{\tilde{l}}-1\right)\left(1-{e}^{2}\right)^{-2}},\label{eq:E4}\end{eqnarray}
for which we use the positive case. For ${Q}=0$, equation (\ref{eq:E4})
simplifies to

\begin{eqnarray}
\tilde{E} & = & \sqrt{1-\frac{\left(1-{e}^{2}\right)}{{\tilde{l}}}\left(1-{\frac{{X}^{2}\left(1-{e}^{2}\right)}{{\tilde{l}}^{2}}}\right)},\label{eq:E4Q0}\end{eqnarray}
which is the expression for ${\tilde{E}}$ in an equatorial orbit.

It is interesting to observe that for an inclined orbit around an
SBH (${\tilde{S}}=0$), the equation for ${\tilde{E}}$ (equation
(\ref{eq:E4})) reduces to \begin{equation}
{\tilde{E}}=\sqrt{{\frac{\left(\tilde{l}-2\left(1+{e}\right)\right)\left(\tilde{l}-2\left(1-{e}\right)\right)}{\tilde{l}\left(\tilde{l}-3-{e}^{2}\right)}}},\end{equation}
which is the expression for orbital energy of a test-particle in orbit
around an SBH, (Cutler, Kennefick, and Poisson (see equation (2.5)
in \cite{PhysRevD.50.3816})). Further, this equation for ${\tilde{E}}$
shows no dependence on ${Q}$. This property is expected since the
orbital energy must be independent of orientation in the spherically
symmetric coordinate system of an SBH.

In the general case, we observe that ${\tilde{E}}$ is a function
of ${X}^{2}$ (see equation (\ref{eq:E4})); and thus it is not in
explicit form because ${X}={\tilde{L}}_{z}-{\tilde{S}}{\tilde{E}}$.
Therefore the use of ${X}^{2}$ in place of $\left({\tilde{L}}_{z}-{\tilde{S}}{\tilde{E}}\right)^{2}$
simplifies the analysis by avoiding an unending recursive substitution
of ${\tilde{E}}$ into the equation. Although one may derive a formula
for ${\tilde{E}}$ in explicit form, it is better to perform the analysis
using equation (\ref{eq:E4}).

To calculate an analytical expression for ${X}^{2}$, we substitute
equation (\ref{eq:E4}) into our original quartic (equation (\ref{eq:Orbital_Equation_Q_Turning_Point}))
and evaluate it at either ${r}_{p}$ or ${r}_{a}$ (the two simplest
choices of the four roots) to yield:\setlength{\mathindent}{0cm}\begin{eqnarray}
\left({1+{e}}\right)^{-2}\biggl({\tilde{l}^{2}\left({\tilde{S}^{2}-l}\right)+2\, QS^{2}\left({1+{e}^{2}}\right)+\tilde{l}\left({X^{2}+Q}\right)\left({\tilde{l}-3-{e}^{2}}\right)}\\
{+2\tilde{S}X\sqrt{l\left({X^{2}+Q}\right)\left({1-{e}^{2}}\right)^{2}-Q\tilde{S}^{2}\left({1-{e}^{2}}\right)^{2}+\tilde{l}^{3}\left({\tilde{l}+{e}^{2}-1}\right)}}\biggr) & = & 0.\nonumber \end{eqnarray}
\setlength{\mathindent}{2.5cm}By eliminating the square root, and
solving for ${X}^{2}$ in the resulting quadratic, one obtains:

\begin{eqnarray}
{X}_{\pm}^{2} & = & {\frac{{Z}_{5}+{Z}_{6}{Q}\pm2{\tilde{S}}\,\sqrt{{Z_{7}}{Z_{8}}{Z_{9}}}}{\tilde{l}\left(\tilde{l}\left(3-\tilde{l}+{e}^{2}\right)^{2}-4\,{\tilde{S}}^{2}\left(1-{e}^{2}\right)^{2}\right)}},\label{eq:X2}\end{eqnarray}
where\[
{Z}_{5}={\tilde{l}}^{3}\left\{ \left(\tilde{l}+3\,{e}^{2}+1\right){\tilde{S}}^{2}-\tilde{l}\left(3-\tilde{l}+{e}^{2}\right)\right\} ,\]

\begin{eqnarray*}
{Z}_{6} & = & -2\,\left(1-{e}^{2}\right)^{2}{\tilde{S}}^{4}+2\,\tilde{l}\left(2\,{e}^{4}+\left(2-\tilde{l}\right){e}^{2}+4-\tilde{l}\right){\tilde{S}}^{2}\\
 &  & -{\tilde{l}}^{2}\left(3-\tilde{l}+{e}^{2}\right)^{2},\end{eqnarray*}

\[
{Z_{7}}={\tilde{S}^{2}\left({1+{e}}\right)^{2}+\tilde{l}\left({\tilde{l}-2(1+{e})}\right)},\]
\[
{Z_{8}}={\tilde{S}^{2}\left({1-{e}}\right)^{2}+\tilde{l}\left({\tilde{l}-2(1-{e})}\right)},\]
and

\begin{equation}
{Z_{9}}={\left({\tilde{l}^{5}+\tilde{S}^{2}Q^{2}\left({1-{e}^{2}}\right)^{2}+Q\tilde{l}^{3}\left({3-\tilde{l}+{e}^{2}}\right)}\right)}.\label{eq:Z9}\end{equation}
${X}_{\pm}^{2}$ has a minus and a plus solution, which we will carefully
describe in the next section. We have avoided the analytical difficulties
that would arise by working with ${\tilde{L}}_{z}$ directly. Indeed,
the advantage of using ${X}={\tilde{L}}_{z}-{\tilde{S}}{\tilde{E}}$
is more than a simple change of variables, but rather an essential
step in solving these equations.

\subsection{Prograde and Retrograde Descriptions of ${X}^{2}$ }

The expression for ${X}_{\pm}^{2}$ (see equation (\ref{eq:X2}))
contains the square root, $\pm2{\tilde{S}}\,\sqrt{{Z_{7}}{Z_{8}}{Z_{9}}}$,
for which, ${Q}$ is found only in ${Z}_{9}$ as a quadratic. Therefore
it is easy to derive an expression for ${Q}$ for which ${Z}_{9}=0$
and thus determine where the minus and plus equations for ${X}_{\pm}^{2}$
meet or abut (\textit{viz.} ${Z}_{9}=0$). This information is important
for determining which form of equation (\ref{eq:X2}) to use. We will
call the set of general orbits for which ${Z}_{9}=0$, the abutment,
to avoid confusing it with the result for the boundary between the
plus and minus forms of ${\tilde{L}}_{z}$ at the LSO.

A prograde orbit has an ${\tilde{L}}_{z}>0$. Correspondingly, when
${\tilde{L}}_{z}<0$ the orbit is retrograde; and if ${\tilde{L}}_{z}=0$,
the orbit is polar. When using the minus and plus forms of the equation
for ${X}_{\pm}^{2}$ (equation (\ref{eq:X2})), one must recognise
that ${X}_{-}^{2}$ governs all of the prograde orbits, the polar
orbits, and the near-polar retrograde orbits up to the abutment; and
${X}_{+}^{2}$ governs the remaining retrograde orbits. If one considers
an SBH system then the abutment will be comprised only of polar orbits
(and it is only then that ${X}_{\pm}^{2}$ cleanly separates the prograde
and retrograde orbits). If ${\tilde{S}}>0$, the abutment will always
consist of retrograde orbits. 

\begin{figure}
\caption{\label{fig:Sign-Switch-by} \label{fig:X2plot}A plot of ${X}_{-}^{2}$
with respect to ${Q}$ for a circular orbit (${\tilde{l}}=6.25$)
about a KBH of spin ${\tilde{S}}=0.5$. The slope of ${X_{-}}$ can
be assumed to have no discontinuities; therefore, the point at ${Q}_{switch}=11.26$
indicates that if ${Q}>{Q}_{switch}$ then ${X}_{-}=-\sqrt{{X}_{-}^{2}}$. }

\includegraphics[scale=0.3]{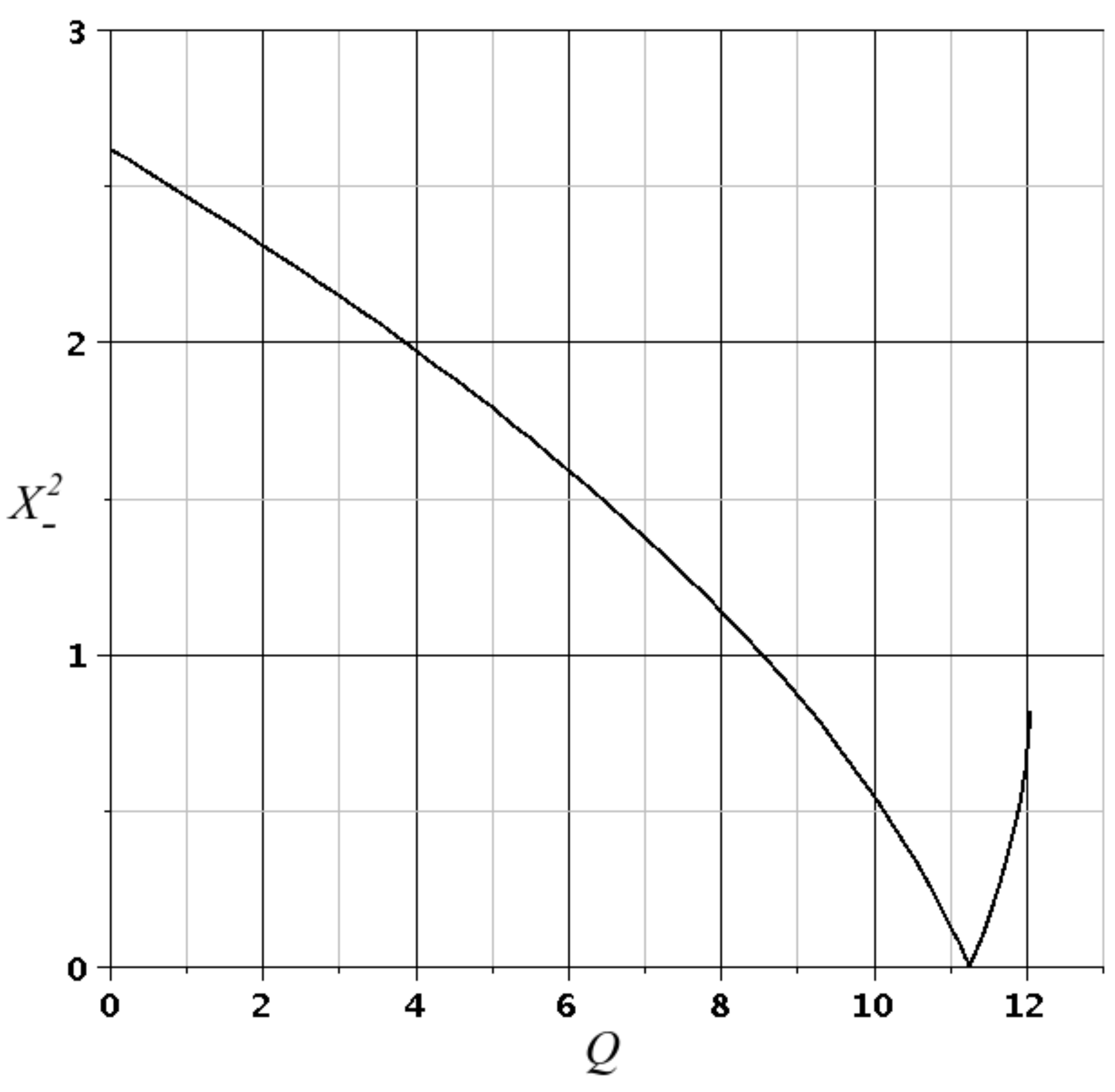}

\end{figure}

For orbits governed by ${X}_{+}^{2}$, ${X}_{+}=-\sqrt{{X}_{+}^{2}}$;
but for those governed by ${X}_{-}^{2}$, the choice of sign depends
on the value of ${Q}$. The plot of ${X}_{-}^{2}$ with respect to
${Q}$ for a circular orbit with ${\tilde{l}}=6.25$ about a KBH of
${\tilde{S}}=0.5$ (Figure \ref{fig:X2plot}) shows that for $\partial{X}_{-}/\partial{Q}$
to remain continuous over the range of real values of ${Q}$, the
minus sign must be chosen when evaluating ${X}_{-}$ for ${Q}>{Q}_{switch}$.
An analytical formula for ${Q}_{switch}$ can be found by solving
${X}_{-}^{2}=0$ for ${Q}$. The general solution is\begin{eqnarray}
{Q}_{switch} & = & {\tilde{l}}^{2}\left({\tilde{l}}-{\tilde{S}}^{2}\right)\left({\tilde{l}}\left({\tilde{l}}-{e}^{2}-3\right)+2\left(1+{e}^{2}\right){\tilde{S}}^{2}\right)^{-1},\label{eq:Q_switch}\end{eqnarray}
and for an SBH\begin{eqnarray}
{Q}_{switch} & = & {\tilde{l}}^{2}\left({\tilde{l}}-{e}^{2}-3\right)^{-1}.\label{eq:Q_switch_SBH}\end{eqnarray}
For large orbits (${\tilde{l}}\rightarrow\infty$) equation (\ref{eq:Q_switch})
can be converted to its asymptotic form by first factoring out ${\tilde{l}}$
from each term to obtain,\begin{eqnarray}
{Q} & = & {\tilde{l}}\left(1-\frac{{\tilde{S}}^{2}}{{\tilde{l}}}\right)\left(1-\frac{3+{e}^{2}}{{\tilde{l}}}+\frac{2\left(1+{e}^{2}\right){\tilde{S}}^{2}}{{\tilde{l}}^{2}}\right)^{-1},\end{eqnarray}
for which the denominator may be brought up to the numerator to yield\begin{eqnarray}
{Q} & \cong & {\tilde{l}}\left(1-\frac{{\tilde{S}}^{2}}{{\tilde{l}}}\right)\left(1+\frac{3+{e}^{2}}{{\tilde{l}}}-\frac{{\left(3+{e}^{2}\right)}^{2}+2\left(1+{e}^{2}\right){\tilde{S}}^{2}}{{\tilde{l}}^{2}}\right).\end{eqnarray}
In the limit as ${\tilde{l}}\rightarrow\infty$, \begin{eqnarray}
{Q}_{switch} & = & {\tilde{l}}+3+{e}^{2}-{\tilde{S}}^{2}.\label{eq:Q_switch_large_l}\end{eqnarray}
Equation (\ref{eq:Q_switch_large_l}) describes the locus of points
at which ${\tilde{L}}_{z}={\tilde{S}}{\tilde{E}}$ (which is effectively
constant for large ${\tilde{l}}$); therefore, ${Q}_{switch}$ does
not describe a trajectory. 

We have developed two formulae: one for the ${\tilde{L}}_{z}^{2}$
at the LSO (equation (\ref{eq:LzSquared})) and the other for the
${X}_{\pm}^{2}$ of general circular and elliptical orbits (equation
(\ref{eq:X2})). For each, there is an expression that describes where
the plus and minus forms are equal. For ${\tilde{L}}_{z}^{2}$, the
boundary between the plus and minus forms is described by\begin{eqnarray}
{R}_{p}^{5}-{Q}{R}_{p}^{4}+3{Q}{R}_{p}^{3}+{Q}^{2}{S}^{2} & = & 0,\label{eq:RpericentreLSO_Abutment}\end{eqnarray}
where ${R}_{p}$ represents the pericentre. And for ${X}_{\pm}^{2}$,
the abutment is described by

\begin{eqnarray}
{\tilde{l}}^{5}+{\tilde{S}}^{2}{Q}^{2}\left(1-{e}^{2}\right)^{2}+{Q}{\tilde{l}}^{3}\left(3-{\tilde{l}}+{e}^{2}\right) & = & 0.\label{eq:LatusRectumGeneralAbutment}\end{eqnarray}
As equations (\ref{eq:RpericentreLSO_Abutment}) and (\ref{eq:LatusRectumGeneralAbutment})
describe the boundaries (where the plus and minus forms are equal)
that pertain to different quantities (${\tilde{L}}_{z}$ and ${X}_{\pm}^{2}$)
they will not in general coincide. If one substitutes ${R}_{p}={\tilde{l}}/\left(1+{e}\right)$
into equation (\ref{eq:RpericentreLSO_Abutment}) one obtains:

\begin{eqnarray}
{\tilde{l}}^{5}-{Q}{\left(1+{e}\right)}{\tilde{l}}^{4}+3{Q}{\left(1+{e}\right)}^{2}{\tilde{l}}^{3}+{Q}^{2}{\tilde{S}}^{2}{\left(1+{e}\right)}^{5} & = & 0,\end{eqnarray}
which equals equation (\ref{eq:LatusRectumGeneralAbutment}) when
${e}=0$, (i.e. for a circular orbit). If ${\tilde{S}}=0$ (SBH case),
then ${X}_{\pm}^{2}={\tilde{L}}_{\pm}^{2}$ and the two boundaries
must be identical. If one substitutes ${\tilde{l}}=6+2{e}$ into equation
(\ref{eq:LatusRectumGeneralAbutment}) and solves for ${Q}$, then
the same expression as in equation (\ref{eq:Q_Lz_Schwarzschild})
is obtained.

\section{\label{sec:The-characteristics-of-the-Carter-Constant-Equations}The
Characteristics of the Carter Constant Equations and the Domain of
the Orbital Parameters}

\subsection{Introduction}

\begin{sidewaysfigure}
\caption{\label{fig:S=00003D0.99}A sequence of ${Q}$ - ${\tilde{l}}$ maps
for various values of ${e}$ for a KBH system with ${\tilde{S}}=0.99$.
The short-dashed lines represent ${Q}_{X}$ (equation (\ref{eq:Q_Abutment})),
the long-dashed lines represent ${Q}_{polar}$ (equation (\ref{eq:Q_polar_physical})),
and the solid lines represent ${Q}_{LSO}$ (equation (\ref{eq:Q_LSO})).}
\subfigure[Circular Orbit]{\includegraphics[scale=0.25]{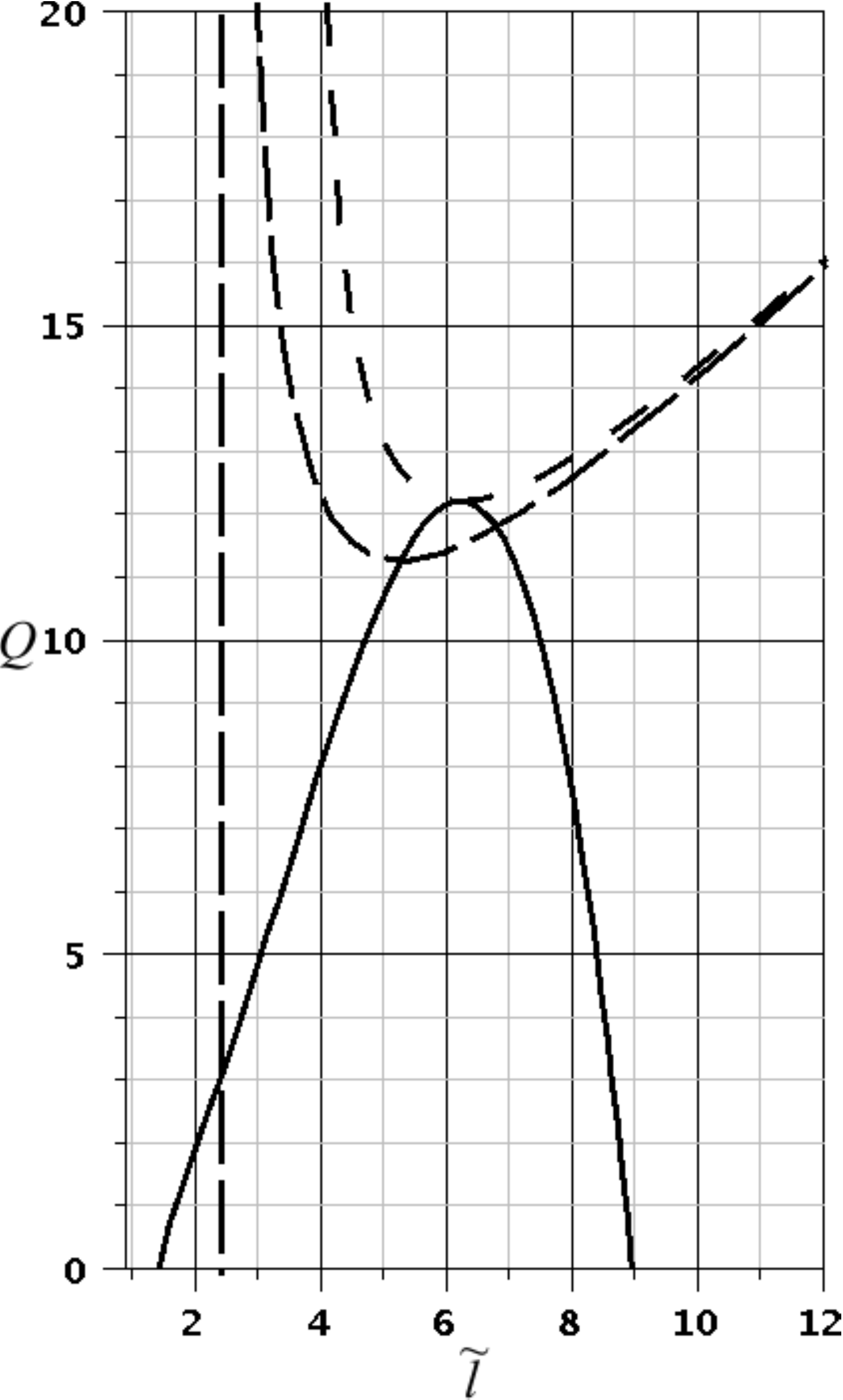}}\subfigure[e = 0.25]{\includegraphics[scale=0.25]{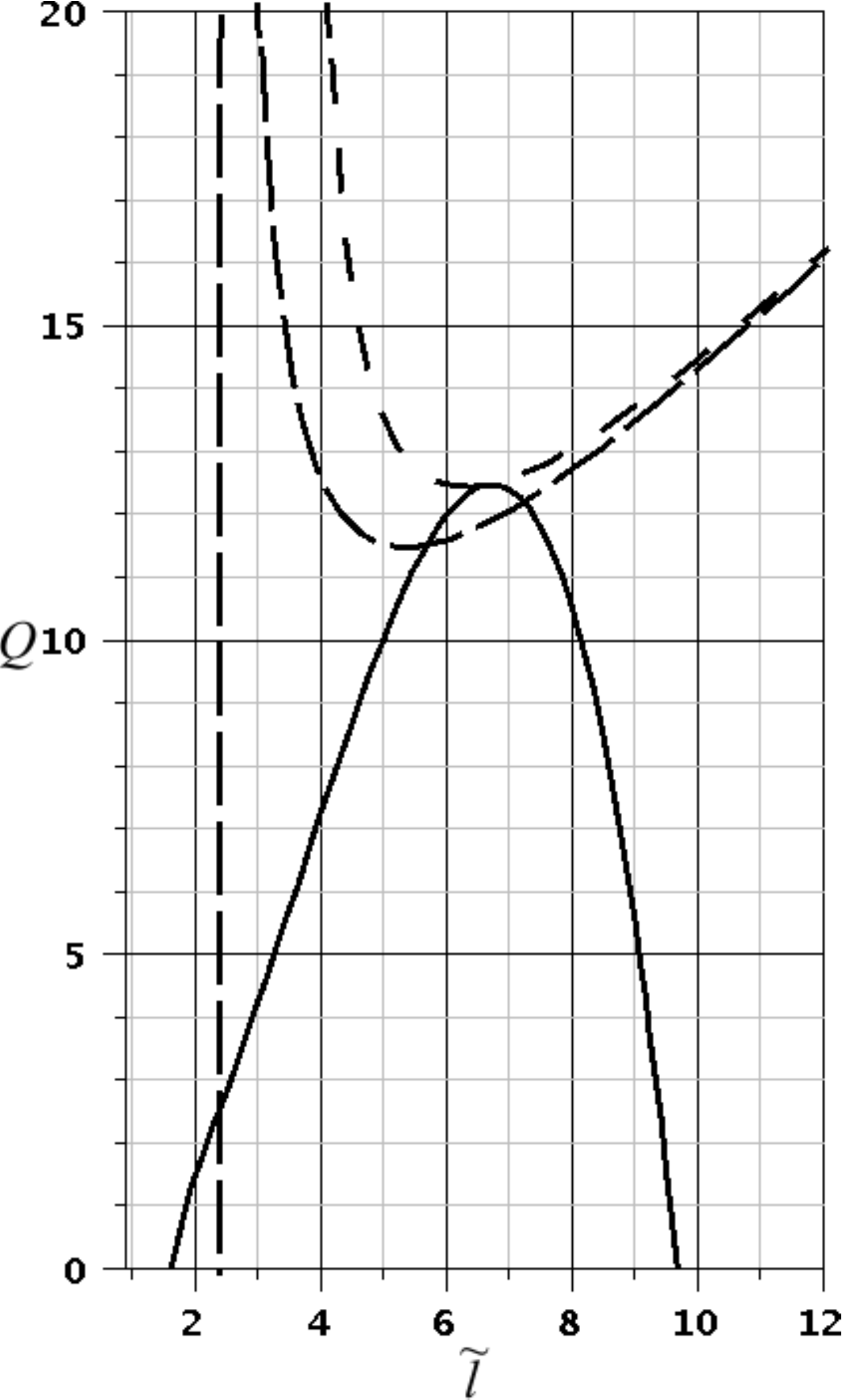}}\subfigure[e = 0.5]{\includegraphics[scale=0.25]{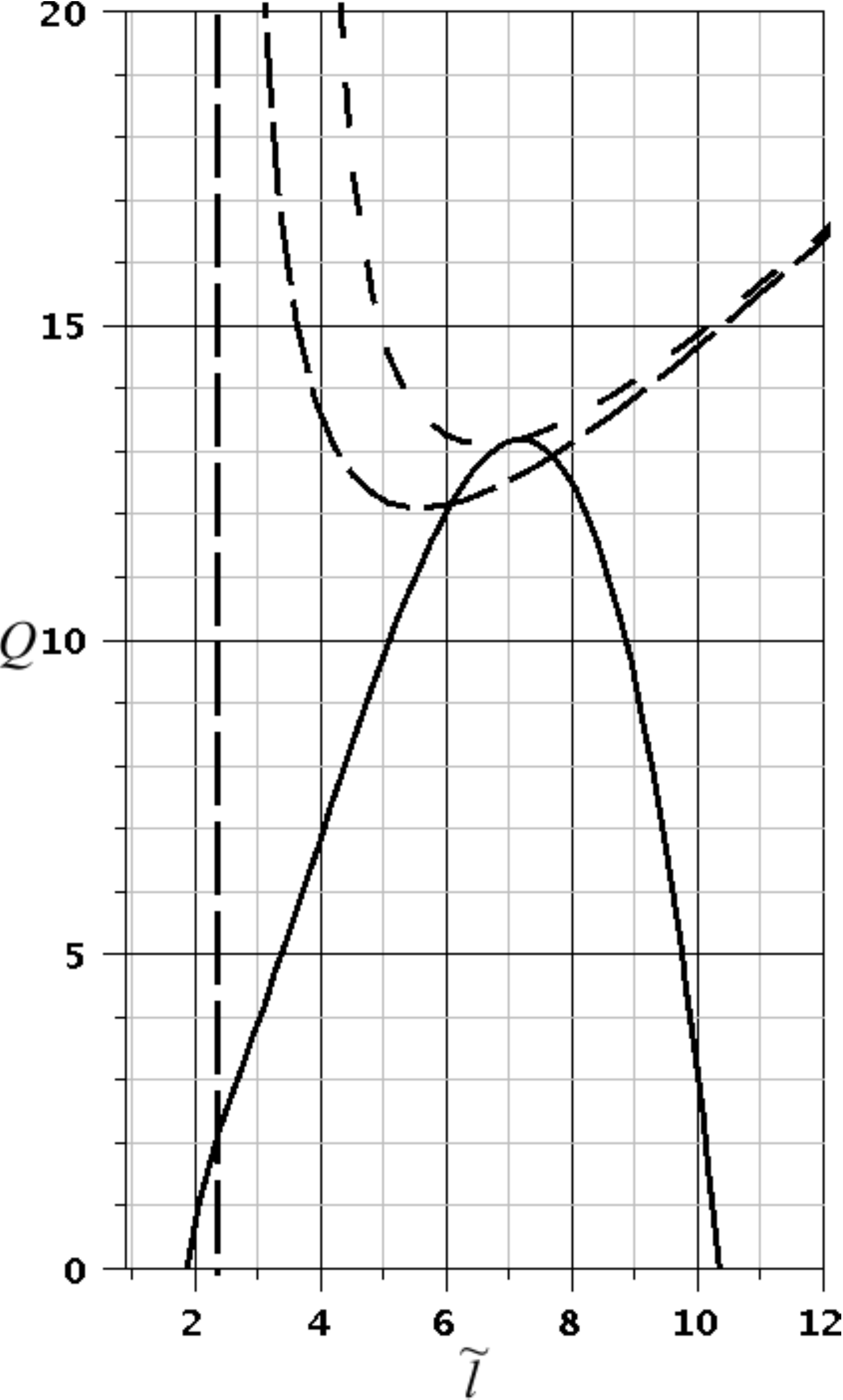}}\subfigure[e = 0.75]{\includegraphics[scale=0.25]{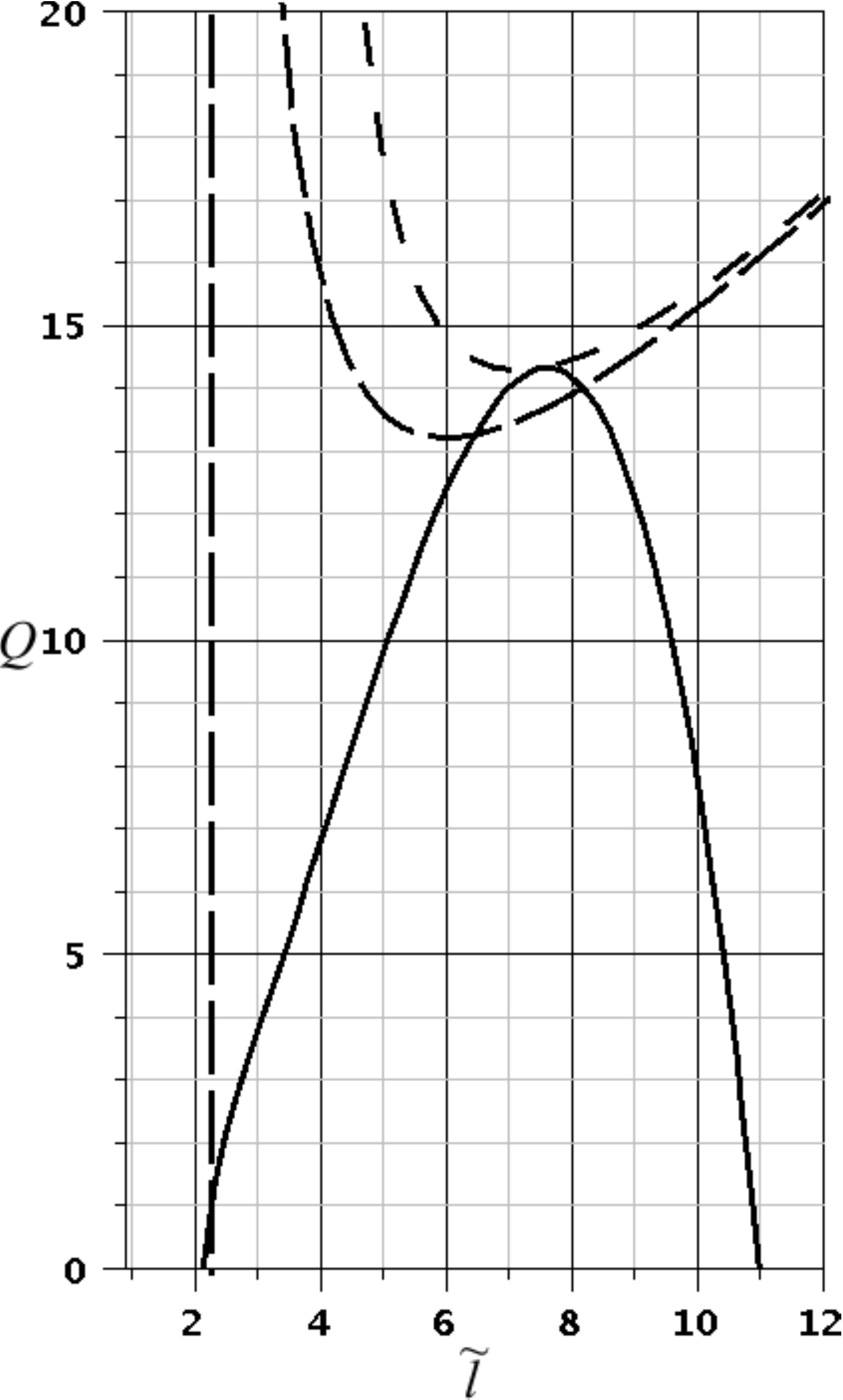}}\subfigure[Parabolic Orbit]{\includegraphics[scale=0.25]{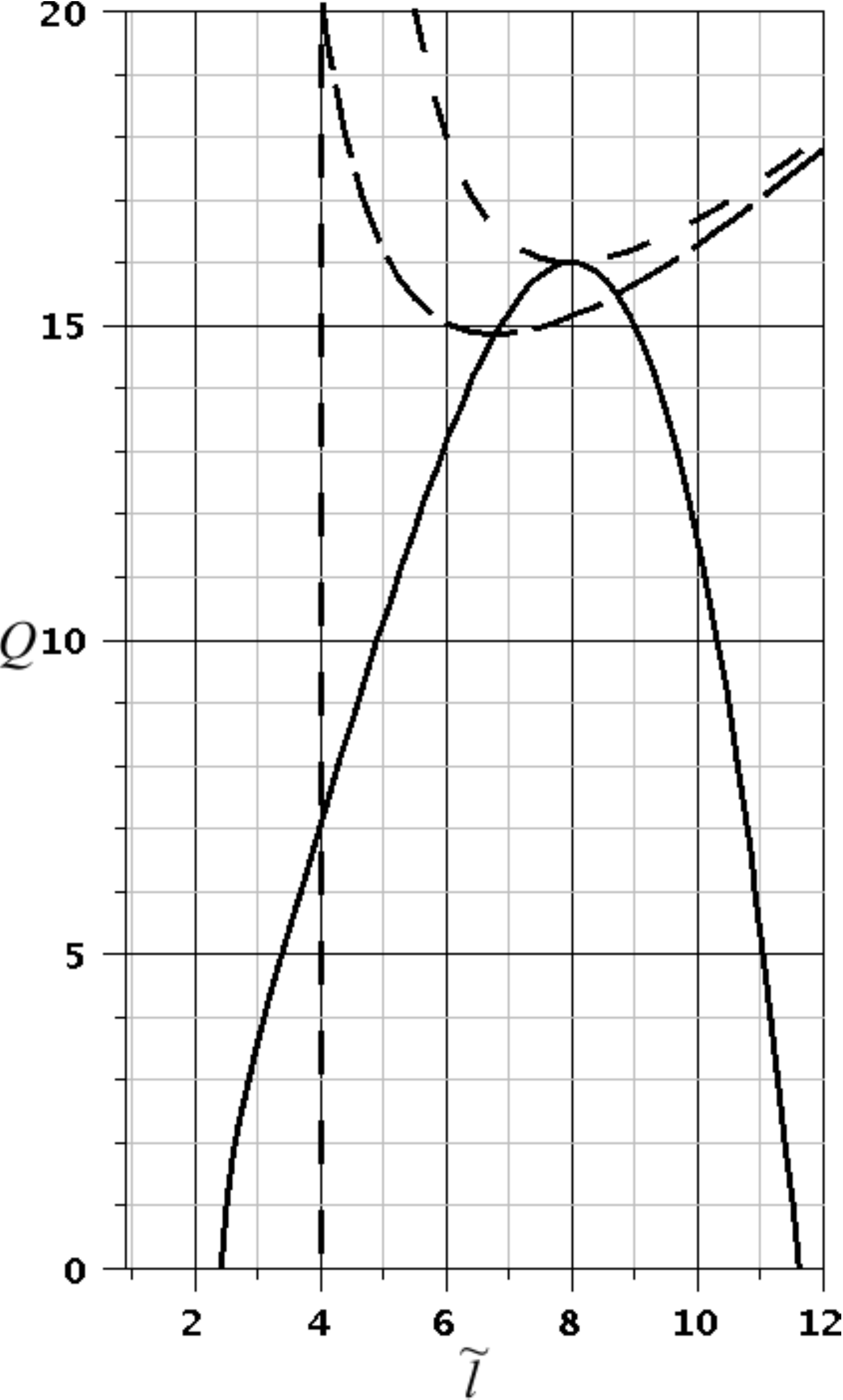}}
\end{sidewaysfigure}

In describing an arbitrary orbit, it must be recognised that each
parameter (${\tilde{l}}$, ${e}$, and ${Q}$) has a domain. The value
of ${e}$ lies between 0, for a circular orbit, and 1 for a parabolic
orbit. Although ${\tilde{l}}$ has no upper limit, its minimum value
is ${\tilde{l}}_{LSO}$; while ${Q}$, which is non-negative, does
have an upper limit that depends on the size of ${\tilde{l}}$. The
complicated interrelationships between these parameters can be better
understood if we derive a set of analytical formulae to describe the
behaviour of ${Q}$ with respect to ${\tilde{l}}$ and ${e}$. In
the sections that follow, we shall examine the LSO, abutment, and
polar orbits. A representative plot of these curves is shown in Figure
\ref{fig:S=00003D0.99} for ${e}=\left\{ 0.0,\,0.25,\,0.50,\,0.75,\,1.0\right\} $
and a KBH spin of ${\tilde{S}}=0.99$.

\subsection{Last Stable Orbit}

In \cite{Komorowski:2008wc} a new analytical formula for the latus
rectum of an elliptical equatorial LSO was developed. We can perform
a similar treatment for inclined LSOs; but the polynomial that results
is of ninth order in ${\tilde{l}}$, currently making the derivation
of an analytical solution infeasible. The use of the companion matrix
\cite{1996maco.book.....G,Edelman:1995rz,Komorowski:2008wc} simplifies
the numerical calculation of the prograde and retrograde ${\tilde{l}}_{LSO}$.

We refer back to equations (\ref{eq:Root_Pericentre}), (\ref{eq:Root_Apocentre}),
(\ref{eq:r3}), and (\ref{eq:r4}); but because ${\tilde{E}}$ will
equal the maximum value of ${\tilde{V}}_{+}$ (closest to the event
horizon) we can specify ${r}_{3}={r}_{p}$ as an additional condition
\cite{1983mtbh.book.....C,PhysRevD.50.3816,Komorowski:2008wc}. Therefore
the remaining root, ${r}_{4}$, can be easily solved \textit{viz.}
\begin{equation}
{\frac{r_{{4}}{\tilde{l}}^{3}}{\left(1+{e}\right)\left(1-{e}^{2}\right)}}={\frac{Q{\tilde{S}}^{2}}{1-{\tilde{E}}^{2}}},\end{equation}
to yield,\begin{equation}
{r}_{4}={\frac{Q{\tilde{S}}^{2}\left(1+{e}\right)\left(1-{e}^{2}\right)}{\left(1-{\tilde{E}}^{2}\right){\tilde{l}}^{3}}}.\end{equation}
Substituting the four roots (two of which are equal) into equation
(\ref{eq:4rs}) yields \begin{equation}
{\frac{Q{\tilde{S}}^{2}\left(1+{e}\right)\left(1-{e}^{2}\right)}{\left(1-{\tilde{E}}^{2}\right){\tilde{l}}^{3}}}+2\,{\frac{\tilde{l}}{1+{e}}}+{\frac{\tilde{l}}{1-{e}}}=2\,\left(1-{\tilde{E}}^{2}\right)^{-1}.\label{eq:LSO_Roots_Equation}\end{equation}
Substituting the expression for ${\tilde{E}}$ \textit{(}equation
(\ref{eq:E4})) into equation (\ref{eq:LSO_Roots_Equation}) and cross
multiplying, we obtain:\begin{equation}
{\tilde{l}}^{4}\left(1+{e}\right)^{2}\left(1-{e}\right)^{3}Z_{{10}}Z_{{11}}=0\label{eq:LSO_Raw_1}\end{equation}
where\[
Z_{10}=\left(-3\,\tilde{l}-2\,\tilde{l}{e}+{e}^{2}\tilde{l}\right){X}_{\pm}^{2}+4\, Q{\tilde{S}}^{2}+4\, Q{\tilde{S}}^{2}{e}-2\, Q\tilde{l}{e}+Q{e}^{2}\tilde{l}-3\, Q\tilde{l}+{\tilde{l}}^{3}\]
and\[
Z_{11}=\left(1-{e}^{2}\right)\left({\tilde{l}}{X}_{\pm}^{2}+Q\tilde{l}-Q{\tilde{S}}^{2}\right)-{\tilde{l}}^{3}.\]

By setting ${Z}_{10}=0$ and substituting the formula for ${X}_{\pm}^{2}$
(\textit{viz.} equation (\ref{eq:X2})) one obtains a polynomial,
${p}\left({\tilde{l}}\right)$, in terms of ${\tilde{l}}$ (to order
9), ${e}$, ${Q}$ (to second order), and ${\tilde{S}}$ (see equation
(\ref{eq:CP_LSO}) in Appendix B). The companion matrix of ${p}\left({\tilde{l}}\right)$
provides a powerful method to calculate the values of ${\tilde{l}}$
for both a prograde and retrograde LSO by numerically evaluating the
eigenvalues of the matrix. Optimised techniques for solving for eigenvalues
are available \cite{ISI:000074991300004,2005JPhCS..16..425Y}. Such
a numerical analysis was performed to obtain representative values
of ${\tilde{l}}_{LSO}$. The corresponding values of ${\tilde{l}}_{LSO}$,
which we derived from ${Z}_{11}=0$ (\textit{ceteris paribus}) were
smaller and thus not physically reachable by a test-particle. This
result demonstrates that ${Z}_{10}=0$ is the appropriate solution. 

Because ${p}\left({\tilde{l}}\right)$ is a quadratic in terms of
${Q}$, an alternative way to analyse the behaviour of orbits as they
approach their LSO is available. One solves for ${Q}$, analytically,
to obtain:

\begin{eqnarray}
{Q}_{LSO} & = & \frac{1}{4}\left({Z_{12}-{\tilde{S}}^{2}{Z}_{13}\sqrt{Z_{14}}}\right)\left({\tilde{S}}^{4}Z_{15}\right)^{-1},\label{eq:Q_LSO}\end{eqnarray}
where\begin{eqnarray*}
{Z}_{12} & = & {\tilde{l}}^{4}\left({e}+1\right)\left({\tilde{l}}^{2}-\left(2\,{e}^{2}+{e}+3\right)\tilde{l}+2\,\left({e}+1\right)\left(2\,{e}^{2}-{e}+3\right)\right){\tilde{S}}^{2}\\
 &  & -{\tilde{l}}^{3}\left({e}+1\right)^{2}\left(2\,{e}^{3}+{e}^{2}+\tilde{l}\left(3-{e}\right)+1\right){\tilde{S}}^{4},\end{eqnarray*}
\begin{eqnarray*}
Z_{13} & = & \left(3-{e}\right)\left(\left({e}+1\right)^{2}{\tilde{S}}^{2}-\tilde{l}\left(2\,{e}+2-\tilde{l}\right)\right),\end{eqnarray*}
\begin{eqnarray*}
Z_{14} & = & {\tilde{l}}^{5}\left({e}+1\right)^{3}\left(\left({e}-1\right)^{2}{\tilde{S}}^{2}+\tilde{l}\left(\tilde{l}+2\,{e}-2\right)\right),\end{eqnarray*}
 and\begin{eqnarray*}
{Z}_{15} & = & \left({e}+1\right)^{2}\left(\left({e}-1\right)^{2}\left({e}+1\right)^{3}{\tilde{S}}^{2}\right.\\
 &  & \left.-\tilde{l}\left({\tilde{l}}^{2}-\left({e}+1\right)\left({e}^{2}-2\,{e}+3\right)\tilde{l}+\left(2\,{e}^{2}-4\,{e}+3\right)\left({e}+1\right)^{2}\right)\right).\end{eqnarray*}

We now have, through equation (\ref{eq:Q_LSO}), the means to plot
the value of ${Q}_{LSO}$ with respect to ${\tilde{l}}_{LSO}$. For
a KBH, one may use ${\tilde{l}}_{LSO}$ in the domain $\left[{\tilde{l}}_{LSO,\: prograde},{\tilde{l}}_{LSO,\: retrograde}\right]$
as the independent variable.

\subsection{The Abutment}

The roots of equation (\ref{eq:Z9}) allow us to calculate the value
of ${Q}$ along the abutment (${Q}_{X}$) in terms of the orbital
values of ${\tilde{l}}$, ${e}$, and the KBH spin, ${\tilde{S}}$;
and we obtain, taking the minus sign of the quadratic solution:

\begin{eqnarray}
{Q}_{X} & = & \frac{{\tilde{l}}^{2}}{2{\tilde{S}}^{2}{\left(1-{e}^{2}\right)}^{2}}\left(\tilde{l}\left(\tilde{l}-{e}^{2}-3\right)\right.\nonumber \\
 &  & \left.-\sqrt{{\tilde{l}}^{2}\left(\tilde{l}-{e}^{2}-3\right)^{2}-4\,\tilde{l}\left(1-{e}^{2}\right)^{2}{\tilde{S}}^{2}}\right).\label{eq:Q_Abutment}\end{eqnarray}
Equation (\ref{eq:Q_Abutment}) appears to have poles at ${\tilde{S}}=0$
and ${e}=1$; but we can demonstrate that it reduces to a well behaved
function at these values of ${\tilde{S}}$ and ${e}$. We first factor
out the term, ${\tilde{l}}\left(\tilde{l}-{e}^{2}-3\right)^{}$, to
obtain\begin{eqnarray}
Q_{X} & = & \frac{{\tilde{l}}^{3}\left({\tilde{l}}-{e}^{2}-3\right)}{2\left(1-{e}^{2}\right)^{2}{\tilde{S}}^{2}}\left(1-\sqrt{1-\frac{4\left(1-{e}^{2}\right)^{2}{\tilde{S}}^{2}}{{\tilde{l}}\left({\tilde{l}}-{e}^{2}-3\right)^{2}}}\right).\label{eq:Q_Abutment_Intermediate_Form}\end{eqnarray}
Equation (\ref{eq:Q_Abutment_Intermediate_Form}) can be simplified
by making a binomial expansion of its square root\textit{ }to yield,\setlength{\mathindent}{0cm}\begin{eqnarray}
Q_{X} & = & \frac{{\tilde{l}}^{3}\left({\tilde{l}}-{e}^{2}-3\right)}{2\left(1-{e}^{2}\right)^{2}{\tilde{S}}^{2}}\Bigl\{1-1+\frac{2\left(1-{e}^{2}\right)^{2}{\tilde{S}}^{2}}{{\tilde{l}}\left({\tilde{l}}-{e}^{2}-3\right)^{2}}\label{eq:Q_Abutment_Intermediate_Form_2}\\
 &  & +2\left[\frac{\left(1-{e}^{2}\right)^{2}{\tilde{S}}^{2}}{{\tilde{l}}\left({\tilde{l}}-{e}^{2}-3\right)^{2}}\right]^{2}+4\left[\frac{\left(1-{e}^{2}\right)^{2}{\tilde{S}}^{2}}{{\tilde{l}}\left({\tilde{l}}-{e}^{2}-3\right)^{2}}\right]^{3}+10\left[\frac{\left(1-{e}^{2}\right)^{2}{\tilde{S}}^{2}}{{\tilde{l}}\left({\tilde{l}}-{e}^{2}-3\right)^{2}}\right]^{4}\ldots\Bigr\}.\nonumber \end{eqnarray}
\setlength{\mathindent}{2.5cm}Equation (\ref{eq:Q_Abutment_Intermediate_Form_2}),
simplifies to a power series in terms of ${\left(1-{e}^{2}\right)}^{2}{\tilde{S}}^{2}$;
therefore, only the first term will remain when ${\tilde{S}}=0$ or
${e}=1$. We now have a greatly simplified expression, which applies
to elliptic orbits around an SBH and parabolic orbits about a KBH,
i.e. \begin{eqnarray}
Q_{X} & = & {\tilde{l}}^{2}\left({\tilde{l}}-{e}^{2}-3\right)^{-1}\nonumber \\
 & = & {\tilde{l}}^{2}\left({\tilde{l}}-4\right)^{-1}.\label{eq:Q_Abutment_Limiting_Form}\end{eqnarray}
Further, as ${\tilde{l}}\rightarrow\infty$ in elliptical orbits,\begin{eqnarray}
\frac{4\left(1-{e}^{2}\right)^{2}{\tilde{S}}^{2}}{{\tilde{l}}\left({\tilde{l}}-{e}^{2}-3\right)^{2}} & \rightarrow0 & ,\end{eqnarray}
therefore, a similar binomial treatment may be performed on equation
(\ref{eq:Q_Abutment_Intermediate_Form}) to yield ${Q}_{X}\cong{\tilde{l}}+{e}^{2}+3$
for large ${\tilde{l}}$. This result is consistent with the fact
that as ${\tilde{l}}\rightarrow\infty$, the spacetime looks more
Schwarzschild in nature. As in the case of equation (\ref{eq:Q_Lz_Schwarzschild}),
${Q}_{X}$ (equation (\ref{eq:Q_Abutment})) also defines an upper
limit on ${Q}$ for specific values of ${\tilde{l}}$ and ${e}$.
The points above the ${Q}_{X}$ curve are inaccessible; and this property
can be shown by direct calculation.

This result confirms the choice of the minus sign in solving the quadratic
equation (\ref{eq:LatusRectumGeneralAbutment}) for ${Q}_{X}$. Recall
that equation (\ref{eq:X2}) applies to bound orbits in general and
is not restricted to LSOs (as is the case for ${L}_{z}^{2}$ (see
equation (\ref{eq:LzSquared}))); therefore, the value of ${Q}_{X}$
can be evaluated for all values of ${\tilde{l}}\geq{\tilde{l}}_{LSO,\: prograde}$;
and it will apply to both sets of orbits governed by ${X}_{-}^{2}$
or ${X}_{+}^{2}$.

\subsection{Polar Orbit}

We shall consider polar orbits in this paper; although orbits of arbitrary
inclination are also important. The polar orbit (governed by ${X}_{-}^{2}$)
is precisely defined by setting ${\tilde{L}}_{z}=0$ from which we
obtain ${X}_{-}^{2}={\tilde{S}}^{2}{\tilde{E}}^{2}$; therefore, we
can derive an analytical formula for ${Q}$ of a polar orbit (${Q}_{polar}$)
of arbitrary ${\tilde{l}}$ and ${e}$. 

We have the formula for ${X}_{-}^{2}$ in terms of ${\tilde{l}}$,
${e}$, and ${Q}$ (see equation (\ref{eq:X2})); and ${\tilde{E}}$
can also be expressed in these terms (see equation (\ref{eq:E4})).
Therefore ${X}_{-}^{2}-{\tilde{S}}^{2}{\tilde{E}}^{2}=0$ can be simplified
and factored to yield:\begin{eqnarray}
{A}{\left({B}_{1}{Q}-{\tilde{l}}^{2}{B}_{2}\right)}\left({C}_{1}{Q}-{\tilde{l}}^{2}{C}_{2}\right) & = & 0\label{eq:Q_Polar_raw}\end{eqnarray}
where

\begin{eqnarray*}
A & = & {\tilde{l}}^{3}-\left(2\,{e}^{2}+6\right){\tilde{l}}^{2}+\left({e}^{2}+3\right)^{2}{\tilde{l}}-4\,{S}^{2}\left(1-{e}^{2}\right)^{2},\end{eqnarray*}
\begin{eqnarray*}
{B}_{1}={\tilde{l}}^{5}{b}_{1} & = & {\tilde{l}}^{5}\Bigl\{1-\left(3+{e}^{2}\right){\tilde{l}}^{-1}+2\,{\tilde{S}}^{2}\left(1+{e}^{2}\right){\tilde{l}}^{-2}-2\,{\tilde{S}}^{2}\left(1-{e}^{2}\right)^{2}{\tilde{l}}^{-3}\\
 &  & +{\tilde{S}}^{4}\left(1-{e}^{2}\right)^{2}{\tilde{l}}^{-4}+{\tilde{S}}^{4}\left(1-{e}^{2}\right)^{3}{\tilde{l}}^{-5}\Bigr\},\end{eqnarray*}
\begin{eqnarray*}
{B}_{2}={\tilde{l}}^{4}{b}_{2} & = & {\tilde{l}}^{4}\left\{ 1+2\,{\tilde{S}}^{2}\left(1+{e}^{2}\right){l}^{-2}\right.\\
 &  & -4\,{\tilde{S}}^{2}\left(1-{e}^{2}\right){\tilde{l}}^{-3}+{\tilde{S}}^{4}\left(1-{e}^{2}\right)^{2}{\tilde{l}}^{-4}\Bigr\},\end{eqnarray*}

\begin{eqnarray*}
{C}_{1}={\tilde{l}}^{5}{c}_{1} & = & {\tilde{l}}^{5}\Bigl\{1-\left(3+{e}^{2}\right){\tilde{l}}^{-1}+2\,{\tilde{S}}^{2}\left(1+{e}^{2}\right){\tilde{l}}^{-2}+2\,{\tilde{S}}^{2}\left(1-{e}^{2}\right)^{2}{\tilde{l}}^{-3}\\
 &  & -3{\tilde{S}}^{4}\left(1-{e}^{2}\right)^{2}{\tilde{l}}^{-4}+{\tilde{S}}^{4}\left(1-{e}^{2}\right)^{3}{\tilde{l}}^{-5}\Bigr\},\end{eqnarray*}
and\begin{eqnarray*}
{C}_{2}={\tilde{l}}^{4}{c}_{2} & = & {\tilde{l}}^{4}\left\{ 1-4{\tilde{S}}^{2}{\tilde{l}}^{-1}+2\,{\tilde{S}}^{2}\left(3-{e}^{2}\right){l}^{-2}\right.\\
 &  & -4\,{\tilde{S}}^{2}\left(1-{e}^{2}\right){\tilde{l}}^{-3}+{\tilde{S}}^{4}\left(1-{e}^{2}\right)^{2}{\tilde{l}}^{-4}\Bigr\}.\end{eqnarray*}

The factor, ${A}$, offers no physically meaningful results. It does
not provide a solution for ${Q}$; and for $0\leqq{\tilde{S}}<1.0$
and $0\leqq{e}\leqq1.0$, we find that $3\leq{\tilde{l}}\leq4$, which
lies beyond the LSO. The factor, ${B}_{1}{Q}-{\tilde{l}}^{2}{B}_{2}=0$,
yields the result\begin{eqnarray}
{Q}_{polar} & = & {\tilde{l}}^{2}{B}_{2}{B}_{1}^{-1}.\label{eq:Q_polar_physical}\end{eqnarray}
And the factor, ${C}_{1}{Q}-{\tilde{l}}^{2}{C}_{2}=0$, yields the
result\begin{eqnarray}
{Q}_{polar} & = & {\tilde{l}}^{2}{C}_{2}{C}_{1}^{-1}.\label{eq:Q_polar_unphysical}\end{eqnarray}

We examine equations (\ref{eq:Q_polar_physical}) and (\ref{eq:Q_polar_unphysical})
to discover which one is physically significant. In the Schwarzschild
limit (${\tilde{S}}\rightarrow0$) we find that they coincide. But
let us consider the weak-field limit (${\tilde{l}}\rightarrow\infty$).
Equation (\ref{eq:Q_polar_physical}), ${Q}_{polar}={\tilde{l}}{b}_{2}{b}_{1}^{-1}$,
can be expanded in powers of ${\tilde{l}}$. And the terms with powers
of ${\tilde{l}}^{-1}$ and lower approach zero as ${\tilde{l}}\rightarrow\infty$
to yield the asymptotic limit of equation (\ref{eq:Q_polar_physical}):
\begin{eqnarray}
{Q}_{polar} & \cong & {\tilde{l}}+3+{e}^{2}.\end{eqnarray}
In a similar treatment of equation (\ref{eq:Q_polar_unphysical})
one obtains: \begin{eqnarray}
{\tilde{l}}{c}_{2}{c}_{1}^{-1} & \cong & {\tilde{l}}+3+{e}^{2}-4{\tilde{S}}^{2},\end{eqnarray}
which we can disregard as unphysical because it incorrectly implies
that the spin of the KBH influences the trajectory of a test-particle
at infinity. This situation differs from that described by equation
(\ref{eq:Q_switch_large_l}), which does not describe a trajectory.
Test calculations confirm that equation (\ref{eq:Q_polar_physical})
is the appropriate choice and equation (\ref{eq:Q_polar_unphysical})
can be disregarded since the values of ${\tilde{l}}_{LSO}$ obtained
by solving equation (\ref{eq:Q_polar_unphysical}) are less than (or
equal to) the results from equation (\ref{eq:Q_polar_physical}) (\textit{ceteris
paribus}). Equation (\ref{eq:Q_polar_physical}) applies to all bound
orbits, hence ${Q}_{polar}$ can be evaluated over a range ${\tilde{l}}\geq{\tilde{l}}_{LSO,\: prograde}$;
but it applies only to orbits governed by ${X}_{-}^{2}$. In the SBH
case (${\tilde{S}}=0$), ${B}_{1}={\tilde{l}}^{5}-\left(3+{e}^{2}\right){\tilde{l}}^{4}$
and ${B}_{2}={\tilde{l}}^{4}$. Therefore

\begin{eqnarray}
{Q}_{polar} & = & {\tilde{l}}^{2}\left({\tilde{l}}-{e}^{2}-3\right)^{-1},\end{eqnarray}
which has an asymptotic behaviour similar to that of ${Q}_{X}$.

\subsection{Some Characteristics of the Carter Constant Formulae}

\subsubsection{The last stable orbit on the abutment.}

\begin{figure}
\caption{\label{fig:The-three-important}The three ${Q}$ formulae derived
in Section \ref{sec:The-characteristics-of-the-Carter-Constant-Equations}
define a map. In zone (A), only prograde orbits are found. And in
zone (B), both prograde and retrograde orbits are found. Above the
${Q}_{polar}$ curve (zones (a) and (b) ) only retrograde orbits can
exist. But in zone (a), the orbits are governed by ${X}_{-}^{2}$;
while in zone (b), they are governed by either ${X}_{-}^{2}$ or ${X}_{+}^{2}$.
The points along ${Q}_{X}$ and ${Q}_{LSO}$ mark the values of ${\iota}$.
In this case the orbit is circular (${e}=0$) and the KBH spin is
${\tilde{S}}=0.99$.}
\includegraphics[scale=0.4]{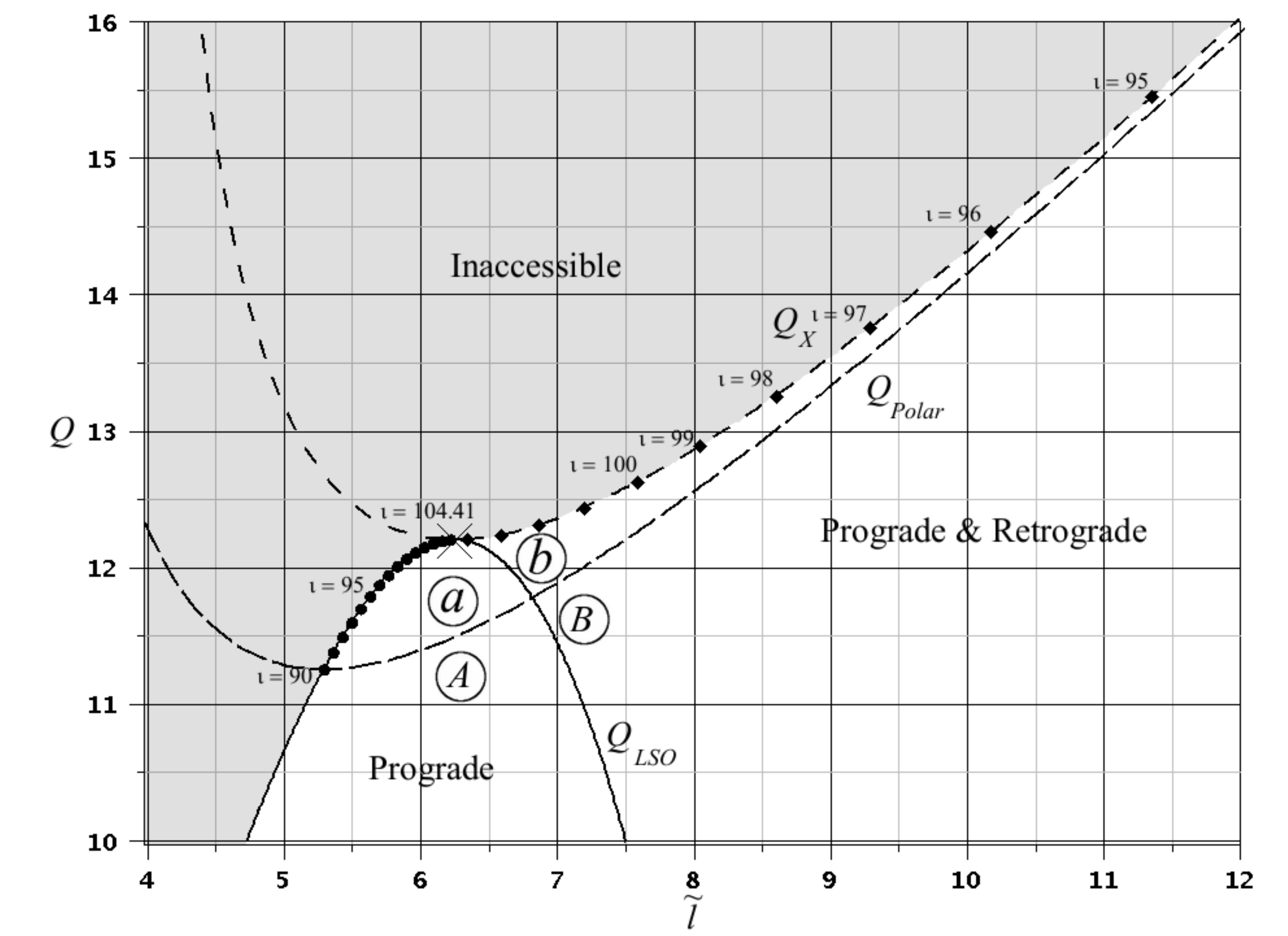}
\end{figure}

In Figures (\ref{fig:S=00003D0.99}) and (\ref{fig:The-three-important})
one observes that the functions for ${Q}_{X}$ and ${Q}_{LSO}$ intersect
at a single tangential point, which represents the value of ${\tilde{l}}$
of an LSO that lies at the abutment described by ${X}_{\pm}^{2}$.
The equation ${Q}_{X}-{Q}_{LSO}=0$ (see equations (\ref{eq:Q_Abutment})
and (\ref{eq:Q_LSO})) can be solved to yield:\begin{eqnarray}
{\tilde{l}}_{LSO,\: abutment} & = & \frac{1}{12}\left({Z}_{16}^{\frac{1}{3}}+{\chi}_{1}+{\chi}_{0}{Z}_{16}^{-\frac{1}{3}}\right),\label{eq:LSO_Abutment}\end{eqnarray}
where\begin{eqnarray*}
{Z}_{16} & = & \left(216\left(1+{e}\right)\left(5+{e}\right)^{2}\left(1-{e}\right)^{2}{\tilde{S}}^{2}+{\chi}_{1}{\chi}_{2}{\chi}_{3}\right)+12\sqrt{3{Z}_{17}},\end{eqnarray*}
\begin{eqnarray*}
{Z}_{17} & = & \left(1+{e}\right)\left(5+{e}\right)^{2}\\
 & \times & \left\{ 108\left(1+{e}\right)\left(5+{e}\right)^{2}\left(1-{e}\right)^{4}{\tilde{S}}^{4}\right.\\
 &  & +{\chi}_{1}{\chi}_{2}{\chi}_{3}\left(1-{e}\right)^{2}{\tilde{S}}^{2}\\
 &  & \left.-\left(1+{e}\right)\left(9-{e}^{2}\right)^{2}{\chi}_{4}^{2}\right\} ,\end{eqnarray*}
\begin{eqnarray*}
{\chi}_{0} & = & {e}^{6}+6{e}^{5}-9{e}^{4}-60{e}^{3}+111{e}^{2}+342{e}+441,\end{eqnarray*}
\begin{eqnarray*}
{\chi}_{1} & = & {e}^{3}+3{e}^{2}+3{e}+33,\end{eqnarray*}
\begin{eqnarray*}
{\chi}_{2} & = & {e}^{3}+3{e}^{2}-9{e}-3,\end{eqnarray*}
\begin{eqnarray*}
{\chi}_{3} & = & {e}^{3}+3{e}^{2}-21{e}-39,\end{eqnarray*}
and\begin{eqnarray*}
{\chi}_{4} & = & {e}^{3}+3{e}^{2}-5{e}+9.\end{eqnarray*}

It is at this tangential point that the ${Q}_{LSO}$ curve is split
into two segments: the minus segment (${\tilde{l}}<{\tilde{l}}_{LSO,\: abutment}$)
that defines the values of ${\tilde{l}}_{LSO}$ (and ${Q}_{LSO}$)
associated with inclined LSOs that are governed by ${X}_{-}^{2}$;
and the plus segment (${\tilde{l}}>{\tilde{l}}_{LSO,\: abutment}$),
which corresponds to inclined LSOs governed by ${X}_{+}^{2}$. Consequently,
the points beneath ${Q}_{LSO}$ define only orbits governed by ${X}_{-}^{2}$.
Further, orbits with ${\tilde{l}}$ and ${Q}$ values that lie to
the left of the minus segment of ${Q}_{LSO}$ are undefined. Therefore
the ${Q}_{LSO}$ curve for ${\tilde{l}}<{\tilde{l}}_{LSO,\: abutment}$
and the ${Q}_{X}$ curve for ${\tilde{l}}>{\tilde{l}}_{LSO,\: abutment}$
define a curve along which the upper limit of ${Q}$ is specified.

Calculations of the value of ${\tilde{l}}_{LSO,\, abutment}$ (equation
(\ref{eq:LSO_Abutment})) were performed for various eccentricities
and two values of KBH spin (${\tilde{S}}=0.5$ and ${\tilde{S}}=0.99$);
they are listed in table \ref{tab:Maxima-Minima_LSO}. We have also
numerically calculated the values of ${\tilde{l}}$ for the maximal
point of the ${Q}_{LSO}$ curve and listed them for comparison. For
both circular and parabolic orbits, \begin{eqnarray}
{\tilde{l}}_{LSO,\, abutment} & = & {\tilde{l}}_{LSO,\, max},\label{eq:LSO_abutment_equal}\end{eqnarray}
irrespective of the value of ${\tilde{S}}$. Otherwise, \begin{eqnarray}
{\tilde{l}}_{LSO,\, abutment} & < & {\tilde{l}}_{LSO,\, max},\label{eq:LSO_abutment_lt}\end{eqnarray}
from which one may infer the curve that specifies the upper limit
of ${Q}$ is monotonically increasing with respect to ${\tilde{l}}$
(with a point of inflection at ${\tilde{l}}_{LSO,\, abutment}$ for
circular and parabolic orbits). The values of ${\tilde{l}}_{LSO,\, abutment}$
show that the upper limit of ${Q}$ also increases monotonically with
respect to ${e}$. Further, ${\tilde{l}}_{LSO,\, abutment}=8.0$ for
a parabolic orbit, which equals the ${\tilde{l}}_{LSO}$ of a parabolic
orbit about an SBH.

\begin{table}
\caption{\label{tab:Maxima-Minima_LSO}Parameters calculated for the tangential
intersection of the ${Q}_{X}$ and ${Q}_{LSO}$ curves for an elliptical
orbit about a KBH of spin ${\tilde{S}}=0.5$ and ${\tilde{S}}=0.99$.
The ${\tilde{l}}$ of the LSO at the maximum point of the ${Q}_{LSO}$
curve is included for comparison.}
\begin{tabular}{cccccccccc}
\hline 
 &  &  & {\footnotesize $\tilde{S}=0.50$} &  &  &  &  & {\footnotesize $\tilde{S}=0.99$} & \tabularnewline
\hline
${e}$ & {\footnotesize ${Q}_{LSO}$ } & {\footnotesize ${\iota}$ (deg.)} & {\footnotesize ${\tilde{l}}_{LSO,\, abutment}$} & {\footnotesize ${\tilde{l}}_{LSO,\, max}$ } &  & ${Q}_{LSO}$  & ${\iota}$ (deg.) & {\footnotesize ${\tilde{l}}_{LSO,\, abutment}$ } & {\footnotesize ${\tilde{l}}_{LSO,\, max}$ }\tabularnewline
\hline
$0.0$ & $12.054503$ & $97.6$ & $6.067459$ & $6.067459$ &  & $12.203171$ & $104.4$ & $6.245480$ & $6.245480$\tabularnewline
$0.1$ & $12.102329$ & $97.2$ & $6.256148$ & $6.268429$ &  & $12.245340$ & $103.8$ & $6.407328$ & $6.449754$\tabularnewline
$0.2$ & $12.237084$ & $96.9$ & $6.445623$ & $6.467077$ &  & $12.366092$ & $103.3$ & $6.570627$ & $6.646723$\tabularnewline
$0.3$ & $12.449596$ & $96.7$ & $6.635958$ & $6.663706$ &  & $12.559674$ & $103.0$ & $6.735976$ & $6.836993$\tabularnewline
$0.4$ & $12.734670$ & $96.6$ & $6.827228$ & $6.858566$ &  & $12.823381$ & $102.8$ & $6.903945$ & $7.020995$\tabularnewline
$0.5$ & $13.090081$ & $96.5$ & $7.019514$ & $7.051863$ &  & $13.156912$ & $102.7$ & $7.075097$ & $7.199015$\tabularnewline
$0.6$ & $13.515997$ & $96.5$ & $7.212909$ & $7.243768$ &  & $13.562022$ & $102.7$ & $7.250004$ & $7.371204$\tabularnewline
$0.7$ & $14.014676$ & $96.6$ & $7.407517$ & $7.434425$ &  & $14.042359$ & $102.9$ & $7.429271$ & $7.537586$\tabularnewline
$0.8$ & $14.590357$ & $96.7$ & $7.603465$  & $7.623953$ &  & $14.603445$ & $103.1$ & $7.613543$ & $7.698050$\tabularnewline
$0.9$ & $15.249308$ & $96.9$ & $7.800900$ & $7.812451$ &  & $15.252772$ & $103.5$ & $7.803527$  & $7.852335$\tabularnewline
$1.0$ & $16.000000$ & $97.1$ & $8.000000$ & $8.000000$ &  & $16.000000$ & $103.9$ & $8.000000$ & $8.000000$\tabularnewline
\end{tabular}
\end{table}

\subsubsection{The last stable polar orbit.}

The polar curve applies to polar orbits, which are governed by ${X}_{-}^{2}$;
therefore, only the intersection of the ${Q}_{polar}$ curve with
the minus segment of the ${Q}_{LSO}$ curve (${\tilde{l}}<{\tilde{l}}_{LSO,\: abutment}$)
needs to be considered. It is at this point that the ${\tilde{l}}_{LSO}$
of a polar orbit of arbitrary ${e}$ is defined. It was found from
numerical calculations of ${\tilde{l}}_{polar,\, LSO}$ (where the
${Q}_{polar}$ curve intersects the minus segment of the ${Q}_{LSO}$
curve) and ${\tilde{l}}$ value at the minimal point of ${Q}_{polar}$
that in the case of circular and parabolic orbits, \begin{eqnarray}
{\tilde{l}}_{polar,\, min} & = & {\tilde{l}}_{polar,\, LSO};\label{eq:polar_LSO_equal}\end{eqnarray}
otherwise, \begin{eqnarray}
{\tilde{l}}_{polar,\, min} & < & {\tilde{l}}_{polar,\, LSO}.\label{eq:polar_LSO_lt}\end{eqnarray}
One may infer from this result that the value of ${Q}_{polar}$ is
monotonically increasing with respect to ${\tilde{l}}$ (and has a
point of inflection at ${\tilde{l}}_{polar,\, LSO}$ for circular
and parabolic orbits).

\subsubsection{The Schwarzschild limiting case.}

The analysis of these formulae in the case where ${\tilde{S}}\rightarrow0$
is an important test. An examination of the analytical formulae for
${Q}_{switch}$ (equation (\ref{eq:Q_switch_SBH})), ${Q}_{X}$ (equation
(\ref{eq:Q_Abutment})), and ${Q}_{polar}$ (equation (\ref{eq:Q_polar_physical})),
show that when ${\tilde{S}}=0$, all three formulae equal\begin{eqnarray}
{Q} & = & {\tilde{l}}^{2}\left({\tilde{l}}-{e}^{2}-3\right)^{-1}.\end{eqnarray}
In the Schwarzschild limit, we find that the abutment and the set
of polar orbits approach one another, as required by the spherical
symmetry of Schwarzschild spacetime.

\section{\label{sec:The-Analysis-of-the-Carter-Constant}The Analysis of the
Carter Constant for an Evolving Orbit}

\subsection{Introduction}

We shall now perform an analysis of the evolution of ${Q}$ in Kerr
spacetime in the domain, which is defined by the three ${Q}$ curves
we have derived (${Q}_{LSO}$, ${Q}_{X}$, ${Q}_{polar}$) in equations
(\ref{eq:Q_LSO}), (\ref{eq:Q_Abutment}), and (\ref{eq:Q_polar_physical}).
The behaviour of ${Q}_{switch}$ (equation (\ref{eq:Q_switch})) will
not be considered here, although it is important in guiding the choice
of sign in $\pm\sqrt{{X}_{-}^{2}}$.

The three equations for ${Q}_{X}$, ${Q}_{LSO}$, and ${Q}_{polar}$
define a map (see Figure \ref{fig:The-three-important}) from which
one might infer the characteristics of a path followed by an inclined
orbit as it evolves. These paths (${Q}_{path}$) fall into two families:
one governed by ${X}_{-}^{2}$ and the other by ${X}_{+}^{2}$. We
conjecture that paths in the same family never cross; therefore, if
${Q}_{path}$ reaches ${Q}_{X}$, then it can do so only once. Let
us concentrate on the behaviour of an evolving orbit at the abutment,
which is where, ${Q}_{path}$, may change from one family to the other.
Therefore there are two modes to consider:

\begin{eqnarray}
{X}_{-}^{2} & \Rightarrow & {X}_{+}^{2}\label{eq:Mode_Fast}\end{eqnarray}
and\begin{eqnarray}
{X}_{+}^{2} & \Rightarrow & {X}_{-}^{2}.\label{eq:Mode_Slow}\end{eqnarray}
The mode represented by equation (\ref{eq:Mode_Fast}) corresponds
to a rapid listing rate, where prograde orbits can cross ${Q}_{polar}$
and intersect ${Q}_{X}$. The mode represented by equation (\ref{eq:Mode_Slow})
corresponds to orbits that: list at a slow rate, have constant ${\iota}$,
or exhibit decreasing ${\iota}$ over time. And a prograde orbit cannot
reach ${Q}_{X}$ since ${Q}_{X}$ lies on the retrograde side of ${Q}_{polar}$.

For this paper we will consider the evolution of a circular orbit
(${e}=0$) because we wish to limit our initial analysis to the relationship
between ${Q}$ and ${\tilde{l}}$. Elliptical orbits will be treated
in a forthcoming paper \cite{2011arXiv1101.0996K}.

\subsection{The Evolutionary Path in the ${Q}$ - ${\tilde{l}}$ Plane}

In Figure \ref{fig:The-three-important} one may imagine a path, ${Q}_{path}$,
that starts at a large value of ${\tilde{l}}$ as both ${Q}$ and
${\tilde{l}}$ monotonically decrease with respect to time. If the
curve reaches ${Q}_{X}$ then it must intersect it tangentially (as
the zone above the ${Q}_{X}$ curve is inaccessible). It is at that
point that the orbit ceases to be governed by ${X}_{\mp}^{2}$ and
is governed by ${X}_{\pm}^{2}$.

At the abutment, $\partial{Q}/\partial{\tilde{l}}$ and $\partial{Q}/\partial{e}$
can be calculated analytically (see Appendix B) irrespective of the
model used to determine the radiation back reaction. Given an ${\dot{\tilde{l}}}=d{\tilde{l}}/d{t}$
and ${\dot{e}}=d{e}/d{t}$ that have been derived according to some
independent model, then according to a linear approximation one may
define \begin{eqnarray}
{\dot{Q}} & = & \frac{\partial{Q}}{\partial{\tilde{l}}}{\dot{\tilde{l}}}+\frac{\partial{Q}}{\partial{e}}{\dot{e}}.\label{eq:Q_dot}\end{eqnarray}
For a circular orbit (${e}=0$) it has been proven that ${\dot{e}}=0$
\cite{1996PhRvD..53.4319K}; therefore, the second term in equation
(\ref{eq:Q_dot}) is zero.

\subsubsection{\label{sub:A-Preliminary-Test}A Preliminary Test at the Abutment.}

Because we can perform an analytical calculation of $\partial{Q}/\partial{\tilde{l}}$
at the abutment, we can estimate ${\dot{\tilde{l}}}$ \textit{viz.}
equation (\ref{eq:Q_dot}),\begin{eqnarray}
\frac{\partial{\tilde{l}}}{\partial{t}} & = & {\dot{Q}}\left(\frac{\partial{Q}}{\partial{\tilde{l}}}\right)^{-1},\label{eq:l_dot}\end{eqnarray}
if the 2PN ${Q}$ flux (see equation (A.3) in \cite{2007PhRvD..76d4007B}
(after equation (56) in \cite{2006PhRvD..73f4037G})) is known, i.e.\begin{eqnarray}
\left(\frac{\partial{Q}}{\partial{t}}\right)_{2PN} & = & -{\sin\left({\iota}\right)}\frac{64}{5}\frac{{m}}{{M}^{2}}{\left(1-{e}^{2}\right)^{3/2}}{\tilde{l}}^{-7/2}\sqrt{{Q}}\nonumber \\
 & \times & \Bigl[{g}_{9}\left({e}\right)-{\tilde{l}}^{-1}{g}_{11}\left({e}\right)+\left({\pi}{g}_{12}\left({e}\right)-{\cos\left({\iota}\right)}{\tilde{S}}{g}_{10}^{b}\left({e}\right)\right){\tilde{l}}^{-3/2}\nonumber \\
 &  & -\left({g}_{13}\left({e}\right)-{\tilde{S}}^{2}{\left({g}_{14}\left({e}\right)-\frac{45}{8}{\sin}^{2}\left({\iota}\right)\right)}\right){\tilde{l}}^{-2}\Bigr],\label{eq:Qdot_2PN}\end{eqnarray}
where the functions ${g}_{9}\left({e}\right)$, ${g}_{10}^{b}\left({e}\right)$,
${g}_{11}\left({e}\right)$, ${g}_{12}\left({e}\right)$, ${g}_{13}\left({e}\right)$,
and ${g}_{14}\left({e}\right)$ are listed in Appendix B (the Carter
constant, $Q$, has been normalised by dividing by $\left(mM\right)^{2}$).
We performed test calculations on circular orbits of ${\tilde{l}}=\left\{ 7.0,\;100.0\right\} $
with KBH spin ${\tilde{S}}=\left\{ 0.05,\;0.95\right\} $, which correspond
to those used by Hughes \cite{PhysRevD.61.084004}. In table \ref{tab:l_dot_compared_to_Hughes},
we compare our results with those of Hughes to find that they are
reasonably consistent, with some deviation for ${\tilde{S}}=0.95$
and ${\tilde{l}}=7.0$. We have adjusted the results in Hughes by
dividing them by $\sin\left(\iota\right)$ (where ${\iota}\cong\pi/3$)
so that they will correspond to our near-polar orbits; but this is
only an approximation since ${\iota}$ appears in other terms in equation
(\ref{eq:Qdot_2PN}).

\begin{table}
\caption{\label{tab:l_dot_compared_to_Hughes}An estimate of $\left(M^{2}/m\right)\partial{\tilde{l}}/\partial{t}$,
based on equation (\ref{eq:l_dot}), for circular orbits. The values
in parenthesis (which were originally calculated at ${\iota}\cong\pi/3$)
are taken from Hughes \cite{PhysRevD.61.084004} and have been adjusted
by dividing by $\sin\left({\iota}\right)$.}

\begin{tabular}{ccccc}
 & \multicolumn{2}{c}{${\tilde{l}}=7$} & \multicolumn{2}{c}{${\tilde{l}}=100$}\tabularnewline
\hline
${\tilde{S}}=0.05$ & $-1.1827\times{10}^{-1}$  & ($-1.2638\times{10}^{-1}$) & $-1.2683\times{10}^{-5}$  & ($-1.4637\times{10}^{-5}$)\tabularnewline
${\tilde{S}}=0.95$ & $-1.3747\times{10}^{-1}$  & ($-5.2540\times{10}^{-2}$) & $-1.2679\times{10}^{-5}$  & ($-1.4553\times{10}^{-5}$)\tabularnewline
\end{tabular}
\end{table}

\subsection{The analysis of $\iota$ on the abutment}

The analysis of $\iota$ and its derivatives with respect to $\tilde{l}$
and $e$ will be treated in detail in a forthcoming paper \cite{2011arXiv1101.0996K};
but it is appropriate to present a short preliminary exploration here.

Unlike $Q_{2PN}$ (which appears in equation (\ref{eq:Qdot_2PN})),
$Q_{X}$ contains no explicit variable, $\iota$; therefore, for circular
orbits this property greatly simplifies the total derivative of $Q_{X}$
with respect to $\tilde{l}$ since $\partial Q_{X}/\partial\iota=0$,
i.e.\begin{eqnarray}
\frac{dQ_{X}}{d\tilde{l}} & = & \frac{\partial Q_{X}}{\partial\tilde{l}}+\frac{\partial Q_{X}}{\partial\iota}\frac{\partial\iota}{\partial\tilde{l}}\nonumber \\
 & = & \frac{\partial Q_{X}}{\partial\tilde{l}}.\label{eq:Qx_Total_Derivative}\end{eqnarray}
Therefore equation (\ref{eq:Qx_Total_Derivative}) demonstrates that
$\partial\iota/\partial\tilde{l}$ is not constrained on the abutment
in the same way as $\partial Q/\partial\tilde{l}$ (see section \ref{sub:A-Preliminary-Test}).
Consider equations (\ref{eq:Q_Abutment}), (\ref{eq:X2}), (\ref{eq:E4}),
and (\ref{eq:Theta-Roots-Equatorial}). They form a calculation sequence,
which on the abutment creates a one to one mapping, $Q_{X}\rightarrow\iota$;
otherwise, there are two possible values of $\iota$ for a given value
of $Q$. Thus $\partial\iota/\partial\tilde{l}$ can be found either
by numerical methods or analytically \cite{2011arXiv1101.0996K}. In the
remainder of this section we shall investigate the behaviour of $\partial\iota/\partial\tilde{l}$
for orbits on the abutment.

\subsubsection{Numerical Analysis of ${\iota}$ on the abutment}

We can numerically estimate the change of ${\iota}$ with respect
to ${\tilde{l}}$ at the abutment by first finding the change in ${Q}$
for an extrapolation of the evolving orbit's path (${Q}_{path}$).
Because both ${Q}_{path}={Q}_{X}$ and ${\partial Q}_{path}/{\partial\tilde{l}}={\partial Q}_{X}/{\partial\tilde{l}}$
at the point where ${Q}_{path}$ intersects ${Q}_{X}$, the equations
of the second-order extrapolation of ${Q}_{path}$ at the abutment
can be written as\begin{eqnarray}
{Q}_{path}\left({\tilde{l}}-{\delta}{\tilde{l}}\right) & = & {Q_{X}}-{\delta}{\tilde{l}}\frac{\partial{Q}_{X}}{\partial{\tilde{l}}}+\frac{{\delta}{\tilde{l}}^{2}}{2}\frac{\partial^{2}{Q}_{path}}{\partial{\tilde{l}}^{2}}\label{eq:Second_Order_Extrapolation_Equations_a}\\
{Q}_{path}\left({\tilde{l}}+{\delta}{\tilde{l}}\right) & = & {Q_{X}}+{\delta}{\tilde{l}}\frac{\partial{Q}_{X}}{\partial{\tilde{l}}}+\frac{{\delta}{\tilde{l}}^{2}}{2}\frac{\partial^{2}{Q}_{path}}{\partial{\tilde{l}}^{2}}.\label{eq:Second_Order_Extrapolation_Equations_b}\end{eqnarray}
These equations are used to calculate ${\iota}_{-}$ (associated with
${\tilde{l}}-{\delta}{\tilde{l}}$) and ${\iota}_{+}$ (${\tilde{l}}+{\delta}{\tilde{l}}$),
where $\delta{\tilde{l}}$ is the small amount (${10}^{-32}$) by
which we extrapolate from the value of ${\tilde{l}}$ at which ${Q}_{X}$
and ${Q}_{path}$ intersect. Equations (\ref{eq:Second_Order_Extrapolation_Equations_a})
and (\ref{eq:Second_Order_Extrapolation_Equations_b}) include the
second derivative of ${Q}_{path}$ with respect to ${\tilde{l}}$,
which warrants further analysis. 

Because $\delta{\tilde{l}}$ is so small we used MATLAB, set to a
precision of 256 digits, to perform the numerical analysis. From these
extrapolated values of ${Q}_{path}$ we obtain\begin{eqnarray}
\frac{\partial{\iota}}{\partial{\tilde{l}}} & \cong & {\left({\iota}_{+}-{\iota}_{-}\right)}{\left(2{\delta}{\tilde{l}}\right)}^{-1}.\label{eq:diotabydl_numerical}\end{eqnarray}
N.B.: Since ${\iota}$ increases as ${\tilde{l}}$ decreases (with
respect to time), $\partial{\iota}/\partial{\tilde{l}}\leqq0$.

\subsubsection{\label{sub:The-first-order-extrapolation}The first-order extrapolation.}

Consider a first-order linear approximation, in which we drop the
second derivatives in equations (\ref{eq:Second_Order_Extrapolation_Equations_a})
and (\ref{eq:Second_Order_Extrapolation_Equations_b}). The derivation
of the corresponding change in ${\iota}$ requires a sequence of calculations
to be performed, which we will briefly outline.

\begin{enumerate}
\item \label{enu:Specify-the-spin}Specify the spin (${\tilde{S}}$) of
the KBH.
\item \label{enu:Select-the-values}Select the values of ${\tilde{l}}$
and ${e}$ for the point of intersection with the abutment. For this
work, ${e}=0$.
\item Calculate ${Q}_{X}$ using equation (\ref{eq:Q_Abutment}) and ${\partial{Q}_{X}/\partial{\tilde{l}}}$
using equation (\ref{eq:Appendix_D_dQXdl}) (given in Appendix B).
\item Calculate ${\tilde{l}}_{LSO,\, abutment}$ using equation (\ref{eq:LSO_Abutment}).
It must be smaller than the value of ${\tilde{l}}$ specified in point
(\ref{enu:Select-the-values}) otherwise the test-particle would be
placed beyond the LSO.
\item \label{enu:Use-equations} Calculate the values of ${Q}_{path}\left({\tilde{l}}-\delta{\tilde{l}}\right)$
and ${Q}_{path}\left({\tilde{l}}+\delta{\tilde{l}}\right)$ according
to a prescribed estimate or extrapolation at ${\tilde{l}}$.
\item Calculate ${X}_{-}$ and ${X}_{+}$(equation (\ref{eq:X2})) from
${Q}_{path}\left({\tilde{l}}+\delta{\tilde{l}}\right)$ (We use $-{\sqrt{{X}_{+}^{2}}}$
and we must use $-{\sqrt{{X}_{-}^{2}}}$ if ${Q}_{path}\left({\tilde{l}}\pm\delta{\tilde{l}}\right)>{Q}_{switch}$).
\item Calculate ${X}_{+}$ and ${X}_{-}$ from ${Q}_{path}\left({\tilde{l}}-\delta{\tilde{l}}\right)$. 
\item Using equation (\ref{eq:E4}), calculate the orbital energies (${\tilde{E}}$)
for each of ${Q}_{path}\left({\tilde{l}}-\delta{\tilde{l}}\right)$
and ${Q}_{path}\left({\tilde{l}}+\delta{\tilde{l}}\right)$.
\item Now that the values of ${X}_{+}$, ${X}_{-}$, ${\tilde{E}}$, ${\tilde{l}}$
are known; the value of ${\iota}$ (${\iota}_{-}$ and ${\iota}_{+}$)
can be calculated \textit{viz.} equation (\ref{eq:Theta-Roots-Equatorial}).
We use the expression, ${\tilde{L}}_{z}={X}+{\tilde{S}}{\tilde{E}}$.
\item We can estimate $\partial{\iota}/\partial{\tilde{l}}$ \textit{viz.}
equation (\ref{eq:diotabydl_numerical}).
\end{enumerate}
We have performed this sequence of first-order calculations for each
mode (equations (\ref{eq:Mode_Fast}) and (\ref{eq:Mode_Slow})) over
a range of ${\tilde{S}}$ and ${\tilde{l}}$ values; and a representative
set is shown in table \ref{tab:The-overestimated-values}. The slow
and fast modes yield $\partial{\iota}/\partial{\tilde{l}}$ values
that are of opposite sign; and these results might suggest that the
slow mode corresponds to orbits for which $\partial{\iota}/\partial{t}<0$.
But let us first assess the validity of the first-order approximation
by testing it in the Schwarzschild limit. The results for small ${\tilde{S}}$
(table \ref{tab:The-overestimated-values}) demonstrate that this
approximation is incomplete since it produces a non-zero result for
${\tilde{S}}\gtrapprox0$, which is unphysical. On each side of the
abutment, the value of ${Q}_{path}$ is underestimated. This observation
warrants the study of a more complete model that includes the higher
derivatives of ${Q}_{path}$. Indeed, we have found that using $\partial{Q}_{X}/\partial{\tilde{l}}$
alone cannot offer a sufficiently accurate mathematical description
of the orbital evolution at the abutment and warrants the development
of second and higher derivatives of ${Q}_{path}$. One reasonably
expects this numerical method to produce an accurate estimate of $\partial{\iota}/\partial{\tilde{l}}$;
but the transition an orbit makes at the abutment from ${X}_{\mp}^{2}$
to ${X}_{\pm}^{2}$ makes its behaviour more complicated.

\begin{sidewaystable}
\caption{\label{tab:The-overestimated-values}The values of $\partial{\iota}/\partial{\tilde{l}}$
(in radians) obtained by a linear extrapolation by ${\delta\tilde{l}}={10^{-32}}$. }
\begin{tabular}{ccccc}
\hline 
 & $\tilde{l}=6.25$ & $\tilde{l}=7.0$ & $\tilde{l}=10.0$ & $\tilde{l}=100$\tabularnewline
\hline
${\tilde{S}}=0.99$ &  &  &  & \tabularnewline
\hline
${\iota}$ (deg) & $104.3865$ & $101.5796$ & $96.1803$ & $90.1719$\tabularnewline
1st-order (slow) & $+8.7969235019\times{10}^{-2}$ & $+6.3997248415\times{10}^{-2}$ & $+2.6730499968\times{10}^{-2}$ & $+2.6454077633\times{10}^{-4}$\tabularnewline
1st-order (fast) & $-2.4667033336\times{10}^{-1}$  & $-1.7158228931\times{10}^{-1}$ & $-6.3094796456\times{10}^{-2}$ & $-3.5520692654\times{10}^{-4}$ \tabularnewline
2nd-order (slow) & $-7.9350549169\times{10}^{-2}$ & $-5.3792520450\times{10}^{-2}$ & $-1.8182148244\times{10}^{-2}$ & $-4.5333075104\times{10}^{-5}$ \tabularnewline
$\left(\frac{\partial{\iota}}{\partial{\tilde{l}}}\right)_{min}$ & $-7.9350549169\times{10}^{-2}$ & $-5.3792520450\times{10}^{-2}$ & $-1.8182148244\times{10}^{-2}$ & $-4.5333075104\times{10}^{-5}$ \tabularnewline
\hline 
${\tilde{S}}=0.5$ &  &  &  & \tabularnewline
\hline
${\iota}$ (deg) & $97.2125$ & $95.8133$ & $93.1100$ & $90.0868$\tabularnewline
1st-order (slow) & $+1.1326853682\times{10}^{-1}$ & $+8.3013385020\times{10}^{-2}$ & $+3.4263245153\times{10}^{-2}$ & $+2.8653789706\times{10}^{-4}$\tabularnewline
1st-order (fast) & $-1.7629701738\times{10}^{-1}$  & $-1.3675258947\times{10}^{-1}$ & $-5.2506108860\times{10}^{-2}$ & $-3.3232275582\times{10}^{-4}$ \tabularnewline
2nd-order (slow) & $-3.9452655089\times{10}^{-2}$ & $-2.6869602227\times{10}^{-2}$ & $-9.1214318535\times{10}^{-3}$  & $-2.2892429382\times{10}^{-5}$ \tabularnewline
$\left(\frac{\partial{\iota}}{\partial{\tilde{l}}}\right)_{min}$ & $-3.9452655089\times{10}^{-2}$ & $-2.6869602227\times{10}^{-2}$ & $-9.1214318535\times{10}^{-3}$  & $-2.2892429382\times{10}^{-5}$ \tabularnewline
\hline 
${\tilde{S}}=0.1$ &  &  &  & \tabularnewline
\hline
${\iota}$ (deg) & $91.4387$ & $91.1602$ & $90.6212$ & $90.0174$\tabularnewline
1st-order (slow) & $+1.4004830192\times{10}^{-1}$ & $+1.0190058051\times{10}^{-1}$ & $+4.1058187759\times{10}^{-2}$ & $+3.0470614607\times{10}^{-4}$\tabularnewline
1st-order (fast) & $-1.5573934281\times{10}^{-1}$ & $-1.1260500157\times{10}^{-1}$ & $-4.4698385942\times{10}^{-2}$ & $-3.1386271509\times{10}^{-4}$ \tabularnewline
2nd-order (slow) & $-7.8455204437\times{10}^{-3}$  & $-5.3522105306\times{10}^{-3}$  & $-1.8200990911\times{10}^{-3}$  & $-4.5782845088\times{10}^{-6}$ \tabularnewline
$\left(\frac{\partial{\iota}}{\partial{\tilde{l}}}\right)_{min}$ & $-7.8455204437\times{10}^{-3}$  & $-5.3522105306\times{10}^{-3}$  & $-1.8200990911\times{10}^{-3}$  & $-4.5782845088\times{10}^{-6}$ \tabularnewline
\hline 
${\tilde{S}}=1.00\times{10}^{-3}$ &  &  &  & \tabularnewline
\hline
${\iota}$ (deg) & $90.0144$ & $90.0116$ & $90.0062$ & $90.0002$\tabularnewline
1st-order (slow) & $+1.4761389187\times{10}^{-1}$ & $+1.0708935531\times{10}^{-1}$ & $+4.2838945741\times{10}^{-2}$ & $+3.0923256836\times{10}^{-4}$\tabularnewline
1st-order (fast) & $-1.4777076382\times{10}^{-1}$ & $-1.0719638097\times{10}^{-1}$ & $-4.2875344203\times{10}^{-2}$ & $-3.0932413388\times{10}^{-4}$ \tabularnewline
2nd-order (slow) & $-7.8435978256\times{10}^{-5}$  & $-5.3512828102\times{10}^{-5}$  & $-1.8199230792\times{10}^{-5}$  & $-4.5782761194\times{10}^{-8}$ \tabularnewline
$\left(\frac{\partial{\iota}}{\partial{\tilde{l}}}\right)_{min}$ & $-7.8435978256\times{10}^{-5}$  & $-5.3512828102\times{10}^{-5}$  & $-1.8199230792\times{10}^{-5}$  & $-4.5782761194\times{10}^{-8}$ \tabularnewline
\hline 
${\tilde{S}}=1.00\times{10}^{-6}$ &  &  &  & \tabularnewline
\hline
${\iota}$ (deg) & $90.0000$ & $90.0000$ & $90.0000$ & $90.0000$\tabularnewline
1st-order (slow) & $+1.4769222926\times{10}^{-1}$ & $+1.0714280363\times{10}^{-1}$ & $+4.2857124658\times{10}^{-2}$ & $+3.0927830473\times{10}^{-4}$\tabularnewline
1st-order (fast) & $-1.4769238613\times{10}^{-1}$ & $-1.0714291066\times{10}^{-1}$ & $-4.2857161056\times{10}^{-2}$ & $-3.0927839630\times{10}^{-4}$ \tabularnewline
2nd-order (slow) & $-7.8435976327\times{10}^{-8}$  & $-5.3512827171\times{10}^{-8}$  & $-1.8199230615\times{10}^{-8}$  & $-4.5782761185\times{10}^{-11}$ \tabularnewline
$\left(\frac{\partial{\iota}}{\partial{\tilde{l}}}\right)_{min}$ & $-7.8435976331\times{10}^{-8}$  & $-5.3512827174\times{10}^{-8}$  & $-1.8199230616\times{10}^{-8}$  & $-4.5782761186\times{10}^{-11}$ \tabularnewline
\end{tabular}
\end{sidewaystable}

\subsubsection{The second-order extrapolation.}

Equations (\ref{eq:Second_Order_Extrapolation_Equations_a}) and (\ref{eq:Second_Order_Extrapolation_Equations_b})
provide a second-order approximation of ${Q}_{path}$ in the vicinity
of the point of tangential intersection between ${Q}_{X}$ and ${Q}_{path}$.
At this formative stage of our work with the abutment, we will use
$\partial^{2}{Q}_{X}/\partial{\tilde{l}}^{2}$ (see equation (\ref{eq:Appendix_D_d2QXdl2})
in Appendix B) in place of $\partial^{2}{Q}_{path}/\partial{\tilde{l}}^{2}$
in equations (\ref{eq:Second_Order_Extrapolation_Equations_a}) and
(\ref{eq:Second_Order_Extrapolation_Equations_b}) as an approximation.
We repeated the ten-point sequence of calculations for the slow mode
(equation (\ref{eq:Mode_Slow})), as outlined in section \ref{sub:The-first-order-extrapolation}.
The results of this numerical analysis are included in table \ref{tab:The-overestimated-values}.
We observe that as ${\tilde{S}}\rightarrow0$, $\partial{\iota}/\partial{\tilde{l}}\rightarrow0$,
as required. Further, the second-order $\partial{\iota}/\partial{\tilde{l}}$
results for the slow mode represent listing orbits (i.e. $\partial{\iota}/\partial{\tilde{l}}<0$);
therefore, the maximum list rate associated with paths that change
from ${X}_{+}^{2}\Rightarrow{X}_{-}^{2}$ (slow) at the abutment will
have an upper limit that corresponds to the minimal value of $\partial{\iota}/\partial{\tilde{l}}$
for that mode.

\subsubsection{The calculation of $\left(\partial{\iota}/\partial{\tilde{l}}\right)_{min}$.}

Let us consider the use of the ${Q}_{X}$ curve itself to estimate
the value of $\left(\partial{\iota}/\partial{\tilde{l}}\right)_{min}$
of an evolving orbit as it intersects the abutment. In this case,\begin{eqnarray}
{Q}_{path}\left({\tilde{l}}-\delta{\tilde{l}}\right) & = & {Q}_{X}\left({\tilde{l}}-\delta{\tilde{l}}\right)\label{eq:Exact_Extrapolation_Equations_a}\\
{Q}_{path}\left({\tilde{l}}+\delta{\tilde{l}}\right) & = & {Q}_{X}\left({\tilde{l}}+\delta{\tilde{l}}\right),\label{eq:Exact_Extrapolation_Equations_b}\end{eqnarray}
where we have assumed that the path followed by the evolving orbit
locally matches the ${Q}_{X}$ curve (equations (\ref{eq:Exact_Extrapolation_Equations_a})
and (\ref{eq:Exact_Extrapolation_Equations_b}) that are used in point
\ref{enu:Use-equations}). This analysis yields the minimum value
of $\left(\partial{\iota}/\partial{\tilde{l}}\right)_{min}$ at ${\tilde{l}}$
for a KBH spin, ${\tilde{S}}$, as specified in point (\ref{enu:Specify-the-spin}).
For the slow mode, the rate of change of ${\iota}$ can be no smaller.
If ${Q}_{path}$ deviates from ${Q}_{X}$ in its second and higher
derivatives, then the actual value of $\partial{\iota}/\partial{\tilde{l}}$
will be greater than $\left(\partial{\iota}/\partial{\tilde{l}}\right)_{min}$.

\subsubsection{Analysis.}

In table \ref{tab:The-overestimated-values} the values of $\left(\partial{\iota}/\partial{\tilde{l}}\right)_{min}$
are in good agreement with the second-order calculations; although
a difference is evident for ${\tilde{S}}={10}^{-6}$.

We calculated $\left(\partial{\iota}/\partial{\tilde{l}}\right)_{min}$
for various KBH spins (${10}^{-18}\leqq{\tilde{S}}\leqq0.99$) over
a wide range of orbit sizes (${10}^{2}\leqq{\tilde{l}}\leqq{10}^{12}$).
It was noted that the results for very large orbits were described
well by an equation of the form\begin{eqnarray}
\left(\frac{\partial{\iota}}{\partial{\tilde{l}}}\right)_{min} & = & -{\kappa}_{1}{\tilde{S}}{\tilde{l}}^{{\varrho}}\label{eq:didl_power_weak_RR}\end{eqnarray}
and that ${\kappa}_{1}$ and ${\rho}$ can be found by performing
a least squares fit on $\left|\left(\partial{\iota}/\partial{\tilde{l}}\right)_{min}\right|$.
For orbits closer to the LSO, we find that \begin{eqnarray}
\left(\frac{\partial{\iota}}{\partial{\tilde{l}}}\right)_{min} & = & -\left({\kappa}_{1}{\tilde{S}}+{\kappa}_{3}{\tilde{S}}^{3}\right){\tilde{l}}^{{\varrho}}.\label{eq:didl_power_strong_RR}\end{eqnarray}

\begin{figure}
\caption{\label{fig:Change-in-angle-VL-l} The values of $\left|\left(\partial{\iota}/\partial{\tilde{l}}\right)_{min}\right|$
plotted for various values of ${\tilde{S}}$ for circular orbits,
where ${\tilde{S}}=0.99$ and ${10}^{2}\leqq{\tilde{l}}\leqq{10}^{12}$. }
\includegraphics[scale=0.4]{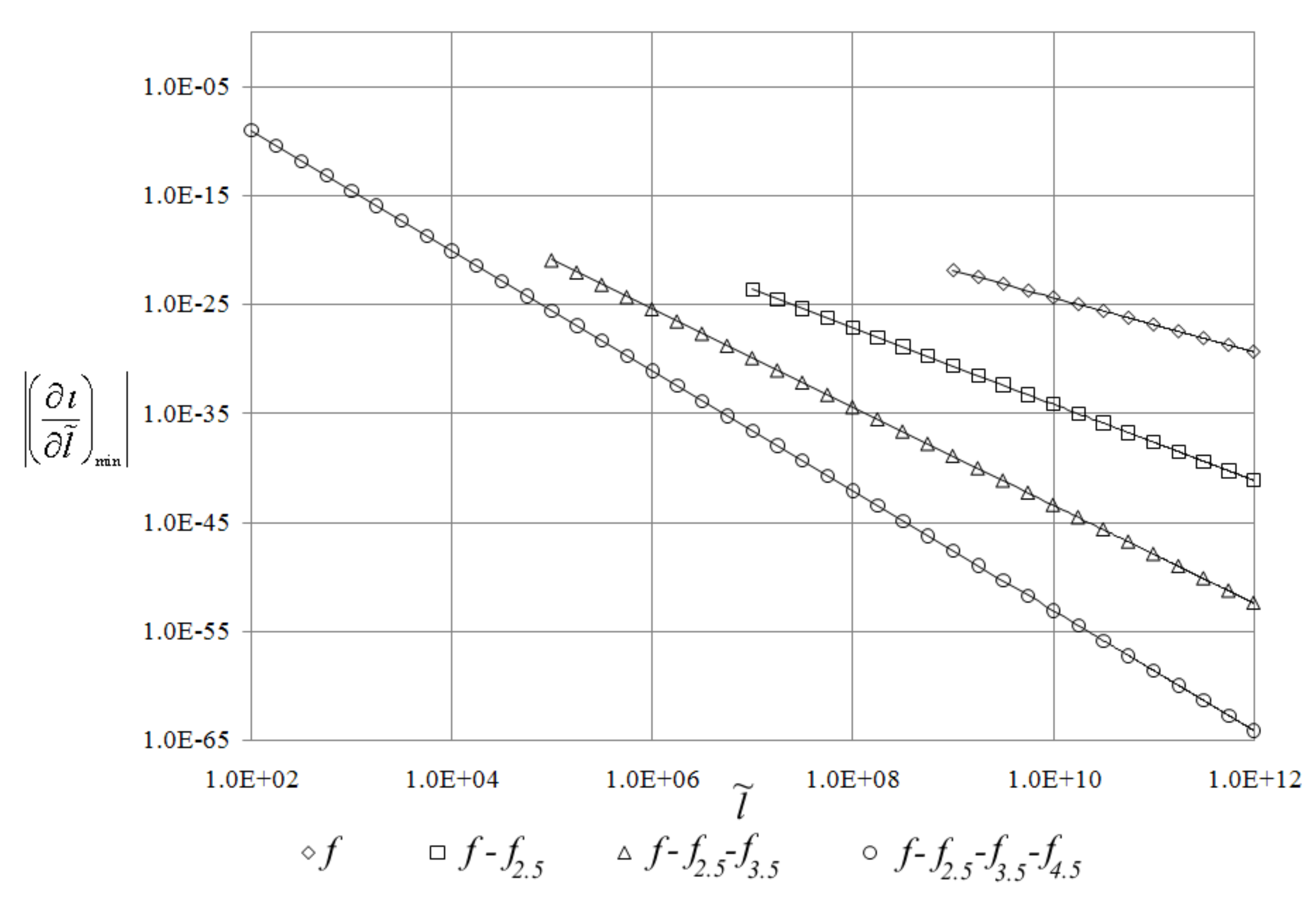}

\end{figure}

In Figure \ref{fig:Change-in-angle-VL-l}, $\left|\left(\partial{\iota}/\partial{\tilde{l}}\right)_{min}\right|$
data for the range ${10}^{9}\leqq{\tilde{l}}\leqq{10}^{12}$ are shown
on a log-log plot. By linear regression analysis, its asymptotic behaviour
(${f}_{2.5}$) can be found (see table \ref{tab:Coeff-and-Powers}).
In successive steps each power of ${\tilde{l}}$ in the series can
be derived as the higher powers are subtracted from the original numerical
data-set. A linear relationship between $\left(\partial{\iota}/\partial{\tilde{l}}\right)_{min}$
and ${\tilde{S}}$ is found for ${f}_{2.5}$ and ${f}_{3.5}$; but
for ${f}_{4.5}$ and ${f}_{5.5}$, which cover ranges of ${\tilde{l}}$
closer to the LSO, an ${\tilde{S}}^{3}$ term appears. The correlation
coefficients (${r}^{2}$) of these regression analyses were better
than $99.9999\%$.

\begin{table}
\caption{\label{tab:Coeff-and-Powers}The coefficients and powers of the series
that describes $\left(\partial{\iota}/\partial{\tilde{l}}\right)_{min}$.}

\begin{tabular}{ccccc}
 & Range & ${\varrho}$ & ${\kappa}_{1}$ & ${\kappa}_{3}$\tabularnewline
\hline
${f}_{2.5}$ & ${10}^{9}\leqq{\tilde{l}}\leqq{10}^{12}$ & $-2.500\pm3\times{10}^{-7}$ & $4.500\pm5\times{10}^{-13}$ & \tabularnewline
${f}_{3.5}$ & ${10}^{7}\leqq{\tilde{l}}\leqq{10}^{12}$ & $-3.500\pm7\times{10}^{-9}$ & $7.500\pm1\times{10}^{-7}$ & \tabularnewline
${f}_{4.5}$ & ${10}^{5}\leqq{\tilde{l}}\leqq{10}^{12}$ & $-4.500\pm2\times{10}^{-7}$ & $31.5\pm4\times{10}^{-9}$ & $8.75\pm1\times{10}^{-4}$\tabularnewline
${f}_{5.5}$ & ${10}^{2}\leqq{\tilde{l}}\leqq{10}^{12}$ & $-5.500\pm2\times{10}^{-4}$ & $122.70\pm2\times{10}^{-8}$ & $-35.98\pm9\times{10}^{-4}$\tabularnewline
\end{tabular}
\end{table}

One may find the specific mode that applies at the abutment by comparing
the results in the first line of table \ref{tab:Coeff-and-Powers}
with the weak-field radiation-reaction post-Newtonian results available
in the literature. Consider the quotient of the formulae presented
in equation (3.9) of Hughes \cite{PhysRevD.61.084004} where ${\iota}\cong\pi/2$.
\begin{eqnarray}
\frac{{\dot{\iota}}_{weak}}{{\dot{r}}_{weak}}=\frac{\partial{\iota}}{\partial{\tilde{l}}} & = & -{\frac{61}{48}}{\tilde{S}}{\tilde{l}}^{-\frac{5}{2}}.\label{eq:Weak-RR}\end{eqnarray}
An identical first-order result can also be derived from equation
(4.3) in Ganz \cite{2007PThPh.117.1041G}. Equation (\ref{eq:Weak-RR})
is of the same form as equation (\ref{eq:didl_power_weak_RR}). Because
$-61/48>-4.5$ in the weak-field radiation-reaction regime, one may
consider ${X}_{+}^{2}\Rightarrow{X}_{-}^{2}$ to be the pertinent
mode. Therefore $\left(\partial{\iota}/\partial{\tilde{l}}\right)_{min}$
describes the lower limit of $\partial{\iota}/\partial{\tilde{l}}$
for all ${\tilde{l}}>{\tilde{l}}_{LSO,\, abutment}$. By numerical
analysis, we found \begin{eqnarray}
\left(\frac{\partial{\iota}}{\partial{\tilde{l}}}\right)_{min} & \cong & -\left(122.7{\tilde{S}}-36{\tilde{S}}^{3}\right){\tilde{l}}^{-11/2}-\left(63/2{\tilde{S}}+35/4{\tilde{S}}^{3}\right){\tilde{l}}^{-9/2}\nonumber \\
 &  & -15/2{\tilde{S}}{\tilde{l}}^{-7/2}-9/2{\tilde{S}}{\tilde{l}}^{-5/2}.\label{eq:didl_Laurent}\end{eqnarray}

\begin{table}
\caption{\label{tab:iota_dot_compared_to_Hughes}An estimate of $\left(M^{2}/m\right)\partial{\iota}/\partial{t}$,
based on equation (\ref{eq:didl_Laurent}), for circular orbits. The
values in parenthesis (which were originally calculated at ${\iota}\cong\pi/3$)
are taken from Hughes \cite{PhysRevD.61.084004} and have been adjusted
by dividing by $\sin\left({\iota}\right)$.}

\begin{tabular}{ccccc}
 & \multicolumn{2}{c}{${\tilde{l}}=7$} & \multicolumn{2}{c}{${\tilde{l}}=100$}\tabularnewline
\hline
${\tilde{S}}=0.05$ & $-9.4541\times{10}^{-5}$  & ($-1.2557\times{10}^{-5}$) & $-2.9301\times{10}^{-11}$  & ($-7.7291\times{10}^{-12}$)\tabularnewline
${\tilde{S}}=0.95$ & $-1.8144\times{10}^{-3}$  & ($-1.3941\times{10}^{-4}$) & $-5.5681\times{10}^{-10}$  & ($-1.3903\times{10}^{-10}$)\tabularnewline
\end{tabular}
\end{table}

In Table \ref{tab:iota_dot_compared_to_Hughes}, we compare the results
of equation (\ref{eq:didl_Laurent}) with those of Hughes \cite{PhysRevD.61.084004};
and although they differ, it is confirmed that the listing of an inclined
orbit in a KBH system proceeds by the slow method. As before (Table
\ref{tab:l_dot_compared_to_Hughes}) we adjust the results in \cite{PhysRevD.61.084004}
by dividing them by $\sin\left(\pi/3\right)$ so that they will correspond
approximately to our near-polar orbits. Although we use equation (\ref{eq:Theta-Roots-Equatorial})
to calculate the value of $\iota$ in this work, and this differs
from the formula used in \cite{2002PhRvD..66f4005G,2006PhRvD..73f4037G};
we recognise that they are sufficiently similar for us to make a general
inference about the relative sizes of $\partial\iota/\partial t$
in \cite{PhysRevD.61.084004} and those calculated here.

\begin{figure}
\caption{\label{fig:Contours-of-constant-Q}Contours of constant Q in the $\tilde{l}-\iota$
plane for a circular (${e}=0$) orbit about a KBH with spin ${\tilde{S}}=0.99$.
Polar orbits are indicated by the short-dashed line on the $\tilde{l}$-axis.
The long-dashed curve corresponds to the abutment. Four curves (solid
lines) of constant $Q=\left\{ 12.25,\,13.0,\,14.0,\,15.0\right\} $
are shown over a range of orbital inclination angles ($90^{o}\leqq\iota\leqq115^{0}$).
The segment of each curve that lies below the abutment is governed
by ${X}_{-}^{2}$; above the abutment, the segment is governed by
${X}_{+}^{2}$. At the points of intersection between the abutment
and the curves of constant $Q$, $\partial\iota/\partial\tilde{l}=\infty$,
which suggests a singularity. The four arrows represent four tangential
intersections on the abutment: (a) corresponds to the case where $\iota$
is constant; (b) corresponds to the evolution of the orbit along the
abutment; (c) represents the fast mode, and (d) the slow mode. N.B.:
the four cases cannot occur together; they are shown on a single plot
for illustrative purposes.}

\includegraphics[scale=0.4]{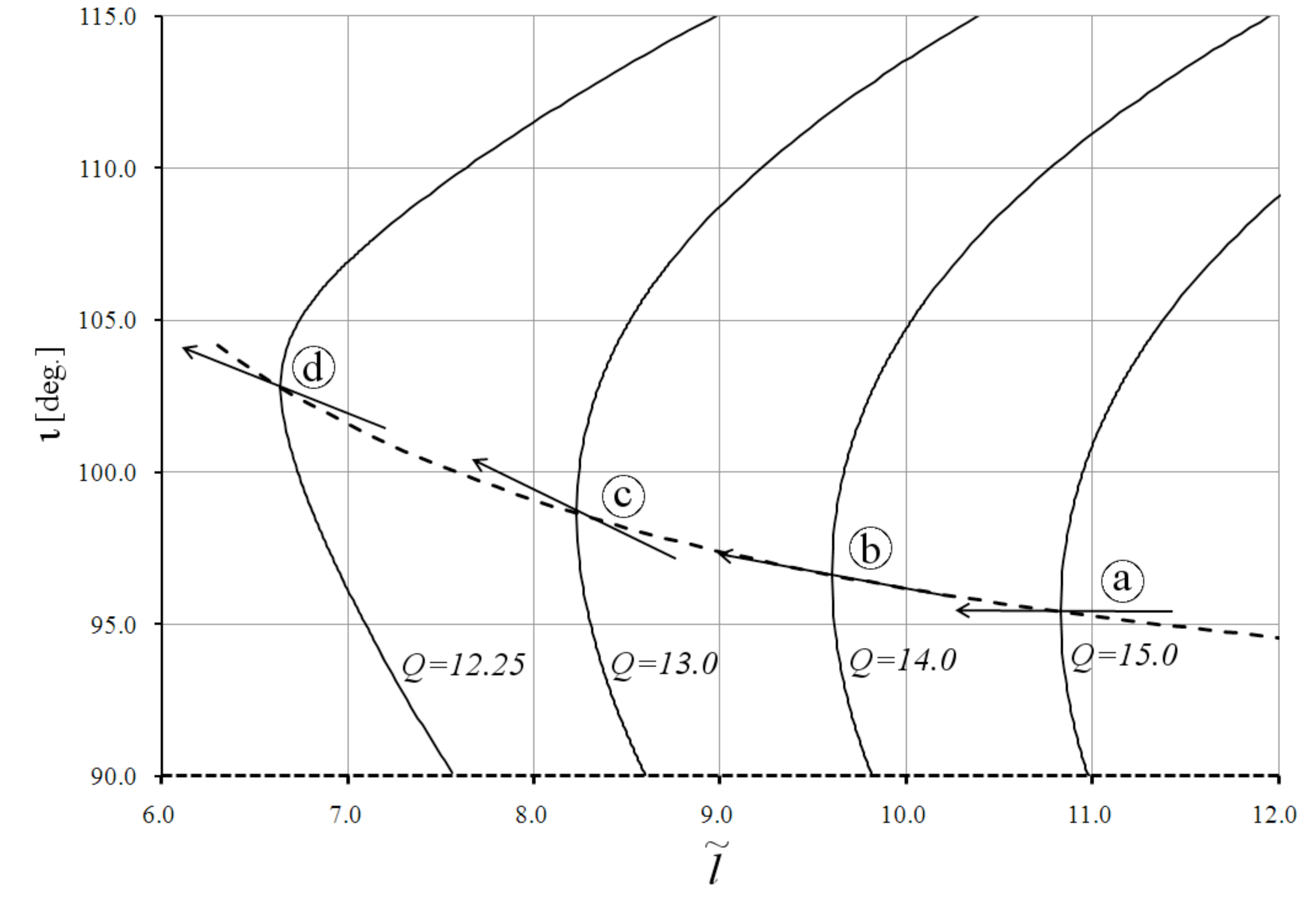}
\end{figure}

Figure \ref{fig:Contours-of-constant-Q} shows the contours of constant
$Q$ on an $\tilde{l}-\iota$ plane for circular orbits ($e=0$) about
a KBH of spin $\tilde{S}=0.99$. One of the important features of
$d\iota/d\tilde{l}$ on the abutment is the suggestion of a coordinate
singularity ($d\iota/d\tilde{l}\rightarrow\infty$); but this is for
the specialized case in which the orbit evolves with a constant value
of $Q$. It has been confirmed that $\partial Q/\partial\tilde{l}>0$
on the abutment (see Section \ref{sub:A-Preliminary-Test}); hence,
such a singularity for the $d\iota/d\tilde{l}$ of an evolving orbit
is not physically manifested. The arrows labelled (a), (b), (c), and
(d) show some important examples of how $\partial\iota/\partial\tilde{l}$
can vary at the abutment. One observes that $\partial\iota/\partial\tilde{l}$
is not uniquely fixed by $\partial Q/\partial\tilde{l}$. Nevertheless,
Figure \ref{fig:Contours-of-constant-Q} provides an important picture
of the behaviour of $\iota$ as the orbit tangentially intersects
the abutment; and it warrants further study.

\section{\label{sec:Conclusions}Conclusions}

In our study of inclined elliptical orbits about a Kerr black hole
(KBH), we found that the minus form of ${X}_{\pm}^{2}$ (${X}={\tilde{L}}_{z}-{\tilde{S}}{\tilde{E}}$)
is shifted away from the polar orbit position to encompass near-polar
retrograde orbits. The abutment (which is a set of orbits that lie
at the junction between the minus and plus forms of ${X}_{\pm}^{2}$)
is shifted the greatest near to the LSO of the KBH and asymptotically
becomes more polar with increasing latus rectum (${\tilde{l}}$) .

We developed a set of analytical formulae that characterise the behaviour
of the Carter constant (${Q}$) at the last stable orbit (LSO), abutment,
and polar orbit. Further, the curves that describe ${Q}$ for an LSO
and ${Q}$ at the abutment (between the minus and plus forms of ${X}_{\pm}^{2}$)
intersect at a single tangential point, for which we derived an analytical
formula. From these equations one can define the domain of ${Q}$
for an evolving orbit (${Q}_{path}$). The two families of curves
defined by ${Q}_{path}$ are governed by either ${X}_{+}^{2}$ or
${X}_{-}^{2}$; and the curves within each family never cross. Therefore,
at the abutment, ${Q}_{path}$ can either change from ${X}_{-}^{2}\Rightarrow{X}_{+}^{2}$
or from ${X}_{+}^{2}\Rightarrow{X}_{-}^{2}$. This result aids in
the investigation of the listing of an orbit at the abutment. 

We have used the abutment as an analytical and numerical laboratory
for the study of the evolution of ${Q}$ for inclined circular orbits.
The first derivative of ${Q}_{X}$ with respect to ${\tilde{l}}$
($\partial{Q}_{X}/\partial{\tilde{l}}$) allows us to test the consistency
of 2PN Q fluxes with estimated values of $\partial{\tilde{l}}/\partial{t}$.
Further. by converting ${Q}$ to the angle of orbital inclination
(${\iota}$), it was possible to calculate the minimum rate of change
of ${\iota}$ with respect to ${\tilde{l}}$, $\left(\partial{\iota}/\partial{\tilde{l}}\right)_{min}$,
independently of a radiation back reaction model. Comparison with
published weak-field post-Newtonian results show that the ${X}_{+}^{2}\Rightarrow{X}_{-}^{2}$
mode applies; and this mode must apply to the entire range of orbit
size, ${\tilde{l}}\geqq{\tilde{l}}_{LSO,\, abutment}$.

Although ${Q}_{X}$ and $\partial{Q}_{X}/\partial{\tilde{l}}$ are
important, the higher derivatives also display critical behaviour.
The second derivatives of ${Q}_{path}$ warrant more study as it will
improve our understanding to the effect of the radiation back reaction
on the listing behaviour of the orbit. The analysis of elliptical
orbits at the abutment will introduce new elements to the listing
behaviour, which arise from the first derivative of ${Q}_{X}$ and
the second and higher derivatives of ${Q}_{path}$, both with respect
to ${e}$. Such a result might be valuable in our understanding of
current and future radiation back reaction models.

\section*{Acknowledgements}
{PGK thanks Western Science (UWO) for their financial support.
MH's research is funded through the NSERC Discovery Grant, Canada
Research Chair, Canada Foundation for Innovation, Ontario Innovation
Trust, and Western's Academic Development Fund programs.  SRV would like to acknowledge the Faculty of Science for a
UWO Internal Science Research Grant award during the progress of this work.  The authors
thankfully acknowledge Drs. N. Kiriushcheva and S. V. Kuzmin for their
helpful discussions of the manuscript.}

\section*{Appendix A: Terms}

\begin{table}[H]
\caption{\label{tab:Parameters}Orbital Parameters}

\begin{tabular}{lcc}
\hline 
\textbf{Parameter} & \textbf{Symbol} & \textbf{Normalised Symbol{*}}\tabularnewline
\hline
Test-Particle Mass & $m$ & -\tabularnewline
Mass of MBH (typically ${{10}^{6}{M}_{\odot}}$) & ${M}$ & -\tabularnewline
Orbital Radius & ${r}$ & ${R}={r}{M}^{-1}$\tabularnewline
Semi-Major Axis & ${a}$ & ${A}={a}{M}^{-1}$\tabularnewline
Latus Rectum & ${l}$ & ${\tilde{l}}={l}{M}^{-1}$ \tabularnewline
Proper Time & $\tau$ & ${\tilde{\tau}}={\tau}{M}^{-1}$\tabularnewline
Orbital Energy & ${E}$ & ${\tilde{E}}={E}{m}^{-1}$\tabularnewline
Effective Potential Energy & ${V}$ & ${\tilde{V}}={V}{m}^{-1}$\tabularnewline
Spin Angular Momentum (KBH) & ${\mathbf{J}}$ & ${\tilde{S}}=\left|{\mathbf{J}}\right|{M}^{-2}$\tabularnewline
Orbital Angular Momentum (z component) & ${L}_{z}$ & ${\tilde{L}}_{z}={L}_{z}{\left(mM\right)}^{-1}$\tabularnewline
Orbital Angular Momentum ($\theta$ component) & ${L}_{\theta}$ & ${\tilde{L}}_{\theta}={L}_{\theta}{\left(mM\right)}^{-1}$\tabularnewline
 & $\Sigma$ & \begin{tabular}{c}
$\tilde{\Sigma}={\Sigma}{M}^{-2}$\tabularnewline
$=R^{2}+\tilde{S}^{2}{\cos^{2}\left(\theta\right)}$\tabularnewline
\end{tabular}\tabularnewline
 & ${\Delta}$ & \begin{tabular}{c}
${\tilde{\Delta}}={\Delta}{M}^{-2}$\tabularnewline
$={R}^{2}-2\, R+{\tilde{S}}^{2}$\tabularnewline
\end{tabular}\tabularnewline
\multicolumn{2}{l}{\begin{tabular}{l}
Governs prograde, polar, and retrograde orbits up to the abutment\tabularnewline
Governs the retrograde orbits beyond the abutment\tabularnewline
\end{tabular}} & \begin{tabular}{c}
${X}_{-}^{2}$\tabularnewline
${X}_{+}^{2}$\tabularnewline
\end{tabular}=${({\tilde{L}}_{z}-{\tilde{S}}{\tilde{E}})}^{2}$\tabularnewline
 &  & \tabularnewline
\end{tabular}

\begin{raggedright}
\textcolor{black}{\footnotesize {*}We set the speed of light and gravitational
constant to unity (i.e. ${c}=1$ and ${G}=1$); therefore, mass-energy
is in units of time (seconds).}
\par\end{raggedright}
\end{table}

\section*{Appendix B: Ancillary Equations}

\subsection*{The Kerr metric and its Inverse}

The inverse Kerr metric expressed in the Boyer-Lindquist coordinate
system:

\setlength{\mathindent}{0cm}

\begin{equation}
g_{\alpha\beta}\Biggr|_{\mathit{Kerr}}=\left[\begin{array}{cccc}
-{\frac{\Delta-{M^{2}}{\tilde{S}}^{2}\sin^{2}\left(\theta\right)}{{\rho}^{2}}} & 0 & 0 & -2{M}{\frac{{M^{2}}\tilde{S}R\sin^{2}\left(\theta\right)}{{\rho}^{2}}}\\
\noalign{\medskip}0 & {\frac{{\rho}^{2}}{\Delta}} & 0 & 0\\
\noalign{\medskip}0 & 0 & {\rho}^{2} & 0\\
\noalign{\medskip}-2{M}{\frac{{M^{2}}\tilde{S}R\sin^{2}\left(\theta\right)}{{\rho}^{2}}} & 0 & 0 & {\frac{{M^{4}}\left({R}^{2}+{\tilde{S}}^{2}\right)^{2}-{M^{2}}{\tilde{S}}^{2}\Delta\,\sin^{2}\left(\theta\right)}{{\rho}^{2}}\sin^{2}\left(\theta\right)}\end{array}\right].\end{equation}
\setlength{\mathindent}{2.5cm}

To simplify the presentation of the metric, we define the parameter:

\begin{eqnarray}
{\Sigma}={\rho}^{2} & = & {M^{2}}{\left(R^{2}+\tilde{S}^{2}{\cos^{2}\left(\theta\right)}\right)}.\end{eqnarray}

The inverse Kerr metric is:\setlength{\mathindent}{0cm}

{\small \begin{eqnarray}
{g}^{\delta\gamma}\Biggr|_{Kerr}\label{eq:InverseKerrMetric}\\
=\left({\Sigma}\right)^{-1} & \left[{\begin{array}{cccc}
{-\frac{{{\Sigma}\left(R^{2}+\tilde{S}^{2}\right)+2\,\tilde{S}^{2}R-2\,{\cos^{2}\left(\theta\right)}\tilde{S}^{2}R}}{{\left({R^{2}-2\, R+\tilde{S}^{2}}\right)}}} & 0 & 0 & \frac{-2\,{\tilde{S}R}}{{M}{\left({R^{2}-2\, R+\tilde{S}^{2}}\right)}}\\
0 & {R^{2}-2\, R+\tilde{S}^{2}} & 0 & 0\\
0 & 0 & 1 & 0\\
\frac{-2\,{\tilde{S}R}}{{M}{\left({R^{2}-2\, R+\tilde{S}^{2}}\right)}} & 0 & 0 & {\frac{{R^{2}-2\, R+\tilde{S}^{2}{\cos^{2}\left(\theta\right)}}}{{\left({R^{2}-2\, R+\tilde{S}^{2}}\right){\sin^{2}\left(\theta\right)}}}}\end{array}}\right].\nonumber \end{eqnarray}
}\setlength{\mathindent}{2.5cm}

The determinant of the Kerr metric was calculated to be, $Det=-{\Sigma}^{2}{\sin^{2}\left(\theta\right)}.$

\subsection*{Effective Potentials}

The effective potentials \cite{1968PhRv..174.1559C,Bardeen:1972fi,2002CQGra..19.2743S,2007PhRvD..75b4005B}
that appear in equations (\ref{eq:rho-drbydt}), (\ref{eq:rho-dthetabydt}),
(\ref{eq:dPhidTao}), and (\ref{eq:dtdtao}):

\begin{equation}
\tilde{V}_{R}\left(R\right)=T^{2}-{\tilde{\Delta}}\left[{R^{2}+\left({\tilde{L}_{z}-\tilde{S}\tilde{E}}\right)^{2}+Q}\right]\end{equation}
\begin{equation}
\tilde{V}_{\theta}\left(\theta\right)=Q-\cos^{2}\left(\theta\right)\left[{\tilde{S}^{2}\left({1-\tilde{E}^{2}}\right)+\frac{{\tilde{L}_{z}^{2}}}{{\sin^{2}\left(\theta\right)}}}\right]\end{equation}

\begin{eqnarray}
{\tilde{V}}_{\phi} & = & -\left({\tilde{S}\tilde{E}-\frac{{\tilde{L}_{z}}}{{\sin^{2}\left(\theta\right)}}}\right)+\frac{{\tilde{S}}{T}}{{\tilde{\Delta}}}\end{eqnarray}
\begin{eqnarray}
{\tilde{V}}_{t} & = & -\tilde{S}\left({\tilde{S}\tilde{E}\sin^{2}\left(\theta\right)-\tilde{L}_{z}}\right)+\frac{{\left({R^{2}+\tilde{S}^{2}}\right)T}}{{\tilde{\Delta}}}\end{eqnarray}
with

\[
T=\tilde{E}\left({R^{2}+\tilde{S}^{2}}\right)-\tilde{L}_{z}\tilde{S}.\]
The Carter constant (${Q}$) is normalised,\begin{eqnarray}
{Q} & = & \frac{1}{\left({m}{M}\right)^{2}}\left[{\frac{\cos^{2}\left(\theta\right){L_{{z}}}^{2}}{\sin^{2}\left(\theta\right)}}+{L_{{\theta}}}^{2}+\cos^{2}\left(\theta\right){S}^{2}\left({m}^{2}-{E}^{2}\right)\right]\nonumber \\
 & = & {\frac{\cos^{2}\left(\theta\right){\tilde{L}_{{z}}}^{2}}{\sin^{2}\left(\theta\right)}}+{\tilde{L}_{{\theta}}}^{2}+\cos^{2}\left(\theta\right){\tilde{S}}^{2}\left(1-{\tilde{E}}^{2}\right).\end{eqnarray}

\subsection*{Ninth Order Polynomial in ${\tilde{l}}$ for calculating ${\tilde{l}}_{LSO}$}

\setlength{\mathindent}{0cm}

\begin{eqnarray}
{p}\left({\tilde{l}}\right)=\nonumber \\
{\tilde{l}}^{9}-4\left(3+{e}\right){\tilde{l}}^{8}\nonumber \\
-\left(-36-2\,{\tilde{S}}^{2}{e}^{2}+6\,{\tilde{S}}^{2}-24\,{e}+4\,{\tilde{S}}^{2}e-4\,{e}^{2}\right){\tilde{l}}^{7}\nonumber \\
+4\,{\tilde{S}}^{2}\left({e}+1\right)\left(-{e}^{2}+2\, Q-7\right){\tilde{l}}^{6}\nonumber \\
-{\tilde{S}}^{2}\left({e}+1\right)\left(-{\tilde{S}}^{2}{e}^{3}+5\,{\tilde{S}}^{2}{e}^{2}-3\,{\tilde{S}}^{2}{e}-9\,{\tilde{S}}^{2}+16\, Q{e}^{2}+8\, Q{e}+24\, Q\right){\tilde{l}}^{5}\nonumber \\
+8\, Q{\tilde{S}}^{2}\left({e}+1\right)^{2}\left(4\,{e}^{2}+{\tilde{S}}^{2}{e}-2\,{e}-3\,{\tilde{S}}^{2}+6\right){\tilde{l}}^{4}\nonumber \\
+8\, Q{\tilde{S}}^{4}\left({e}+1\right)^{2}\left(-2\,{e}^{3}-{e}^{2}+2\, Q-1\right){\tilde{l}}^{3}-16\,{\tilde{S}}^{4}{Q}^{2}\left({e}^{2}-2\,{e}+3\right)\left({e}+1\right)^{3}{\tilde{l}}^{2}\nonumber \\
+16\,{\tilde{S}}^{4}{Q}^{2}\left(2\,{e}^{2}-4\,{e}+3\right)\left({e}+1\right)^{4}{\tilde{l}}-16\,{\tilde{S}}^{6}{\tilde{Q}}^{2}\left({e}-1\right)^{2}\left({e}+1\right)^{5}\nonumber \\
=0\label{eq:CP_LSO}\end{eqnarray}
\setlength{\mathindent}{2.5cm}

\subsection*{The First and Second Derivatives of ${Q}_{X}$}

Given:\begin{eqnarray*}
{\xi}_{1} & = & {\tilde{l}}{\left({\tilde{l}}-{e}^{2}-3\right)},\\
{\xi}_{2} & = & 4{\tilde{l}}{\tilde{S}}^{2}{\left(1-{e}^{2}\right)^{2}}\\
{\xi}_{3} & = & 2{\tilde{l}}-{e}^{2}-3.\end{eqnarray*}
From equation (\ref{eq:Q_Abutment}) the equation for ${Q}_{X}$ is

\begin{eqnarray}
{Q}_{X} & = & {\frac{2{\tilde{l}}^{3}}{{\xi}_{2}}}\left({\xi}_{1}-\sqrt{{\xi}_{1}^{2}-{\xi}_{2}}\right).\label{eq:Appendix_D_QX}\end{eqnarray}
We obtain the following first and second derivatives of ${Q}_{X}$
with respect to ${\tilde{l}}$ and with respect to ${e}$:

\begin{eqnarray}
\frac{\partial{Q}_{X}}{\partial{\tilde{l}}} & = & {\frac{2{\tilde{l}}^{3}}{{\xi}_{2}}}\left(\left({\tilde{l}}{\xi}_{3}-\frac{1}{2}\frac{\left(2{\tilde{l}}{\xi}_{1}{\xi}_{3}-{\xi}_{2}\right)}{\sqrt{{\xi}_{1}^{2}-{\xi}_{2}}}\right)+2\left({\xi}_{1}-\sqrt{{\xi}_{1}^{2}-{\xi}_{2}}\right)\right),\label{eq:Appendix_D_dQXdl}\end{eqnarray}

\begin{eqnarray}
\frac{\partial{Q}_{X}}{\partial{e}} & = & {\frac{4{e}{\tilde{l}}^{3}}{{\xi}_{2}{\left(1-{e}^{2}\right)}}}{\Biggl[{\left(-{\tilde{l}}{\left(1-{e}^{2}\right)}+{\frac{\left({{\tilde{l}}{\left(1-{e}^{2}\right)}{\xi}_{1}-{\xi}_{2}}\right)}{\sqrt{{\xi}_{1}^{2}-{\xi}_{2}}}}\right)}}\nonumber \\
 &  & +2\left({\xi}_{1}-\sqrt{{\xi}_{1}^{2}-{\xi}_{2}}\right)\Biggr],\label{eq:Appendix_D_dQXde}\end{eqnarray}

\begin{eqnarray}
\frac{{\partial}^{2}{Q}_{X}}{\partial{\tilde{l}}^{2}} & = & {\frac{2{\tilde{l}}}{{\xi}_{2}}}\Biggl[2{{\tilde{l}}^{2}}+\frac{1}{4}{\frac{\left(2{\tilde{l}}{\xi}_{1}{\xi}_{3}-{\xi}_{2}\right)^{2}}{\left({\xi}_{1}^{2}-{\xi}_{2}\right)^{3/2}}}-{\tilde{l}}^{2}\frac{\left(6{\xi}_{1}+{\left(3+{e}^{2}\right)}^{2}\right)}{{\sqrt{{\xi}_{1}^{2}-{\xi}_{2}}}}\nonumber \\
 &  & +4{\left({\tilde{l}}{\xi}_{3}-\frac{1}{2}\frac{\left(2{\tilde{l}}{\xi}_{1}{\xi}_{3}-{\xi}_{2}\right)}{\sqrt{{\xi}_{1}^{2}-{\xi}_{2}}}\right)}+2\left({\xi}_{1}-\sqrt{{\xi}_{1}^{2}-{\xi}_{2}}\right)\Biggr],\label{eq:Appendix_D_d2QXdl2}\end{eqnarray}

\begin{eqnarray}
\frac{{\partial}^{2}{Q}_{X}}{\partial{e}^{2}} & = & {\frac{4{\tilde{l}}^{3}}{{\xi}_{2}{\left(1-{e}^{2}\right)}^{2}}}\Biggl[-{\tilde{l}}{\left(1-{e}^{2}\right)^{2}}+2{e}^{2}\left(\frac{\left({\tilde{l}}\left(1-{e}^{2}\right){\xi}_{1}-{\xi}_{2}\right)^{2}}{\left({\xi}_{1}^{2}-{\xi}_{2}\right)^{3/2}}\right)\nonumber \\
 &  & +{\left(1-{e}^{2}\right)}^{2}\left(\frac{\left({l}^{2}\left({\tilde{l}}-3\left(1+{e}^{2}\right)\right)+4{\tilde{l}}{\tilde{S}}^{2}\left(3{e}^{2}-1\right)\right)}{\sqrt{{\xi}_{1}^{2}-{\xi}_{2}}}\right)\nonumber \\
 &  & +8{e}^{2}\left(-{\tilde{l}}\left(1-{e}^{2}\right)+\left(\frac{\left({\tilde{l}}\left(1-{e}^{2}\right){\xi}_{1}-{\xi}_{2}\right)}{\sqrt{{\xi}_{1}^{2}-{\xi}_{2}}}\right)\right)\nonumber \\
 &  & +2\left(5{e}^{2}+1\right)\left({\xi}_{1}-\sqrt{{\xi}_{1}^{2}-{\xi}_{2}}\right)\Biggr].\label{eq:Appendix_D_d2QXde2}\end{eqnarray}

\subsection*{The 2PN flux for Q}

According to equation (A.3) in \cite{2007PhRvD..76d4007B} (after
equation (56) in \cite{2006PhRvD..73f4037G}):

\begin{eqnarray}
\left(\frac{\partial{Q}}{\partial{t}}\right)_{2PN} & = & -{\sin\left({\iota}\right)}\frac{64}{5}\frac{{m}}{{M}^{2}}{\left(1-{e}^{2}\right)^{3/2}}{\tilde{l}}^{-7/2}\sqrt{{Q}}\nonumber \\
 & \times & \Bigl[{g}_{9}\left({e}\right)-{\tilde{l}}^{-1}{g}_{11}\left({e}\right)+\left({\pi}{g}_{12}\left({e}\right)-{\cos\left({\iota}\right)}{\tilde{S}}{g}_{10}^{b}\left({e}\right)\right){\tilde{l}}^{-3/2}\nonumber \\
 &  & -\left({g}_{13}\left({e}\right)-{\tilde{S}}^{2}{\left({g}_{14}\left({e}\right)-\frac{45}{8}{\sin}^{2}\left({\iota}\right)\right)}\right){\tilde{l}}^{-2}\Bigr],\label{eq:Qdot_2PN_Appendix_D}\end{eqnarray}
where

\begin{eqnarray*}
{g}_{9} & = & 1+\frac{7}{8}{e}^{2}\end{eqnarray*}

\begin{eqnarray*}
{g}_{10}^{b} & = & \frac{61}{8}+\frac{91}{4}{e}^{2}+\frac{461}{64}{e}^{4}\end{eqnarray*}
\begin{eqnarray*}
{g}_{11} & = & \frac{1247}{336}+\frac{425}{336}{e}^{2}\end{eqnarray*}
\begin{eqnarray*}
{g}_{12} & = & 4+\frac{97}{8}{e}^{2}\end{eqnarray*}
\begin{eqnarray*}
{g}_{13} & = & \frac{44711}{9072}+\frac{302893}{6048}{e}^{2}\end{eqnarray*}
and\begin{eqnarray*}
{g}_{14} & = & \frac{33}{16}+\frac{95}{16}{e}^{2}.\end{eqnarray*}
N.B.: the Carter constant has been normalised by dividing by $\left(mM\right)^{2}$.

\section*{Appendix C: An explicit treatment of $\mathbf{d\theta/d\tau}$ in
the effective potential}

For the sake of completeness, we shall demonstrate a method for calculating
a polynomial that describes ${d\theta}/{d\tau}$. 

Recall that:\begin{eqnarray}
\Delta & = & M^{2}\left({R}^{2}-2\, R+{\tilde{S}}^{2}\right)\end{eqnarray}
and

\begin{eqnarray}
{\Sigma}={\rho}^{2} & = & {M^{2}}{\left(R^{2}+\tilde{S}^{2}{\cos^{2}\left(\theta\right)}\right)}.\end{eqnarray}
Therefore one may consider a normalised form of these equations:\begin{eqnarray*}
{\tilde{\Delta}}=\frac{\Delta}{{M}^{2}} & = & \left({R}^{2}-2\, R+{\tilde{S}}^{2}\right)\end{eqnarray*}
and \begin{eqnarray*}
{\tilde{\Sigma}}=\frac{{\rho}^{2}}{{M}^{2}} & = & {\left(R^{2}+\tilde{S}^{2}{\cos^{2}\left(\theta\right)}\right)}.\end{eqnarray*}
N.B.: \begin{eqnarray}
\frac{{\rho}^{2}}{{\Delta}} & = & \frac{{\Sigma}}{{\Delta}}\nonumber \\
 & = & \frac{{\tilde{\Sigma}}}{{\tilde{\Delta}}}.\end{eqnarray}

In working with the quantity, ${dr}/{d\tau}$, in \cite{Komorowski:2008wc}
it did not matter about the division of the radial distance by the
black hole mass, ${M}$, since the proper time, ${\tau}$, would also
have been normalised in the same way i.e., \begin{eqnarray}
\frac{dr}{d\tau} & = & \frac{{d\left({M}\left(\frac{r}{M}\right)\right)}}{{d\left({M}\left(\frac{\tau}{M}\right)\right)}}\nonumber \\
 & = & \frac{d{R}}{d{\tilde{\tau}}}.\end{eqnarray}
Indeed, one ought to consider the normalisation of the proper time
with respect to black hole mass. Although one could escape difficulties
when only considering ${dr}/{d\tau}$, it is mandatory that the normalised
proper time be explicitly considered when evaluating the quantity,
${d\theta}/{d\tau}$. The polar angle, $\theta$, is already dimensionless,
and as such cannot be normalised. Therefore\begin{eqnarray}
\frac{{d\theta}}{{d\tau}} & = & \frac{{d\theta}}{{d\Bigl({M}\left({\frac{\tau}{M}}\right)\Bigr)}}\nonumber \\
 & = & \frac{1}{M}\frac{{d\theta}}{{d{\tilde{\tau}}}}.\end{eqnarray}

We may rewrite the 4-momentum in terms of ${d\theta}/{d\tilde{\tau}}$
in addition to ${dR}/{d\tilde{\tau}}$, where ${X}={\tilde{L}}_{z}-{\tilde{S}}{\tilde{E}}$.

\begin{equation}
P_{\gamma}=\left[-{m}{\tilde{E}},m\frac{{\rho^{2}}}{\Delta}\left(\frac{{dR}}{{d{\tilde{\tau}}}}\right),m\frac{{\rho^{2}}}{M}\left(\frac{{d\theta}}{{d{\tilde{\tau}}}}\right),mM\left({X}+{\tilde{S}}{\tilde{E}}\right)\right].\end{equation}
By evaluating $\vec{P}\cdot\vec{P}$ and substituting the known relation
for ${dR}/{d\tilde{\tau}}$ one obtains,\setlength{\mathindent}{0cm}

\begin{eqnarray}
{\tilde{\Sigma}}^{2}\sin^{2}\left(\theta\right)\left({\left(\frac{{dR}}{{d{\tilde{\tau}}}}\right)^{2}+\tilde{\Delta}\left(\frac{{d\theta}}{{d{\tilde{\tau}}}}\right)^{2}}\right)\label{eq:Big_Derivative}\\
=-\sin^{2}\left(\theta\right)\left(1-{\tilde{E}}^{2}\right){R}^{4}+2\,\sin^{2}\left(\theta\right){R}^{3}\nonumber \\
-\left({X}^{2}+{\tilde{S}}^{2}+2\, X\tilde{E}\tilde{S}-\cos^{4}\left(\theta\right){\tilde{S}}^{2}\left(1-{\tilde{E}}^{2}\right)\right){R}^{2}\nonumber \\
+2\,\left({X}^{2}+{\tilde{S}}^{2}\cos^{2}\left(\theta\right)+2\, X\cos^{2}\left(\theta\right)\tilde{E}\tilde{S}-\cos^{4}\left(\theta\right){\tilde{S}}^{2}\left(1-{\tilde{E}}^{2}\right)\right){R}\nonumber \\
-\left(\left({X}^{2}+{S}^{2}+2\, XES\right)\cos^{2}\left(\theta\right)-\cos^{4}\left(\theta\right)\left(1-{E}^{2}\right){S}^{2}\right){S}^{2}.\nonumber \end{eqnarray}
We know from equation (\ref{eq:Orbital_Equation_Q}) that\begin{eqnarray}
{\tilde{\Sigma}}^{2}{\left(\frac{dR}{d{\tilde{\tau}}}\right)}^{2} & = & -\left({1-{\tilde{E}}^{2}}\right){R}^{4}+2\,{R}^{3}\label{eq:Appendix_C_Vr_Q0_dRdtao}\\
 & - & \left({X}^{2}+{\tilde{S}}^{2}+2\,\tilde{S}\tilde{E}X+Q\right){R}^{2}+2\,\left({X}^{2}+Q\right)R-{Q{\tilde{S}}^{2}}\nonumber \end{eqnarray}
and we can substitute this expression into equation (\ref{eq:Big_Derivative})
to simplify it thus:

\begin{eqnarray}
{\tilde{\Sigma}}^{2}\sin^{2}\left(\theta\right)\left({\tilde{\Delta}\left(\frac{{d\theta}}{{d{\tilde{\tau}}}}\right)^{2}}\right)\label{eq:Simplified_Derivative}\\
=0\times{R}^{4}+0\times{R}^{3}\nonumber \\
-\left(\left({X}^{2}+{\tilde{S}}^{2}+2\, X\tilde{E}\tilde{S}\right)\cos^{2}\left(\theta\right)-\cos^{4}\left(\theta\right){\tilde{S}}^{2}\left(1-{\tilde{E}}^{2}\right)-\sin^{2}\left(\theta\right){Q}\right){R}^{2}\nonumber \\
+2\,\left(\left({X}^{2}+{\tilde{S}}^{2}+2\, X\tilde{E}\tilde{S}\right)\cos^{2}\left(\theta\right)-\cos^{4}\left(\theta\right){\tilde{S}}^{2}\left(1-{\tilde{E}}^{2}\right)-\sin^{2}\left(\theta\right){Q}\right){R}\nonumber \\
-\left(\left({X}^{2}+{\tilde{S}}^{2}+2\, X\tilde{E}\tilde{S}\right)\cos^{2}\left(\theta\right)-\cos^{4}\left(\theta\right){\tilde{S}}^{2}\left(1-{\tilde{E}}^{2}\right)-\sin^{2}\left(\theta\right){Q}\right){\tilde{S}}^{2},\nonumber \end{eqnarray}
which factors to

\begin{eqnarray}
{\tilde{\Sigma}}^{2}\sin^{2}\left(\theta\right)\left({\tilde{\Delta}\left(\frac{{d\theta}}{{d{\tilde{\tau}}}}\right)^{2}}\right)\label{eq:Factored_Derivative}\\
=\tilde{\Delta}\left(\sin^{2}\left(\theta\right){Q}-\left({X}^{2}+{\tilde{S}}^{2}+2\, X\tilde{E}\tilde{S}\right)\cos^{2}\left(\theta\right)+\cos^{4}\left(\theta\right){\tilde{S}}^{2}\left(1-{\tilde{E}}^{2}\right)\right)\nonumber \end{eqnarray}
and simplifies to\begin{eqnarray}
{\tilde{\Sigma}}^{2}\left(\frac{{d\theta}}{{d{\tilde{\tau}}}}\right)^{2} & = & \frac{1}{\sin^{2}\left(\theta\right)}\left(\sin^{2}\left(\theta\right){Q}-\left({X}^{2}+{\tilde{S}}^{2}+2\, X\tilde{E}\tilde{S}\right)\cos^{2}\left(\theta\right)+\cos^{4}\left(\theta\right){\tilde{S}}^{2}\left(1-{\tilde{E}}^{2}\right)\right)\nonumber \\
 & = & {Q}-{\tilde{L}}_{z}^{2}\left(\frac{\cos^{2}\left(\theta\right)}{\sin^{2}\left(\theta\right)}\right)-{\tilde{S}}^{2}\left(1-{\tilde{E}}^{2}\right)\cos^{2}\left(\theta\right).\label{eq:Simplified_Derivative_1}\end{eqnarray}
\setlength{\mathindent}{2.5cm}One can equate:

\begin{eqnarray}
{m}{M}{\tilde{L}}_{\theta} & = & m\frac{{\rho^{2}}}{M}\left(\frac{{d\theta}}{{d{\tilde{\tau}}}}\right)\end{eqnarray}
and thus obtain:\begin{eqnarray}
{\tilde{L}}_{\theta} & = & {\tilde{\Sigma}}{\frac{d{\theta}}{d{\tilde{\tau}}}},\label{eq:Ltheta_1}\end{eqnarray}
which \textit{viz.} equation (\ref{eq:Simplified_Derivative}) confirms
the relationship for the Carter constant. Further, we note that ${\tilde{L}}_{z}^{2}={Q}$
at the orbital nodes and ${\tilde{L}}_{z}=0$ at the zenith or nadir
of the orbit. It is in these respects that the Carter constant possesses
a physical meaning for a KBH. 

\newpage{}

\bibliographystyle{unsrt}

\newpage{}

\begin{table}[H]
\caption{\label{tab:SBH}Numerical values of ${R}_{LSO}$ estimated from Figure
\ref{fig:SBH} for a circular LSO around an SBH. There is no circular
LSO for ${Q}>12$.}

\begin{tabular}{ccccccc}
\hline 
$Q$ & label & ${R}_{LSO}$ & (Prograde) & label & ${R}_{LSO}$ & (Retrograde)\tabularnewline
\hline 
$0.000000$ & \textit{a} & \multicolumn{2}{c}{$6.000$} & ${A}$ & \multicolumn{2}{c}{$6.000$}\tabularnewline
$5.000000$ & \textit{b} & \multicolumn{2}{c}{$6.000$} & \textit{${B}$} & \multicolumn{2}{c}{$6.000$}\tabularnewline
$10.000000$ & \textit{c} & \multicolumn{2}{c}{$6.000$} & \textit{${C}$} & \multicolumn{2}{c}{$6.000$}\tabularnewline
$12.000000$ & \textit{d,e} & \multicolumn{2}{c}{$6.000$} & \textit{${D}$, ${E}$} & \multicolumn{2}{c}{$6.000$}\tabularnewline
$13.000000$ & $f$ & \multicolumn{2}{c}{} & ${F}$ & \multicolumn{2}{c}{}\tabularnewline
$16.000000$ & \textit{g} & \multicolumn{2}{c}{} & \textit{G} & \multicolumn{2}{c}{}\tabularnewline
\end{tabular}
\end{table}
\newpage{}

\begin{table}[H]
\caption{\label{tab:KBH50}Numerical values of ${R}_{LSO}$ estimated from
Figure \ref{fig:KBH_S_0.5} for a circular LSO around a KBH of spin
${\tilde{S}}=0.50$. There is no circular LSO for ${Q}>12.0545$.}

\begin{tabular}{ccccccc}
\hline 
$Q$ & label & ${R}_{LSO}$ & (Governed by ${X}_{-}^{2}$) & label & ${R}_{LSO}$ & (Governed by ${X}_{+}^{2}$)\tabularnewline
\hline 
$0.000000$ & \textit{a} & \multicolumn{2}{c}{$4.233$} & ${A}$ & \multicolumn{2}{c}{$7.555$}\tabularnewline
$5.000000$ & \textit{b} & \multicolumn{2}{c}{$4.709$} & \textit{${B}$} & \multicolumn{2}{c}{$7.227$}\tabularnewline
$10.000000$ & \textit{c} & \multicolumn{2}{c}{$5.366$} & \textit{${C}$} & \multicolumn{2}{c}{$5.366$}\tabularnewline
$11.828365$ & \textit{d} & \multicolumn{2}{c}{$5.842$} & \textit{${D}$} & \multicolumn{2}{c}{$6.287$}\tabularnewline
$12.054503$ & \textit{e} & \multicolumn{2}{c}{$6.067$} & \textit{${E}$} & \multicolumn{2}{c}{$6.068$}\tabularnewline
$13.000000$ & \textit{f} & \multicolumn{2}{c}{} & ${F}$ & \multicolumn{2}{c}{}\tabularnewline
$16.000000$ & \textit{g} & \multicolumn{2}{c}{} & \textit{G} & \multicolumn{2}{c}{}\tabularnewline
\end{tabular}
\end{table}

\newpage{}

\begin{table}[H]
\caption{\label{tab:KBH99}Numerical values of ${R}_{LSO}$ estimated from
Figure \ref{fig:KBH_S_0.99} for a circular orbit around a KBH of
spin ${\tilde{S}}=0.99$. There is no circular LSO for ${Q}>12.203171$}

\begin{tabular}{ccccccc}
\hline 
$Q$ & label & ${R}_{LSO}$ & (Governed by ${X}_{-}^{2}$) & label & ${R}_{LSO}$ & (Governed by ${X}_{+}^{2}$)\tabularnewline
\hline 
$0.000000$ & \textit{a} & \multicolumn{2}{c}{$1.455$} & ${A}$ & \multicolumn{2}{c}{$8.972$}\tabularnewline
$5.000000$ & \textit{b} & \multicolumn{2}{c}{$3.074$} & \textit{${B}$} & \multicolumn{2}{c}{$8.403$}\tabularnewline
$10.000000$ & \textit{c} & \multicolumn{2}{c}{$4.730$} & \textit{${C}$} & \multicolumn{2}{c}{$7.501$}\tabularnewline
$11.252920$ & \textit{d} & \multicolumn{2}{c}{$5.280$} & \textit{${D}$} & \multicolumn{2}{c}{$7.091$}\tabularnewline
$12.203171$ & \textit{e} & \multicolumn{2}{c}{$6.245$} & \textit{${E}$} & \multicolumn{2}{c}{$6.245$}\tabularnewline
$13.000000$ & \textit{f} & \multicolumn{2}{c}{} & ${F}$ & \multicolumn{2}{c}{}\tabularnewline
$16.000000$ & \textit{g} & \multicolumn{2}{c}{} & \textit{G} & \multicolumn{2}{c}{}\tabularnewline
\end{tabular}
\end{table}

\newpage{}

\newpage{}

\newpage{}

\end{document}